\documentclass[12pt,a4paper]{report}

\usepackage{geometry}
\usepackage{bbm,yfonts}
\usepackage[nottoc]{tocbibind} 
\usepackage{xspace}	

\usepackage{ifthen}	

\usepackage{titlesec}					

\titleformat{\section}{\vspace*{1em}\normalfont\Large\bfseries}
{\thesection}{1em}{\MakeUppercase} 
\titlespacing*{\section}{0em}{0em}{7em}   

\titleformat{\subsection}{\normalfont\large\bfseries}
{\thesubsection}{1em}{\MakeUppercase} 
\titlespacing*{\subsection}{0em}{7em}{3em}   

\titleformat{\subsubsection}{\normalfont\normalsize\bfseries}
{\thesubsubsection}{1em}{\MakeUppercase} 
\titlespacing*{\subsubsection}{0em}{7em}{3em}   

\usepackage{amsmath,amssymb}
\usepackage[bookmarks=false,colorlinks=true]{hyperref}
\newcommand{\be}{\begin{equation}}
\newcommand{\ee}{\end{equation}}

\newcommand{\bea}{\begin{eqnarray}}
\newcommand{\eea}{\end{eqnarray}}

\def\bu{{\bar u}}

\newcommand{\ba}{\begin{array}}
\newcommand{\ea}{\end{array}}

\newcommand{\ch}{{\tt h}}

\usepackage{tikz}

\begin{document}
	
\thispagestyle{empty}
\vspace*{0.5cm}
\begin{center}
{\large A.I.ALIKHANYAN NATIONAL SCIENCE LABORATORY\newline
(YEREVAN PHYSICS INSTITUTE) }\\
\vspace*{5cm}
{\bf\Large Hovhannes Shmavonyan}\\
\vspace{1cm}
{\bf\Large Superintegrable deformations of oscillator and Coulomb systems}\\
\vspace{1cm}
{\Large PhD Thesis}\\
\vspace{1cm}
{\large 01.04.02 - Theoretical Physics}\\
{\large  Adviser: Armen Nersessian}\\
\vspace{5cm}
{\large Yerevan-2019}
\end{center}

\newpage

\tableofcontents

\newpage

\section*{Acknowledgments}

The success and final outcome of this thesis required a lot of guidance and assistance from many people and I am extremely privileged to have got this all along the completion of my thesis. All that I have done is only due to such supervision and assistance and I would not forget to thank them.

First of all I have to mention my Academic supervisor, Professor Armen Nersessian. I would like to thank him for his constant motivation,  guidance  and encouragement during this project.  It would never have been possible for me to take
this work to completion without his incredible support.

I am greatly thankful  to the permanent co-author of many papers   Professor Tigran Hakobyan.   My grateful thanks are
due to him for his suggestions and guidance at every stage of my
research work. I would like to express my gratitude also to the co-authors from e Joint Institute for Nuclear Research
(Dubna). Namely, I would like to express my sincere thanks to Professor   Evgeny Ivanov and Stepan Sidorov.
I would like also to express many thanks to Professor Ulf Meissner and  Dr. Akaki Rusetsky from Bonn University.

I am so much grateful to  Volkswagen Foundation, Foundation for Armenian Science and Technology (FAST), Science Committee of Armenia, Armenian National Science and Education Fund (ANSEF) as well as  International Centre for Theoretical Physics (ICTP) for their support during my research.

My greatest thanks go to  A. Alikhanyan National Laboratory, namely the stuff of the theoretical department. My special thanks to Ruben Manvelyan, Hrachya Asatryan, David Karakhanyan and Armen Allahverdyan.

I would like also to thank I. Javakhishvili High Energy Physics Institute (Tbilisi), as a leading organization.

Last but not least I thank the reviewers of my thesis  Rubik Poghossian,  Ruben Mkrtchyan, Hrachya Babujyan and Levon Mardoyan
who carefully read it and made many useful comments.

It is my great pleasure to acknowledge the contribution of
respectable personalities without whose support it was not possible for
me to complete my work in a meaningful way. At last I wish to thank many other people whose names are not
mentioned here but this does not mean that I have forgotten their help.

\newpage

\chapter{Introduction}
\setcounter{equation}{0}
\setcounter{section}{0}

Integrable models are crucial for modern theoretical and mathematical physics.
 Due to the fact that different physical phenomena can have similar mathematical description, exactly solvable models can be used in many different areas.
 One can see that using these models huge amount of (both macroscopic and microscopic) physical phenomena can be described.
 Moreover integrable models can have applications even in other disciplines, due to the fact that system of integrable differential equations arise in other subjects e.g. mathematics, computer science, biology etc.
The thesis is devoted to superintegrable extensions of oscillator and Coulomb models with an inverse square potential.
Integrable models with inverse square potential are studied for few decades. Due to this fact they are well studied and there are many important results about these systems.
 Namely the Calogero-model has unique properties and due to that nowadays this is an important system in mathematical physics.
On the other hand projective spaces have also interesting properties .
Due to the fact that they are maximally symmetric spaces it is important to consider physical systems on these spaces.
 Unfortunately these two branches of mathematical physics are disconnected now.
Complex analogs of Calogero model are not studied well and attempts to construct complexification of Calogero-like models haven't succeeded yet.

Possible applications of this work should be highlighted. Namely in condensed matter physics  models on complex projective spaces are strongly related with the quantum Hall effect.
 In High energy physics their role cannot be overestimated. These systems can be viewed as simplified toy models for field theoretical complicated models in high energy research.
Our particular example of  Calogero model is an example of conformal mechanics. It is well known that conformal symmetry has a crucial role in modern high energy research.
In this context supersymmetrization of these systems is also important. Moreover Calogero-like  models are strongly related with $AdS_2/CFT_1$ correspondence \cite{ads}.
Particularly four-dimensional Hall effect can be related with the systems in $\mathbb{CP}^3$  \cite{hall}.

This chapter is devoted to the basic introductory information about Hamiltonian formalism, K\"ahler manifolds, and supersymmetric mechanics, which is widely used in the current work.

In {\it Section 1.1} we discuss the basic examples of maximally superintegrable models (oscillator, Coulomb). Then we consider the Hamiltonian approach for the interaction with an external magnetic field. Finally we present important information about action-angle variables.

In {\it Section 1.2} we present information about K\"ahler manifolds and consider the examples of maximally symmetric K\"ahler spaces which will be used in the next parts.

In {\it Section 1.3}  We focus on  the Hamiltonian approach for the supersymmetric classical mechanics, since the last chapter of this thesis is devoted to that subject.

The second chapter of this thesis is based on the three articles \cite{shmlob,shm2,shmhol}. The material of the third chapter can be found in \cite{shmsw}. The fourth and the fifth parts are based on \cite{shmros} and on another paper which  is in progress and will be published soon and is done  with coauthors Armen Nersessian, Evgeny Ivanov and Stepan Sidorov .

\section{Integrability and Hamiltonian \newline mechanics}

$N$-dimensional mechanical system (system with N degrees of freedom) will be called, {\sl integrable} if it has N  mutually commuting and functionally independent constants of motion\cite{Arnold,perelomov}.
 In addition to these constants of motion the system can have additional  ones. In that case we will say that the system is {\sl superintegrable}.
Particularly if $N$-dimensional mechanical system has $2N-1$ functionally independent constants of motion it will be called {\sl maximally superintegrable}.
In case the system has N+1  conserved quantities it is called {\sl minimally superintegrable}.\
 While integrable models possess separation of variables in one coordinate system, superintegrability guarantees separation of variables in many coordinate systems.
 For example two-dimensional oscillator is superintegrable, which allows us to separate variables in Cartesian and polar coordinates. In classical mechanics maximal superintegrability guarantees the closeness of trajectories.
Quantum mechanically energy spectrum of integrable models depend on $N$ quantum numbers. If the system has $K$ additional  conserved quantities (superintegrable) energy spectrum depends on $N-K$ quantum numbers.
 For maximal superintegrability we have that the energy spectrum contains only one quantum number. So we can conclude that superintegrability leads to  degeneracy of energy spectrum in quantum level.
 Well known examples of maximally superintegrable models are $N$-dimensional Coulomb system and $N$-dimensional harmonic oscillator. Another important but recently discovered model is the Calogero model which is discussed in this thesis later.

\subsection{Oscillator}
Harmonic oscillator is well known and maybe the most important example of a maximally superintegrable  model \cite{fradkin65}.
 Due to its simplicity and unique properties it plays a crucial role in all areas of modern physics.
Techniques developed for harmonic oscillator can be used in all areas of physics, e. g. in  condensed matter physics and quantum field theory.
There are several extensions and generalizations of harmonic oscillator, namely non -harmonic oscillator, oscillator with additional potential. In current work oscillator is the key system.
We will consider superintegrable generalizations of oscillator in curved spaces, for instance on spherical and pseudospherical spaces, Euclidean and projective complex manifolds.
Extensions with additional potential will also be discussed, namely we will focus on superintegrable generalizations with an inverse square potential.
Before discussing this generalizations it is important to discuss the standard harmonic oscillator.

$N$-dimensional harmonic oscillator is the system with quadratic potential and standard Poisson brackets.
\be
H=\sum_{i=1}^{N}\frac{p_i^2}{2}+\frac{\omega^2x_i^2}{2},\qquad \{p_i,x_j\}=\delta_{ij},\quad \{p_i,p_j\}=\{x_i,x_j\}=0 \label{oscham}
\ee

Since the system has rotational symmetry angular momentum is conserved. As is known the symmetry these conserved quantities is $SO(N)$.
\be
L_{ij}=p_ix_j-p_jx_i, \qquad \{L_{ij},L_{kl}\}=\delta_{il}L_{kj}-\delta_{kj}L_{il}+\delta_{jl}L_{ik}-\delta_{ik}L_{jl}
\ee
Moreover oscillator has additional conserved quantities quadratic on momenta
\be
I_{ij}=p_{i}p_{j}+\omega^2x_ix_j
\ee
This is the so called Fradkin tensor and together with angular momentum the system of conserved quantities of harmonic oscillator has $U(N)$ symmetry.
We have to highlight that there are functional relations between these conserved quantities and due to that the number of functionally independent conserved quantities is $2N-1$.
$U(N)$ symmetry is more obvious if we introduce complex quantities, which can be viewed as classical analog of creation and annihilation operators.
\be
u=\frac{p_i+ix_i}{\sqrt2}, \quad \bar{u}=\frac{p_i-ix_i}{\sqrt2} ,\quad \{\bar{u}_i,u_j\}=i\delta_{ij}
\ee
In these coordinates Hamiltonian will have manifest $U(N)$ invariance and we can write down conserved quantities as generators of this symmetry.

\be
H=\sum_{i=1}^N u_i\bar u_i, \qquad M_{ij}=u_i\bar u_j
\ee
Energy spectrum can be written down and as was mentioned it depends only one quantum number ($n$) \cite{landau}.
\be
E=\hbar\omega(n+\frac{N}2)
\ee

\subsection{Coulomb problem}
Coulomb problem is another well known example of superintegrable model. It plays an important role in celestial mechanics and that's why it is known for few centuries.
Symmetries of this system are also known for centuries namely the angular momentum conservation (Kepler's second law) and Laplace-Runge-Lenz or simply Runge-Lenz vector conservation.
In this thesis we again consider superintegrable extensions of a Coulomb system on spherical and pseudospherical spaces with an inverse square potential.
Investigation of complex generalizations of Coulomb system is quite challenging and are not discussed by us, since Coulomb problem has orthogonal symmetries,
while complex structure requires unitary symmetry.

The Hamiltonian  of N-dimensional Coulomb problem is  as follows
\be
H=\sum_{i=1}^{N}\frac{p_i^2}{2}-\frac{\gamma}{r},   \qquad  r=\sqrt{\sum_{i} x_i^2}
\ee
Poisson brackets are the same as given in\eqref{oscham}. Again we have $SO(N)$ rotational  symmetry and due to that angular momentum is a conserved quantity.
\be
L_{ij}=p_ix_j-p_jx_i, \qquad \{L_{ij},L_{kl}\}=\delta_{il}L_{kj}-\delta_{kj}L_{il}+\delta_{jl}L_{ik}-\delta_{ik}L_{jl}
\ee
 We have additional constants of motion, which is called Runge-Lenz vector
\be
A_i=L_{ij}p_j+\frac{\gamma x_i}{r}
\ee
Together with angular momentum the system of conserved quantities has $SO(N+1)$ symmetry \cite{fradkin67}.
 $N$-dimensional Coulomb problem can be obtained via reduction from free particle moving on $N+1$ dimensional sphere
Since the symmetry of this system is obviously $SO(N+1)$, the symmetry of $N$-dimensional Coulomb problem is not surprising.

Again the number of independent constants of motion is $2N-1$. So the $N$-dimensional Coulomb system is maximally superintegrable.
So the energy spectrum depends on one quantum number
\be
E=-\frac{\gamma}{2\hbar^2(n+\frac{N-3}{2})^2}
\ee

\subsection{Interaction with external magnetic field}

In this chapter we see that in many cases inclusion of an external constant magnetic field does not violate integrability properties. For this purpose we can discuss the Hamiltonian approach for systems interacting with an external magnetic field.
Hamiltonian formalism allows to introduce magnetic field without changing the form of the Hamiltonian.
The price we pay is the modification of the symplectic structure \cite{SupHam}.
 Here we consider this approach and from now on we will introduce magnetic field via modification of the basic Poisson brackets.

Consider particle moving on $N$-dimensional  Riemannian manifold. Hamiltonian and basic non-zero Poisson brackets are as follows.
\be
H=\frac12g^{ab}p_ap_b+U(q) ,\qquad \{p_a,q^b\}=\delta_a^b
\ee
One can additionally include an  external magnetic field. As is known this interaction modifies the momenta (minimal coupling)
\be
H=\frac12g^{ab}(p_a-A_a)(p_b-A_b)+U(q) ,
\ee
where $A_a$ is the magnetic vector potential.
 It is worth to mention that although for general Riemannian manifold with non-trivial topology introduction of magnetic potential is not possible globally, it is at least possible locally (for a chosen chart).
We can redefine momenta and introduce new (non-canonical) ones $\pi_a=p_a-A_a$.
 In terms of these momenta Hamiltonian will have the usual form, but the basic Poisson brackets i. e. the symplectic structure will be modified.
\be
H=\frac12g^{ab}\pi_a\pi_b+U(q) ,\qquad \{\pi_a,q^b\}=\delta_a^b, \qquad \{\pi_a,\pi_b\}= F_{ab}
\ee
where  $F_{ab}$  consists of the components of magnetic strength.
\be
F_{ab}=\partial_a A_b-\partial_b A_a
\ee

\subsection{Action-angle variables}

As was mentioned integrable system has $N$ functionally independent constants of motion. In this case we can choose these variables to be canonical momenta.
 They will be called action variables. Moreover one can compute canonically conjugate coordinates corresponding to these variables, which will be called angle variables.
 This approach is very important in the theory of integrability and it is one of the most effective ways to deal with integrable models.
After change of variables it is obvious that Hamiltonian will also depend only on action variables, because it for closed systems Hamiltonian is always a conserved quantity and so there is a functional relation between action variables and the Hamiltonian.
So the angle variable in this context is  cyclic.
It is important to highlight that action angle variables are highly effective even for exactly solvable field theories, such as sin-Gordon theory and non-linear Schr\"odinger equation \cite{takht}.
On the other hand quantum mechanically these variables can be used in Bohr-Sommerfeld quantization. Moreover due to the adiabatic invariance these variables can be used in perturbation theory if one considers system which is a small perturbation on an integrable system.
Another crucial fact about these variables is that they can indicate whether two integrable models are equivalent or not.

Now let us discuss another important result related to the action angle variables, namely the Arnold-Liouville theorem \cite{Arnold} . Suppose that we have an integrable system and we fixed the conserved quantities.
 Then  on a phase space the motion is restricted on an $N$-dimensional manifold ($M$).
If this manifold is connected than it is diffeomorphic to , $M\cong\mathbb{R}^p\times T^q$  where $p+q=N$ and $p$ is the number of  non-compact coordinates, while $q$ is  to the compact coordinates.
 We will manly focus on compact motion so we can write that the manifold is diffeomorphic to the $N$-dimensional torus $M\cong T^N$. This theorem can be viewed as a geometric interpretation of action angle variables.
Action variables can be viewed as the conserved quantities which are fixed, while the angle variables are the coordinates on the torus. In this context these mutually commuting constants of motion are sometimes called Liouville integrals of motion.
In this context superintegrability also has an interesting geometrical interpretation. Each additional constant of motion puts restriction an the torus and reduces the dimensionality by one. Incase of the maximal superintagrability we have that the dimension of the $N$-dimensional torus is reduced by $N-1$ and consequently it is diffeomorphic to $S^1$, since it is the only one-dimensional compact manifold.
 This corresponds  to closeness of the classical trajectory.
Action and angle variables can be found via computing the following relations
\be
I_a=\frac{1}{2\pi}\oint p_adq_a, \qquad \Phi_a=\frac{\partial S}{\partial I_a}
\ee
As was mentioned they are canonically conjugated ($\{I_a,\Phi_a\}=\delta_{ab}$) and due to that canonical quantization is straightforward \cite{phdSagh}
\be
\hat{I}_a \Psi_a(\Phi)=I_a \Psi_{\Phi}, \quad \hat{I}_a=-i\hbar\frac{\partial}{\partial \Phi_a},\quad  \Psi=\frac{1}{(2\pi)^{N/2}}e^{-in_a \Phi_a},   \quad  I_a=\hbar n_a
\label{quantact}
\ee

It will be beneficial to briefly consider the simplest example of one-dimensional oscillator. Hamiltonian can be chosen as an action variable, so the energy levels will correspond to $M$.
It is obvious that energy levels on the space correspond are circles , which can be considered as one-dimensional torus ($T^1=S^1$).
 Quantum mechanically solution in energy picture corresponds to canonical quantization via action angle variables  \eqref{quantact}.

\section{K{\"a}hler manifolds}

K{\"a}hler manifolds play an important role in modern theoretical physics and mathematics \cite{SupHam,Kob2,Freedman}.
In algebraic geometry a class of algebraic varieties are K{\"a}hler manifolds.
In supersymmetry the target space can be sometimes viewed as a K{\"a}hler manifold.
Moreover,  in string theory some compactification schemes are based on K{\"a}hler manifolds , e.g  Calabi-Yau manifolds is a compact K{\"ahler} manifold with vanishing first Chern class, that is also Ricci flat.
 We will mainly focus on the role of K{\"a}hler spaces in Hamiltonian mechanics.  K{\"a}hler manifolds  have three mutually compatible structures, namely complex structure, Riemannian structure and symplectic structure.
K{\"a}hler manifold is a private case of more general Hermitian manifold ($g_{a\bar{b}}dz^a d\bar{z}^b $). For any  Hermitian metric one can define a 2-form

\be
\omega=ig_{a\bar{b}} dz^a\wedge d\bar{z}^b
\ee
This 2-form is called a fundamental form.
 Hermitian manifold is called {\sl  K{\"a}hler} if this 2-form is symplectic (closed and non-degenerate). This requirement  is quite restrictive and due to that K{\"a}hler metric can be written as a second derivative of a function called  K{\"a}hler potential.

\be
g_{a \bar{b}}=\frac{\partial^2K(z,\bar{z})}{\partial z^a \partial \bar{z}^b}
\ee
It is worth to mention that  this function is not uniquely determined and one can add holomorphic or antiholomorphic function to it.

 Due to the symplectic structure K{\"a}hler manifolds have natural symplectic structure and can be equipped with Poisson brackets.
\be
\{f,g\}_0=ig^{a\bar b}\Big(\frac{\partial f}{\partial z^a} \frac{\partial g}{\partial \bar{z}^b}-\frac{\partial g}{\partial z^a} \frac{\partial f}{\partial\bar{z}^b}\Big), \quad g^{a\bar{ b}}g_{\bar{b}c}=\delta^a_c
\ee
Since the symplectic structure relates functions (Hamiltonian) and  vector fields (Hamiltonian vector fields), we can introduce functions, which generate Killing vector fields.
\be
\bold{V}_{\mu}=\{h_{\mu},\}_0=V_{\mu}^a\frac{\partial}{\partial z^a}+\bar{V}_{\mu}^{\bar{a}}\frac{\partial}{\partial \bar{z}^a}, \qquad V^{a}_{\mu}=-ig^{a\bar{b}}\partial_{\bar{b}}h_{\mu}
\ee
Such functions will be called Killing potentials. Using Killing Equations one can derive restrictions on Killing potentials. They should be real and they have to fulfill the following equation.
\be
\frac{\partial^2 \ch_\mu}{\partial z^a \partial z^b} -
\Gamma^c_{ab}\frac{\partial \ch_\mu}{\partial z^c}=0
\ee

  These functions are extremely useful for studying systems on K{\"a}hler manifolds in presence of a constant magnetic field.
Due to the fact that any 2-form is closed in two (real) dimensions, one-dimensional orientable complex manifold (Riemann surface) can always be equipped with a K{\"ahler} structure.
Many components of the Christoffel symbols and Riemann tensor will vanish.
\be
\Gamma^a_{bc}=g^{a\bar{d}}g_{b\bar{d},c}  ,  \qquad  R^a_{bc\bar{d}}=-(\Gamma^a_{bc})_{,\bar{d}}
\ee

In this thesis some superintegrable  models on maximally symmetric K{\"ahler} manifolds are discussed, namely on $\mathbb{C}^N$ (complex Euclidean space) and $\mathbb{CP}^N$ (complex projective space)

\subsection{$\mathbb{C}^N$ as a K{\"a}hler manifold }

The metric of the N-dimensional complex Euclidean space is well known.
\be
ds^2=dzd\bar{z} , \qquad g_{a\bar{b}}=\delta_{a\bar{b}}.
\ee
It is easy to note the K{\"a}hler potential and the symplectic structure is as follows
\be
K(z,\bar{z})=z\bar{z}, \qquad \omega=-idz\wedge d\bar{z}, \qquad \{ z^a,\bar{z}^b \}_{0}=i\delta^{a\bar{b}}
\ee
will lead to this well known metric. All the components of Christoffel symbols and Riemann tensor vanish. Finally we present the results for Killing potentials and corresponding Killing vector fields.
\be
h_{a\bar{b}}=\bar{z}^{a}z^b,\qquad \bold{V}_{a\bar{b}}=-i(z^b\partial_{a}+\bar{z}^a\partial_{\bar b})
\ee
\be
h_{a}^{+}=\bar{z}^a, \qquad    \bold{V}^{-}_{a}=-i\partial_{a} , \qquad
h_{a}^{-}=z^a, \qquad    \bold{V}^{+}_{a}=-i\partial_{\bar{a}}
\ee
$V_{a\bar{b}}$ vector fields generate rotations, while $V^{-}_{a}$ and  $V^{+}_{a}$ are the generators of translation. Although $h_{a\bar{b}}$, $h^+_a$ and $h_{a}^{-}$ are not real, one can take real combinations using these functions.
The number of real Killing potentials is $N(2N+1)$, so as is mentioned $\mathbb{C}^N$ is maximally symmetric space.

 \subsection{$\mathbb{CP}^N$ as a K{\"a}hler manifold}

The $N$-dimensional complex projective space is a space of complex rays in the $(N+1)$-dimensional complex Euclidian space
$(\mathbb{C}^{N+1},\; \sum_{i=0}^N du^i d\bar u^i)$, with $u^i$ being homogeneous coordinates of the complex projective space.
 Equivalently, it can be defined as the quotient $\mathbb{S}^{2N+1}/U(1)$, where  $\mathbb{S}^{2N+1}$ is the $(2N+1)$-dimensional sphere  embedded
 in $\mathbb{C}^{N+1}$ by the constraint $\sum_{i=1}^N u^i\bar u^i =1$.
One  can solve the latter by introducing locally ``inhomogeneous" coordinates $z^a_{(i)}$
\be
z^a_{(i)}=\frac{u^a}{u^i},\qquad{\rm with}\quad  a\neq i, u^i\neq 0.
\ee
Hence,  the full complex projective space can be covered by  $N+1$ charts marked by the indices $i=0,\ldots, N$, with the following
transition functions on the intersection  of $i$-th and $j$-th charts:
\be
z^a_{(i)}=\frac{z^a_{(j)}}{z^i_{(j)}}.
\label{transition}\ee
Let us endow $\mathbb{C}^{N+1}$ with the canonical Poisson brackets $\{u^i,{\bar u}^j\}=\imath\delta^{i\bar j}$, and
define, with respect to them, the $u(N+1)$ algebra formed by the generators
\be
h_{i\bar j}= {\bar u}^i u^j\,.
\ee
Reducing the manifold $\mathbb{C}^{N+1}$ by the action of the $U(1)$ group with the generator  $h_0= \sum_{i=0}^N u^i{\bar u}^i$,
 we  arrive at the $SU(N+1)$-invariant Ka\"hler structure  defined by the Fubini-Study metrics
 \begin{equation}
  g_{a\bar b}dz^ad{\bar z}^b =\frac{\partial^2\log(1+z\bar z)}{\partial z^a\partial\bar z^b}dz^ad{\bar z}^b =  \frac{dzd{\bar z}}{1+z \bar z}-\frac{({\bar z}dz) (zd\bar z)}{(1+z \bar z)^2 }, \quad K=\log(1+z\bar z).
 \label{FS} \end{equation}
This metrics is obviously invariant under the passing from one chart to another. Hence, we can omit the  indices marking charts and assume,
without loss of generality,  that we are dealing with  $0$-th chart, so that the indices $a,b,c$ run from $1$ to $N$.

Being K\"ahler manifold, the complex projective space is  equipped with the Poisson brackets $\{z^{a},{\bar z}^b\}_0=\imath g^{ a\bar b}$,
where $g^{ a\bar b}=(1+z\bar z)(\delta^{ a\bar b}+z^a \bar z^b)$ is the inverse Fubini-Study metrics.
 The $su(N+1)$ isometry of $\mathbb{CP}^N$ is generated by the  holomorphic Hamiltonian vector fields defined as the
  following momentum maps  (Killing potentials).
\be
h_{a\bar b}=\frac{\bar{z}^a z^b}{1+z\bar z},\qquad h_a^{-}=\frac{{\bar z}^a}{1+z\bar z},\qquad h_a^{+}=\frac{ z^a}{1+z\bar z}
\label{suKilling}\ee
Like for the Euclidean case the number of independent Killing vector fields indicates that this space is again maximally superintegrable.
Finally we can compute the components of Christoffel symbol and Riemann tensor.
\be
\Gamma^{a}_{bc}=-\frac{\delta^a_b\bar z^c+\delta^a_c \bar z^b}{1+z\bar z},
\quad
R_{a\bar b c \bar d}=g_{a\bar b}g_{c\bar d}+g_{c\bar{b}}g_{a\bar d},
\ee

\section{Supersymmetric mechanics}

Now we consider the Hamiltonian approach to the classical supersymmetric mechanics. Although initially supersymmetry was introduced in quantum field theory, further development of supersymmetry showed that supersymmetric mechanical models themselves are also interesting for modern physics.
First of all, since mechanics can be viewed as one-dimensional field theory this models can be  viewed as simple "toy" models for supersymmetric field theories and superstring theory.
 But as is known there is no any evidence for existence of supersymmetry in high energy physics yet.
In contrast to this supersymmety can be found in many physical quantum mechanical phenomena. For instance, the well known Landau problem can be viewed as a supersymmetric model \cite{superqm}.

The last chapter of this thesis is devoted to supersymmetric generalizations of some integrable models on K\"ahler manifolds so it is useful to present basic information about supersymmetric mechanics.
It should be highlighted that K\"ahler structures play crucial role in supersymmetric field theoretical models and for instance supersymmetric Lagrangians can be composed out of chiral superfields using the K\"ahler potential\cite{Freedman}.
 First of all we should extend the notion of Poisson brackets for odd Grassmann quantities.
This structure will be called supersymplectic structure.
First of all Poisson brackets for two odd-Grassmann quantities is symmetric and is analogous to anticommutator for operators in quantum mechanics\cite{SupHam}.
 Moreover Jacobi identity must be also extended.

\be
\{f^{(a)},g^{(b)}\}=-(-1)^{ab}\{g^{(b)},f^{(a)}\}
\ee
\be
 (-1)^{ac}\{f^{(a)},\{g^{(b)},h^{(c)}\}\}+(-1)^{ab} \{g^{(b)},\{h^{(c)},f^{(a)}\}\}+(-1)^{bc}\{h^{(c)},\{f^{(a)},g^{(b)}\}\}=0
\ee
where $a,b,c$ take values $0$ for even Grassmann variables and $1$ for odd Grassmann variables.

So we say that we have ${\cal N}=n$ supersymmetric mechanics if there exist $n$ odd-Grassmann variables $Q_i$ (supercharges), which satisfy the following relation
\be
\{Q_i,Q_i\}=\delta_{ij}H, \qquad \{Q_i,H\}=0
\ee

Since the field theoretical context is that here we deal with a one-dimensional field theory   our superspace consists of time and  Grassmann variables  ($t, \theta_i$) , which can be called supertime.
It is obvious that this supersymmetry will be the ${\cal N}=n,d=1$ SuperPoincare algebra.

 Consider the simplest  example, namely the ${\cal N}=1$ supersymmetric mechanics. In this case any odd Grassmann variable can be chosen and its square can be identified with the Hamiltonian
Since this case is quite trivial, it is not very interesting.

The next example is ${\cal N}=2$ supersymmetric mechanics. In this case supercharges can be redefined ($Q^{\pm}=(Q_1\pm iQ_2)/\sqrt{2}$) and the supersymmetric algebra will have the following form
\be
\{Q^{+},Q^{-}\}=H,  \qquad \{Q^{+},Q^{+}\}= \{Q^{-},Q^{-}\}=0
\ee
One can see that, if we discuss particle on a Riemannian manifold, supercharges and the symplectic structure can be chosen in the following form
\be
Q^{\pm}=(p_a\pm iW_{,a})\eta^a_{\pm}, \qquad \omega=dp_a\wedge dx^a+\frac{1}{2}R_{abcd}\eta^a_{+} \eta_{-}^b dx^c\wedge dx^d +g_{ab} D\eta_{+}^a\wedge D\eta_{-}^{b}
\ee
where $D\eta^a_{\pm}=d\eta^a_{\pm}+\Gamma^a_{bc}\eta^b_{\pm}dx^c$ and $W$ is called  superpotential. One can compte the Hamiltonian
\be
H=\frac12 g^{ab}(p_ap_b+W_{,a}W_{,b})+W_{a;b}\eta^a_{+}\eta^b_{-}+R_{abcd}\eta^a_{-}\eta^{b}_{+}\eta_{-}^c\eta^d_{+}
\ee
We should highlight that introduction of the external magnetic field breaks the standard ${\cal N}=2$ supersymmetry and later we will call this  "weak" supersymmetry.
\be
\{Q^{+},Q^{-}\}=H+iF_{ab}\eta_{+}^a\eta_{-}^{b},  \qquad \{Q^{\pm},Q^{\pm}\}=F_{ab}\eta_{\pm}^{a}\eta_{\pm}^{b}
\ee
The last part of this thesis is devoted to discussion of higher supersymmetries (${\cal N}>2$).

\newpage

\newpage

\chapter{Deformations of oscillator/Coulomb\newline systems (holomorphic factorization) }
\setcounter{equation}{0}
\setcounter{section}{0}
\section{Introduction}

This chapter is  based on three papers\cite{shmlob,shm2,shmhol} written with Armen Nersessian and Tigran Hakobyan.

The $N$-dimensional oscillator and Coulomb problem play special role among other integrable systems by many reasons.
One of the main reasons, due to which these models continue to attract permanent interest during the last centuries,
 is their maximal superintegrability.  Another important example of superintegrable system is Calogero model.
The rational  Calogero model and its generalizations, based on arbitrary Coxeter root systems,  are  highlighted  among  the non-trivial unbound
superintegrable systems.
This property was established for the classical  \cite{woj83,algebra,algebra1} and quantum \cite{kuznetsov,gonera}
rational  Calogero model, which is described by the Hamiltonian \cite{calogero1,calogero2}
\be
H_0=
 \sum_{i=1}^N \frac{p_i^2}{2} + \sum_{i<j} \frac{g^2}{(x_i-x_j)^2}.
\label{calo}
\ee
Its generalization, associated
with an arbitrary finite Coxeter group, is defined by the Hamiltonian \cite{algebra,algebra1}
\be
\label{coxeter}
H_0=
 \sum_{i=1}^N \frac{p_i^2}{2} + \sum_{\alpha\in\mathcal{R}_+}
\frac{g^{2}_\alpha (\alpha\cdot\alpha)}{2(\alpha\cdot x)^2}.
\ee
Let us mention that the Coxeter group is described
 as a finite group  generated
 by a set of orthogonal reflections  across the hyperplanes $\alpha\cdot x=0$
 in the $N$-dimensional Euclidean space,
where the vectors $\alpha$ from the set $\mathcal{R}_+$  (called the system of positive roots)
 uniquely characterize the reflections.
The coupling constants $g_\alpha$ form a reflection-invariant discrete function.
The original Calogero potential in \eqref{calo} corresponds to the $A_{N-1}$ Coxeter system with
the positive roots, defined in terms of the standard basis by $\alpha_{ij}=e_i-e_j$
for $i<j$. The reflections become the coordinate
permutations in this particular case.

 The oscillator and Coulomb systems admit obvious separation
of the radial and angular variables, which is useful to formulate in terms of conformal algebra $so(1,2)\equiv sl(2,\mathbb{R})$   defined by
the following Poisson bracket relations
\be
 \{  \mathcal{H}_0 , \mathcal{D}\}=2 \mathcal{H}_0  ,\qquad\{
\mathcal{H}_0 , \mathcal{K}\}=\mathcal{D},\qquad  \{ \mathcal{K},\mathcal{D}\}= -2\mathcal{K}.
\label{ca}
\ee
The generators $\mathcal{H}_0, \mathcal{K},\mathcal{D}$ could be  identified, respectively, with the Hamiltonian of some $N$-dimensional mechanical system,
and with the generators  of conformal boost and dilatation.This system is usually called "conformal mechanics", and $so(1,2)$ symmetry appears as its dynamical symmetry \cite{fubini}.
Introduce the effective "radius" and conjugated momentum,
 \be
r=\sqrt{2\mathcal{K}},\qquad p_r=\frac{\mathcal{D}}{\sqrt{2\mathcal{K}}},\qquad \{p_r, r\}=1,
\ee
and  define a Casimir of conformal algebra
\begin{equation}
{\cal I}=2{\cal H}_0{\cal K}-\frac12 {\cal D}^2:
\qquad
\{\mathcal{I},\mathcal{H}_0\}=\{\mathcal{I},\mathcal{K}\}=\{\mathcal{I},\mathcal{D}\}=0.
\label{ml}
\end{equation}
It  is obviously   a constant of motion  independent on radial coordinate and momentum, and thus
could be expressed via appropriate   angular coordinates $\phi_a $ and
canonically conjugate momenta $\pi_a $ which are independent on radial ones: ${\cal I}={\cal I}(\phi_a,\pi_a)$.
 In these terms the generators of conformal algebra read:
\begin{equation}
 {\cal H}_0=\frac{{p}^2_r}{2}+
\frac{{\cal I}}{r^2},
\qquad
 {\cal D}=r p_r ,
\qquad
{\cal K}=\frac{r^2}{2}.
\label{so2}
\end{equation}
Hence, such a separation of angular and radial parts could be defined for any system with dynamical conformal symmetry, and for those with additional potentials be function of  conformal boost $\mathcal{K}$.
In particular, such a generalized oscillator and Coulomb systems assume adding of potential
\be
V_{osc}=\omega^2\mathcal{K},\qquad V_{Coul}=-\frac{\gamma}{\sqrt{2\mathcal{K}}},
\ee
 so that their Hamiltonian takes the form
 \be
\mathcal{H}_{osc/Coul}=\frac{p^2_r}{2}+\frac{\mathcal{I}}{r^2}+V_{osc/Coul}(r).
\label{planarHam}\ee
Well-known  generalizations  of oscillator and Coulomb systems to $N$-dimensional spheres and two-sheet hyperboloids (pseudospheres) \cite{higgs,leemon} can be described in a similar way.

 In Refs.~\cite{hlnsy,rapid} a separation of "radial" and "angular" variables has been used for constructing the integrable deformations of oscillator and Coulomb systems (and of their (pseudo)spherical generalizations) via replacement of the spherical part of pure oscillator/Coulomb
 Hamiltonians (quadratic casimir of $SO(N)$ algebra)  by some other integrable system
formulated  in terms of the action-angle variables.
Analyzing these deformations in terms of action-angle variables, it was found that they are superintagrable iff
the spherical part has the form
\be
\mathcal{I}=\frac12\left(\sum_{a=1}^{N-1}k_a I_a + c_0 \right)^2
\label{angular0}
\ee
with $c_0$ be arbitrary constant and $k_a$ be rational numbers.
Moreover, it was demonstrated, by the use of the results of Ref.~\cite{flp}, that the angular part of rational Calogero model   belongs to this set of systems. Thus, it was concluded that rational Calogero model with Coulomb potential (Calogero-Coulomb system) is superintegrable system. Besides,  superintegrable generalizations of the rational Calogero models with oscillator/Coulomb potentials on the $N$-dimensional spheres and two-sheet hyperboloids have been suggested there. The explicit expressions of their symmetry generators and respective algebras have been given in
Refs.~\cite{runge,rapid1,rapid2}. An integrable  two-center  generalization  of the  Calogero-Coulomb systems (and those in the presence of Stark term, which was called Calogero-Coulomb-Stark model) has been also revealed \cite{Calogero-Stark}.

The goal of this chapter is to present "holomorphic factorization" to the superintegrable generalizations of oscillator and Coulomb systems on $N$-dimensional Euclidean space, sphere and  two-sheet hyperboloid (pseudosphere).
For this purpose we parameterize the phase spaces of that system by the
 complex variable $Z=p_r+\imath\sqrt{2\mathcal{I}}/r$
  identifying the radial phase subspace with the Klein model of
Lobachevsky plane, and by the
 complex variables $u_a=\sqrt{I_a}{\rm e}^{\imath\Phi_a}$  unifying action-angle variables of the angular part of the systems.
We formulate, in these terms, the constants of motion of the systems under consideration and  calculate their  algebra.
Besides, we  extend to these systems the known oscillator-Coulomb duality transformation.

This chapter is organized as follows:

In {\it Section 2.2} we review the classical properties of  Tremblay-Turbiner-Winternitz and Post-Wintenitz systems
and their relation  with $N$-dimensional rational Calogero model with oscillator and Coulomb potentials,
paying special attention to their hidden symmetries. Then we show that combining the radial coordinate
and momentum in a single complex coordinate in proper way,
we get an elegant description for the hidden and dynamical symmetries in these systems
related with action-angle variables.

In {\it Section 2.3} we introduce the appropriate complex coordinates unifying radial and angular variables and formulate the Poisson brackets and  generators of conformal algebra in these terms. Then we give "holomorphic factorization formulation" of the constants of motion of  higher-dimensional superintegrable conformal mechanics, and calculate their algebra.

In {\it Section 2.4} we formulate in these terms,  the higher-dimensional superintegrable generalizations of oscillator and Coulomb systems given by \eqref{so2},\eqref{angular0} and calculate the
algebra of their constants of motion.

In {\it Section 2.5} we  formulate, in this terms, the well-known oscillator-Coulomb duality transformation.

In {\it Section 2.6} we  extend the results of Section 2 to the systems  on $N$-dimensional sphere and two-sheet hyperboloid (pseudosphere).

Finally, in the {\it Section 2.7} we discuss examples  of angular part of these systems.

\section{TTW and PW systems}
The   Trembley-Turbiner-Wintenitz (TTW) system, invented a few years ago \cite{TTW},
is a particular case of the Calogero-oscillator system.
 It is defined by the Hamiltonian of two-dimensional oscillator, with the angular part replaced by a P\"oschl-Teller system on circle:
\be
\mathcal{H}_{TTW}=\frac{p^2_r}{2}+\frac{\mathcal{I}_{PT}}{r^2}+\frac{\omega^2r^2}{2},
\label{TTW}
\ee
\be
\mathcal{I}_{PT}=\frac{p^2_\varphi}{2}+\frac{k^2\alpha^2}{\sin^2k\varphi}+\frac{k^2\beta^2}{\cos^2k\varphi},
\label{PT}\ee
where $k$ is an integer.
It  coincides with the two-dimensional rational Calogero-oscillator model associated with the
dihedral group $D_{2k}$ \cite{lny}
and was initially considered as a new superintegrable model. The superintegrability was observed by  numerical simulations.
Later an  analytic expression for the
additional constant of motion was presented  \cite{kalnins}.

The two-dimensional Calogero-Coulomb system,  associated with dihedral group, is known as  a Post-Winternitz (PW) system.
It was  constructed
from the TTW system by performing the well-known Levi-Civita transformation,
 which maps the two-dimensional oscillator
into the Coulomb problem \cite{pw}. The PW system was also suggested as a new (independent from Calogero)
superintegrable  model.
It is also expressed via the P\"oschl-Teller Hamiltonian \eqref{PT},
\be
\mathcal{H}_{PW}=\frac{p^2_r}{2}+\frac{\mathcal{I}_{PT}}{r^2}-\frac{\gamma}{r}.
\label{PW}
\ee
In Ref.~\cite{gonera1}, the superintegrability of the TTW-system  was explained from the  viewpoint of action-angle variable
formulation, while in
Ref.~\cite{hlnsy},  using the same (action-angle) arguments,  the superintegrable generalizations of the TTW and PW systems   on sphere and
hyperboloid  were suggested. Below we briefly describe the constructions.

Consider an integrable $N$-dimensional system with the following Hamiltonian in action-angle variables:
\be
{\cal H}={\cal H}(nI_1+mI_2,I_3,\ldots, I_N),\qquad \{I_i,\Phi_j\}=\delta_{ij},\quad \Phi_i\in [0,2\pi),
\ee
where $n$ and $m$ are integers. The  Liouville integrals are expressed via the action variables $I_i$.
The system has a hidden symmetry, given by the additional constant of motion
\be
K_{hidden}={\rm Re}\; A(I_i){\rm e}^{\imath (m\Phi_1-n\Phi_2)},
\ee
where $A(I_i)$ is an arbitrary complex function on Liouville integrals.
Respectively, for the Hamiltonian
\be
{\cal H}={\cal H}(n_1I_1+n_2I_2+\ldots n_N I_N),
\ee
where $n_1,\ldots, n_N$ are integer numbers,
all  the functions
\be
K_{ij}={\rm Re}\; A_{ij}(I){\rm e}^{\imath(n_j\Phi_i-n_i\Phi_j)}.
\ee
are constants of motion, which are distinct  from the Liouville integrals.
The Liouville integrals  together with the additional integrals  $I_{i\,i+1}$ with $i=1,\ldots N-1$
constitute a set of $2N-1$  functionally independent constants of motion, ensuring the maximal superintegrability.

In Ref.~\cite{hlnsy}  the integrable deformations of the   $N$-dimensional oscillator and  Coulomb systems
have been proposed on Euclidean space, sphere and  hyperboloid by replacing their
 angular part by an $(N-1)$-dimensional integrable system,
formulated in action-angle variables:
\be
H=\frac{p^{2}_r}{2}+\frac{{\cal I}(I_a)}{r^2}+V(r),
\qquad
\{p_r, r\}=1,
\qquad
\{I_a,\Phi^0_b\}=\delta_{ab},
\label{2}
\ee
where $ a,b=1,\ldots,N-1$ and
\be
 V_{osc}(r)=\frac{\omega^2r^2}{2}, \qquad V_{Coulomb}(r)=-\frac{\gamma}{r}.
\label{5}\ee
In other words, we obtain the deformation of the $N$-dimensional oscillator and Coulomb systems
by replacing the $SO(N)$ quadratic Casimir element $\mathbf{J}^2$, which defines the  kinetic part of the system on sphere
$\mathbb{S}^{N-1}$, with the Hamiltonian
 of some $(N-1)$-dimensional integrable system written in terms of the action-angle variables.

Next we have performed   similar analyses for the systems on $N$-dimensional sphere and (two-sheet) hyperboloid
with the oscillator and Coulomb potentials. These models were introduced, respectively, by Higgs \cite{higgs} and Schr\"odinger \cite{sch,sch1},
\be
\mathbb{S}^{N}:\quad H=\frac{p_{\chi}^2}{2 r_0^2}+\frac{{\cal I}}{ r_0^2 \sin^2\chi}+V(\tan\chi),
\quad \{p_\chi, \chi\}=1,
\label{3}\ee
\be
\mathbb{H}^{N}:\quad H=\frac{p_{\chi}^2}{2 r_0^2}+\frac{{\cal I}}{ r_0^2\sinh^2\chi}+V(\tanh\chi), \quad \{p_r, r\}=1
\label{4}
\ee
with  ${\cal I}$ depending on the (angular) action variables.
The exact forms for the potential are:
\begin{align}
\mathbb{S}^{N}:& \quad V_{Higgs}(\tan\chi)=\frac{r^2_0\omega^2\tan^2\chi}{2}, \qquad V_{Sch-Coulomb}(\tan\chi)=-\frac{\gamma}{r_0}\cot\chi,
\label{6}
\\
\mathbb{H}^{N}:& \quad V_{Higgs}(\tanh\chi)=\frac{r^2_0\omega^2\tanh^2\chi}{2}, \qquad V_{Sch-Coulomb}(\tanh\chi)=-\frac{\gamma}{r_0}\coth\chi.
\label{7}
\end{align}
 The following expressions for the Hamiltonians of oscillator-like systems had been derived:
\be
{\cal H}_{osc}={\cal H}_{osc}(2 I_r + \sqrt{2{\cal I}})=\left\{\begin{array}{ccc}
\omega (2 I_r + \sqrt{2{\cal I}})& {\rm for }& \mathbb{R}^N,\\
\frac{1}{2}(2I_\chi+ \sqrt{2{\cal I}}+\omega)^2- \frac{\omega^2}{2}&{\rm for}& \mathbb{S}^N,\\
-\frac{1}{2}(2I_\chi +\sqrt{2{\cal I}}-\omega)^2+ \frac{\omega^2}{2}&{\rm for}&  \mathbb{H}^N.
\end{array}
\right.
 \ee
 Respectively, the Hamiltonians of the  Coulomb-like systems read:
 \be
{\cal H}_{Coulomb}={\cal H}_{Coulomb}( I_r + \sqrt{2{\cal I}})=\left\{\begin{array}{ccc}
-\frac{\gamma^2}{2}(I_r + \sqrt{2{\cal I}})^2& {\rm for }& \mathbb{R}^N,\\
-\frac{\gamma^2}{2}(I_\chi + \sqrt{2{\cal I}})^2+\frac12(I_\chi+\sqrt{2{\cal I}})^2&{\rm for}& \mathbb{S}^N,\\
-\frac{\gamma ^2}{2} {(I_\chi-\sqrt{2 {\cal I}})^2} -\frac12(I_\chi-\sqrt{2 {\cal I}})^2&{\rm for}&  \mathbb{H}^N.
\end{array}
\right.
 \ee
Thus, it is easy to deduce that  for the angular Hamiltonian
\be
{\cal I}_{SphCalogero}=\frac 12\Big(\sum_{a=1}^{N-1} k_aI_a+\text{const}\Big)^2, \qquad k_a\in \mathbb{N},
\label{app}
\ee
the deformations of the oscillator and Coulomb systems become superintegrable.
In particular, the P\"oschl-Teller Hamiltonian has the same form \cite{lny}:
\be
\mathcal{I}_{PT}=\frac{k^2(I+\alpha+\beta)^2}{2}.
\ee
Hence, choosing $N=2$ and  $\mathcal{I}=\mathcal{I}_{PT}$,
we obtain the generalizations of the TTW and PW systems on sphere and hyperboloid
with additional constants of motion given by
 \be
\mathcal{K}_{TTW}={\rm Re}\; A(I){\rm e}^{\imath (k\Phi_r-2\Phi_\varphi)},\qquad \mathcal{K}_{PW}={\rm Re}\; A(I){\rm e}^{\imath (k\Phi_r-\Phi_\varphi)}.
 \ee
Here $\Phi_{\varphi}$ is the angle variable in the P\"oschl-Teller system,
and $\Phi_r$ is the angle variable  associated with $r$ and $p_r$. For explicit expressions,
see Ref.~\cite{hlnsy}.

 Note that the angular part of the $N$-dimensional rational Calogero model has  the form \eqref{app} as well.
 This is a reason for the superintegrabilty of the Calogero-oscillator and Calogero-Coulomb problems.
 It also suggests  that their superintegrable generalizations on the $N$-dimensional spheres and hyperboloids \cite{sphCal,sphCal1,sphCal2,sphCal3,sphCal4}.
Although the TTW and PW systems are  particular cases of the Calogero-type  models,
they continue to attract enough interest due to their simplicity.
In particular, a couple of years ago, Ranada suggested a specific  representation for the  constants of motion of the
TTW and PW systems (including those on sphere and hyperboloid) \cite{ranada,ranada1,ranada2},
called a "holomorphic factorization". For the TTW system it reads
\be
{\cal R}_{TTW}=(\bar M_0)^{k}N^2,
\label{ranadaTTWlint}
\ee
where
\be M_{0}=\frac{2 p_r}{r}\sqrt{2\cal{I}_{PT}}+2\imath \mathcal{H}_{TTW},
\label{ranadaTTW}
\ee
and
\be
 N=k(\beta-\alpha)+2\mathcal{I}_{PT}\cos{2k\varphi}+\imath\sqrt{2\mathcal{I}_{PT}}p_{\varphi}\sin 2k\varphi.
\label{ranadaN}\ee
A similar expression  exists in case of the (pseudo)spherical TTW system.
The additional constant of motion of PW system in Ranada's representation reads:
\be
\mathcal{M}_{PW}=({\bar M}_0)^{k}N,
\ee
and $N$ is given by Eq.~\eqref{ranadaN}, and
 \be
 M_0=p_r\sqrt{2\mathcal{I}_{PT}}+\imath\Big(\gamma-\frac{2\mathcal{I}_{PT}}{r}\Big) .
\ee
Such forms of the hidden constants of motion have a visible relation with their expressions in terms of the action-angle variables, which will be discussed below.
Hence,  the TTW and PW systems possess a natural  description  in spherical coordinates, where the "radial" part is separated
from the "angular" one. On the other hand, the radial parts are expressed via the generators of conformal algebra,
which can be  viewed as generators of isometries of the K\"ahler structure  of    Klein model of the Lobachevsky space \cite{lobach}.
Hence, we can represent phase spaces of the TTW and PW systems as a (semidirect) product of Lobachevsky space with cotangent bundle
of circle, and expect that the reformulation in these coordinates will help us to extend the expressions of hidden constants of motion to
higher dimensions. Similarly, phase spaces of the  $N$-dimensional oscillator and Coulomb systems and their Calogero-deformations
could be represented as a  semidirect product of Lobachevsky space and cotangent bundle on $(N-1)$-dimensional sphere \cite{sigma}.
One can expect, that Ranada's representation of hidden symmetries of the TTW and PW systems in these terms will take
a more transparent and elegant form. Furthermore, having in mind the relation of the TTW  and PW systems with rational Calogero models,
one can expect that the hidden symmetries of  Calogero model could be represented in a similar way.

\subsection{One-dimensional systems}
 Since the middle of seventies with Ref.~\cite{fubini}  in the field-theoretical
literature
much attention has been paid  to a simple one-dimensional mechanical system
 given by the Hamiltonian
 \begin{equation}
   H_0 =\frac{{p}^2}{2}+\frac{g^2}{2x^2}.
\label{h}
\end{equation}
The reason was that it  forms
 the conformal algebra $so(1,2)$ \eqref{ca} together with the generators:
\begin{equation} D=px,\qquad K=\frac{x^2}{2} \label{dk}. \end{equation}

In Ref.~\cite{lobach}  the following formulation of this is suggested.
Its   phase space   is parameterized by a single complex coordinate and identified with the Klein model
of the Lobachevsky plane:
\begin{equation} { z}=\frac{p}{x}+\frac{\imath g}{x^2}, \qquad {\rm
Im}\;{ z}>0: \qquad \{{ z},\bar{z}\}=-\frac{\imath}{g}\left({
z}-\bar {z} \right)^2. \label{px} \end{equation}

In this   parametrization,  the $so(1,2)$ generators (\ref{h}), (\ref{dk})
define the Killing potentials (Hamiltonian generators of the
isometries of the K\"ahler structure) of Klein model:
\begin{equation}
 H_0= g\frac{{ z} {\bar z}}{\imath({\bar z}-{ z})},
\qquad
D=g\frac{{ z}+\bar{ z}}{\imath ({\bar z}-{ z})},
\qquad
K= g\frac{1}{\imath ({ \bar z}- { z})}.
\label{uk}
\end{equation}
Let us remind, that the K\"ahler structure  is
\begin{equation} ds^2=-\frac{g d{ z} d\bar
{z}}{({\bar z}- { z})^2}. \label{klein}
\end{equation}
It is invariant under the
 discrete transformation
\begin{equation} { z}\to -\frac{1}{z}, \label{simt} \end{equation}
whereas the Killing
potentials (\ref{uk}) transform as follows:
 \begin{equation}
 H_0\to  K,\qquad K\to H_0,\qquad D\to -D.
\label{t}\end{equation}
Thus, it  maps $H_0$ to the free one-dimensional particle system. This
can be viewed as a one-dimensional analog of the decoupling transformation
of the Calogero Hamiltonian, considered in Refs.~\cite{decoupling,dec1,dec2}.

In order to construct a similar construction for higher-dimensional systems,
first, we introduce an appropriate "radial"
coordinate and  conjugated momentum, so that the higher-dimensional
system looks very similar to the one-dimensional conformal
mechanics. In that picture, the remaining "angular" degrees of
freedom are packed in the  Hamiltonian system on the
$(N-1)$-dimensional sphere, which replaces the coupling constant $g^2$
in the one-dimensional conformal mechanics. The angular Hamiltonian defines the constant
of motion of the initial conformal mechanics.  Then
we relate the radial part of the $N$-dimensional conformal
mechanics with the Klein model of the Lobachevsky space, which is
completely similar to the aforementioned one-dimensional case.

\subsection{Higher-dimensional systems}
Let us consider the $N$-dimensional conformal mechanics, defined by
the  following Hamiltonian and symplectic structure:
\begin{equation}
\omega={d{\bf p} }\wedge {d{\bf x}},
\qquad
\mathcal{H}_0=\frac{{\bf p}^2}{2}+V({\bf x}),
\qquad
{\rm where}
\qquad ({\bf x}\cdot\nabla) V({\bf x} )=-2V({\bf x}).
\label{h1}
\end{equation}
 This Hamiltonian together with the generators
\begin{equation}
\mathcal{D}={\bf p}\cdot{\bf x},\qquad  \mathcal{K}=\frac{{\bf x}^2}{2} \label{dk1}
\end{equation}
 forms the
conformal algebra $so(1,2)$  (\ref{ca}). Here $\mathcal{D}$ defines the
dilatation and $\mathcal{K}$ defines  the conformal boost, ${\bf
x}=(x^1,\ldots, x^N)$, ${\bf p}=(p_1,\ldots, p_N)$.

Extracting the radius $r=|{\bf x}|$,  we can present the above
generators  in the following form:
\begin{equation} {\cal D}=p_r r ,
\qquad
{\cal K}=\frac{r^2}{2},
\qquad {\cal H}_0=\frac{{p}^2_r}{2}+
\frac{{\cal I}}{r^2},
\qquad {\cal I}\equiv \frac{{\bf J}^2}{2}+ U,
\qquad U\equiv r^2V({\bf r}).
\label{so2}
\end{equation}
 Here  $p_r=({\bf p}\cdot{\bf x})/r $ is the momentum,
conjugate to the radius:  $\{p_r, r\}=1$.
It is easy to check that $\mathcal{I}$ is the Casimir element  of conformal algebra $so(1.2)$:
\begin{equation}
4{\cal H}{\cal K}-{\cal D}^2= 2{\cal I}:
\qquad
\{\mathcal{I},\mathcal{H}_0\}=\{\mathcal{I},\mathcal{K}\}=\{\mathcal{I},\mathcal{D}\}=0.
\label{ml}
\end{equation}
Thus, it defines the constant of motion of the system \eqref{h1} and commutes with $r,p_r$ and,
hence, does not depend on them.
Instead, it depends on  spherical coordinates $\phi^a $ and
canonically conjugate momenta $\pi_a $.
As a Hamiltonian, ${\cal I}$ defines
the particle motion  on $(N-1)$-sphere in the potential $U(\phi^\alpha )$. The phase space is the
cotangent bundle  $T^*S^{N-1}$.
%

As in one dimension \cite{lobach} instead of the radial phase space
coordinates $r$ and $p_r$ we introduce  the following complex
variable (for simplicity, we restrict to $\mathcal{I}>1$):
\begin{equation} { z}=\frac{p_r}{r}+\frac{\imath \sqrt{2{\cal I}}}{r^2}\equiv
\frac{{\cal D}+\imath \sqrt {2{\cal I}}}{2{\cal K}},
\qquad
{\rm Im}\;{ z}>0.
 \label{px2}
\end{equation}
It obeys the following Poisson brackets:
\begin{equation}
\{{ z},\bar{ z}\}=-\frac{\imath}{\sqrt{2{\cal I}(u)}}\left({ z}-\bar { z}
\right)^2,
\label{k1}
\end{equation}
\begin{equation}
 \{u^\alpha,u^\beta\}=\omega^{\alpha\beta}(u),\qquad \{u^\alpha , z\}= (z -\bar z
)\frac{V^\alpha(u)}{2{\cal I}},\qquad\{u^\alpha, \bar z\}=( z -\bar z
)\frac{V^\alpha(u)}{2{\cal I}},\qquad
\end{equation}
where $ V^\alpha=\{u^\alpha, {\cal I}(u)\}$ are the
equations of motion of the angular system.

 The symplectic structure of the
conformal mechanics can be represented as follows:
 \begin{equation}
\Omega=-\imath \frac{\sqrt{2{\cal I}(u)} d{ z}\wedge  d\bar { z}}{({\bar z}-
{ z})^2} + \frac{(dz+d\bar z)\wedge d\sqrt{ 2{\cal I}(u)}}{\imath(\bar z- z
)}+ \frac{1}{2}\omega_{\alpha\beta}du^\alpha\wedge {du^\beta},
\end{equation}
  while the local
one-form, defining this symplectic structure, reads
 \begin{equation} \Omega=
d{\cal A},\qquad {\cal A}=\imath \sqrt{2\mathcal{I}(u)}\frac{dz+d\bar
z}{\imath(z-\bar z)} +A_0(u),\qquad dA_0=\frac{1}{2}
\omega_{\alpha\beta}du^\alpha\wedge du^\beta .
\end{equation}
 %
 Taking into account Eq.~(\ref{ml}), we can write:
\begin{equation}
\mathcal{H}_0=\sqrt{2{\cal I}(u)}\frac{{ z}\bar { z}}{\imath({ \bar z}- { z})},
\qquad \mathcal{D}= \sqrt{2{\cal I}(u)}\frac{{ z}+\bar { z}}{\imath ({\bar z}- {
z})}, \qquad \mathcal{K}= \frac{\sqrt{2{\cal I}(u)}}{\imath ({\bar z}- { z})},
\label{uk2}
\end{equation}
The  transformation (\ref{simt}) does not preserve the
symplectic structure, i. e., it is not a canonical transformation
for the  generic  conformal mechanics of dimension $d>1$.


Now we introduce the following generators, which will be used in our further considerations:
\be
M=\frac{z}{\sqrt{\imath({\bar z}- z )}}, \qquad \bar M=\frac{{\bar z}}{\sqrt{\imath(\bar z- z)}}.
\ee
 With the generators of the conformal algebra they form a highly nonlinear algebra:
\be
\{M,{\cal H}_0\}=\frac{\imath}{2}z\sqrt{\imath (\bar z -z)},
\qquad \{M,{\cal K}\}=\frac{2z}{\imath(\bar z-z)}, \qquad \{M,{\cal D}\}=\frac{z}{\sqrt{\imath(\bar z-z)}}=M,
\ee
\be
 \{M,\bar M\}= \frac{z-\bar z}{2\sqrt{2\mathcal{I}}}.
\ee

Let us introduce the angle-like  variable, conjugate with $\sqrt{2\mathcal{I}}$:
\be
\Lambda(u):
\qquad
 \big\{\Lambda,\sqrt{2\mathcal{I}}\big\}=1,
 \qquad \Lambda\in[0,2\pi).
\label{Phi}\ee
Using $M$ and $\Lambda$,  one can easily build a (complex) constant of motion for the conformal mechanics:
\be
{\cal M}=M{\rm e}^{\imath\Lambda},
\qquad \{{\cal M}, \mathcal{H}_0\}=0.
\ee
Evidently, its real part is the ratio of  Hamiltonian and its angular part
and does not contain any new constant of motion.
Nevertheless, such a complex representation seems to be useful not only from an aesthetical viewpoint, but also
for the construction of supersymmetric extensions.

Note that  we can write down the hidden symmetry generators for the conformal mechanics,
modified by the oscillator and Coulomb potentials  as well.
The Hamiltonian of the $N$-dimensional oscillator and its hidden symmetry generators  look
 as follows:
\be
\begin{gathered}
{\cal H}_{osc}={\cal H}_{0} +\omega^2{\cal K},
\quad {\cal M}_{osc}=\frac{z^2+\omega^2}{\imath(\bar z-z)}{\rm e}^{2\imath\Lambda}=\left(M^2+\omega^2{\cal K}\right){\rm e}^{2\imath\Lambda}\; :\quad
\{{\cal M}_{osc}, {\cal H}_{osc}\}=0
\end{gathered}
\ee
The Hamiltonian and  hidden symmetry of the Coulomb problem are defined by
 \be
 \begin{split}
 {\cal H}_{Coul}={\cal H}_0-\frac{\gamma}{\sqrt{2{\cal K}}},
  \quad{\cal M}_{Coul}=\left({{ M}}-\frac{\imath\gamma}{(8\sqrt{2\cal{I}})^{3/2}} \right){\rm e}^{\imath\Lambda}:\quad \{{\cal M}_{Coul}, {\cal H}_{Coul}\}=0,
 \end{split}
 \ee
The absolute values of both integrals do not produce anything new:
\be
|{\cal M}_{osc}|^2=\frac{\mathcal{H}^2_{osc}}{2\mathcal{I}}-\omega^2,
\qquad |{\cal M}_{Coul}|^2=\frac{\mathcal{H}_{Coul}}{\sqrt{2\mathcal{I}}}+\frac{\gamma^2}{2(\sqrt{2\mathcal{I}})^{3}}.
\ee
So, the hidden symmetry is encoded in their phase, depending on
the angular variables $\Phi(u)$.
Assume that the angular system
 is integrable. Hence the Hamiltonian
 and two-form are expressed in terms of the action-angle variables as follows:
 \[
\mathcal{I}=\mathcal{I}(I_a),
\qquad
\Omega=\sum_a dI_a\wedge d\Phi_a.
\]
Then the condition \eqref{Phi} implies the following local solutions   for $\Lambda$:
\be
\Lambda_a=\frac{\Phi_a}{\omega_a(I)}, \quad{\rm where}\quad \omega_a=\frac{\partial\sqrt{2\mathcal{I}}}{\partial I_a}.
\ee
Thus, to provide the global solution for a certain coordinate $a$,
we are forced to set $\omega_a(I)= k_a$ to a rational number:
\be
k_a=\frac{n_a}{m_a} ,\qquad m_a,n_a \in \mathcal{N}.
\ee

Then, taking $k_a$-th power for the locally defined conserved quantity, we get a globally defined
constant of motion for the system.
In this case, the hidden symmetry  of the  conformal mechanics reads:
\be
{\cal M}_a=M^{n_a}{\rm e}^{\imath m_a\Phi_a }.
\ee
Similarly, for the systems with oscillator and Coulomb potentials  one has:
\be
{\cal M}_{(a)osc}=\left(M^2+\omega^2{\cal K}\right)^{n_a}{\rm e}^{2\imath m_a\Phi_a},
\qquad
{\cal M}_{(a)Coul}=\left({{ M}}-\frac{\imath\gamma}{(8\sqrt{2\cal{I}})^{3/2}} \right)^{n_a}{\rm e}^{\imath m_a\Phi_a}.
\ee

To find the expression(s) for $\Phi$, let us remind
that the angular part  of these systems is just the quadratic Casimir element (angular momentum)
of $so(N)$ algebra on $(N-1)$- dimensional sphere, ${\cal I}=L_{N}^2/2$.
It can be decomposed by the eigenvalues of the embedded $SO(a)$ angular momenta $I_a $
as follows:
\be
{\cal I}=\frac12\left(\sum_{a=1}^{N-1}I_a \right)^2.
\ee

Hence, our expressions define
the $N-1$ functionally independent constants of motion
\be
{\cal M}_{(a)osc}=\left(M^2+\omega^2{\cal K}\right){\rm e}^{2\imath\Phi_a},
\qquad
{\cal M}_{(a)Coul}=\left({{ M}}+\imath\gamma \right){\rm e}^{\imath\Phi_a},
\ee
respectively,  for the $N$-dimensional oscillator and Coulomb problems.
Since these  systems have $N$ commuting constants of motion ($I_a$, ${\cal H}$), we
have obtained in this way the full set of their integrals.

To clarify the origin of these  generators, let us consider a particular case of two-dimensional systems.
The angular part   is a circle, and, respectively, $I=|p_\varphi|$, $\Phi=\varphi$ with $\varphi$
being a polar angle.
In this case, the oscillator Hamiltonian and its hidden constant of motion  read
\be
 H_{osc}=|p_\varphi|\frac{z\bar{z}+\omega^2}{\imath(\bar z-{z})},\qquad
\mathcal{M}_{osc}=\frac{i}{z-\bar{z}}(z^2+\omega^2)e^{2\imath\varphi}.
\ee
The latter can  also be presented as follows:
\be
\mathcal{M}_{osc}=\frac{H_1-H_2+2iH_{12}}{|p_{\varphi}|},\quad{\rm with} \quad H_{ab}= p_ap_b+\omega^2x_ax_b.
\ee
Here $H_{ab}$  is a standard  representation of the oscillator's hidden symmetry generators,
 sometimes (Fradkin tensor).

The Hamiltonian of two-dimensional Coulomb problem and its hidden symmetry generator are of the form

 \be
 H_{Coul}=|p_\varphi|\frac{z \bar{z}}{\imath(\bar z-{z})}-{\gamma}\sqrt{
 \frac{\imath(\bar z-{z}) }{2|p_\varphi |}},\qquad
{\cal M}_{Coul} = \Bigg(\frac{z}{\sqrt{\imath(\bar z-{z})}}-\frac{\imath\gamma}{\sqrt{2|p_\varphi|^3}}\Bigg)e^{\imath\varphi}
\ee
The Latter  is related with the components of the two-dimensional Runge-Lenz vector \newline
${\bf A}=(A_x,A_y)$ as follows
\be
 \mathcal{M}_{Coul}=\frac{A_y-\imath A_x}{\sqrt{2|p_\varphi|^3}},\quad {\rm where}\quad
A_{x}=  p_\varphi p_{y}-\gamma \cos{\varphi},
\quad
A_{y}= p_\varphi p_{x}-\gamma \sin{\varphi}.
\ee

Now we are ready to apply this constructions to the TTW and PW systems.
In order to formulate TTW and PW systems in the above terms,  we will use the action-angle formulation of the P\"oshl-Teller
Hamiltonian given in Ref.~\cite{lny}:
\be
\mathcal{I}_{PT}=\frac{k^2{\tilde I}^2}{2}, \qquad {\tilde I}=I+\alpha+\beta,
\ee
where $I$ is an action variable.

The angle variable is related to the initial phase space coordinates as follows:
\be
a \sin(-2 \Phi) =\cos(2k\varphi)+b  ,
\qquad
 a=\sqrt{\Big(1-\frac{2(\alpha+\beta)}{{(k\tilde I)}^2}+b^2\Big)} ,
 \qquad
 b=\frac{\beta-\alpha}{(k{\tilde I})^2}.
\ee

Using the above expressions, we can present the Hamiltonian of TTW system and its hidden symmetry generator
 as follows:
\be
H_{TTW}=k\tilde{I}\frac{z \bar{z}+\omega^2}{\imath(\bar z-{z})} ,
\qquad
{\cal M}_{TTW}=\Big(\frac{z^2+\omega^2}{\imath(\bar{z}-z)}\Big)^k e^{2\imath \Phi}.
\ee
The Ranada's constant of motion is related with the above one:
\be
K= -a^2 \frac{(2k\tilde{I})^{2k+4}}{16}\Big(\frac{\bar{z}^2+\omega^2}{z-\bar{z}}\Big)^{2k}e^{-4\imath \Phi}
=-a^2 \frac{(2k\tilde{I})^{2k+4}}{16}{\cal \bar M}_{TTW}^2.
\ee

 We repeat the same procedure   for the PW system as well.
Using the expressions  for action-angle variables of the P\"oschl-Teller Hamiltonian, we get:
 \be
 \mathcal{H}_{PW}=ik\tilde{I}\frac{\bar{z}z}{z-\bar{z}}-\frac{\gamma}{2k\tilde{I}}\sqrt{i(\bar{z}-z)} ,
 \qquad
  {\cal M}_{PW}=\left(\frac{z}{\sqrt{i(\bar{z}-z)}}-\frac{i\gamma}{k\tilde{I}\sqrt{2k\tilde{I}}}\right)^{k}e^{i\Phi}.
 \ee
Respectively, the Ranada's constant of motion takes the form
\be
 K=-ia(k\tilde{I})^2\left(k\tilde{I}\sqrt{2k\tilde{I}}\frac{z}{\sqrt{i(\bar{z}-z)}}+{i\gamma}\right)^{2k} e^{2i\Phi}=
 -ia(k\tilde{I})^{2k+2}{\cal \bar M}^2_{PW}.
 \ee

\section{Alternative complex notations }

Introduce another complex variable $Z$, identifying the radial phase subspace with the Klein model of
Lobachevsky plane  (compare with  the notations in the previous section),
 and complex variables $u_a$  unifying the action-angle variables:
\begin{equation}
{ Z}=\frac{p_r}{\sqrt2}+\frac{\imath \sqrt{{\cal I}}}{r},
\qquad
u_a=\sqrt{I_a}{\rm e}^{\imath\Phi_a}\qquad{\rm with }\qquad {\rm Im}\,{ Z}>0.
 \label{px2}
\end{equation}
These variables have   the following nonvanishing  Poisson brackets:
\begin{equation}
\{{ Z},\bar{ Z}\}=-\frac{\imath ({ Z}-\bar {Z})^2}{2\sqrt{2{\cal I}}},
\qquad
\{u_a, {\bar u}_b\}=-\imath\delta_{ab},
\ee
\be
 \{Z,u_a\}=-u_a\Omega_a\frac{\imath(\bar{Z}-Z)}{2\sqrt{2{\cal I}}},
 \qquad  \{Z,{\bar u}_a\}= {\bar u}_{a}\Omega_a\frac{\imath(\bar{Z}-Z)}{2\sqrt{2{\cal I}}},
\label{z-z}
\end{equation}
where
\be
\Omega_a=\Omega_a(I)=\frac{\partial\sqrt{2\mathcal{I}}}{\partial I_a}.
\label{I}\ee
In these terms
 the generators of conformal algebra take the form
 \begin{equation}
\mathcal{H}_0=Z\bar Z,
\qquad \mathcal{D}= {\sqrt{2{\cal I}(u_a\bu_a)}}\frac{{ Z}+\bar { Z}}{\imath ({\bar Z}- {
Z})}, \qquad \mathcal{K}= \frac{{2{\cal I}(u_a\bu_a)}}{(\imath ({\bar Z}- { Z}))^2}.
\label{uk2}
\end{equation}
Note that  the action variables $I_a$ complemented with the Hamiltonian
form a set of Liouville integrals of the conformal mechanics \eqref{h1}. They
have a rather simple form while being expressed via the complex variables:
 \begin{gather}\label{Hzz}
\mathcal{H}_0=Z\bar Z,
\qquad
I_a = u_a\bar u_a:
\qquad
\{H_0,I_a\}=\{I_a,I_b\}=0.
\end{gather}

Let us now look for the additional integrals of motion, if any.
It is easy to verify using \eqref{z-z}, \eqref{Hzz}  that
\be
\{Ze^{\imath \Lambda},\mathcal {H}_0\}=0
\qquad {\rm iff }\qquad
\{ \Lambda,\sqrt{2\mathcal{I} }\}=-1.
\ee
To get the single-valued function  we impose $\Lambda\in[0,2\pi )$ .
The   local  solutions of  the above equation  read
\be
\label{wa}
\Lambda_a=\frac{\Phi_a}{\Omega_a},
\ee
where $\Phi_a\in[0,2\pi)$ is angle variable and $I_a$ is given by \eqref{I}.
Therefore, the following local quantities are preserved and generate the set
of $N-1$ additional constants of motion:
\be
\label{Ma}
M_a=Zu_a^\frac{1}{\Omega_a}=ZI_a^\frac{1}{2\Omega_a}e^{\imath\frac{\Phi_a}{\Omega_a}},
\qquad
\{M_a,{\cal H}_0\}=0.
\ee
Using \eqref{px2}, \eqref{z-z}, one can verify that the only nontrivial  Poisson bracket
relations among them occur  between the conjugate $M_a$-s:
\begin{gather}
\label{M-M}
\qquad
\{M_a,M_b  \}=0,
\qquad
\{M_a,\overline{M}_b  \}
=-\frac{\imath\delta_{ab}}{\Omega_a^2} I_a^{\frac{1}{\Omega_a}-1}{\cal H}_0.
\end{gather}
However, for the generic $\Omega_a$, the constant \eqref{Ma} is not still globally well-defined,
since {$\Lambda\in [0,2\pi/\Omega_a)$.
To get the global solution for a certain coordinate $\Phi_a$,
we are forced to set $\Omega_a$ to a rational number:
\be
\label{ka}
\Omega_a=k_a=\frac{n_a}{m_a} ,\qquad m_a,n_a \in \mathbb{N}.
\ee
Then, taking $n_a$-th power for the locally defined conserved quantity, we get a globally defined
constant of motion for the system,
\be
\label{calM}
{\cal M}_a=M_a^{n_a}=Z^{n_a}u_a^{m_a}=I_a^\frac{m_a}{2}Z^{n_a}{\rm e}^{\imath m_a\Phi_a }.
\ee
Although both $M_a$ and ${\cal M}_a$ are complex, their absolute values
are expressed via Liouville integrals, and, hence,  do not produce new constants
of motion:
\be
\label{absM}
|M_a|^2=\mathcal{H}_0 I_a^\frac{1}{k_a},
\qquad
|{\cal M}_a|^2=\mathcal{H}_0^{n_a} I_a^{m_a}.
\ee
So, we have constructed  $2N-1$ functionally independent  constants of motion of the
generic superintegrable conformal mechanics \eqref{h1} with rational frequencies \eqref{wa}.
Therefore,  the conformal mechanics  will be superintegrable provided that the angular Hamiltonian has the form \eqref{angular0}
with  rational numbers $k_a$ \eqref{ka} and arbitrary constant $c_0$.\\

Full  symmetry algebra is given by the relations
\be
\label{M-bMcal}
\{{\cal M}_a,\overline{\mathcal{ M}}_b  \}
=-\imath \delta_{ab}  m_a^2 I_a^{m_a-1}{\cal H}_0^{n_a},\qquad \{H_0,{\cal M}_a\}=\{{\cal M}_a,{\cal M}_b  \}=0.
\ee
Note  that
\be
\{I_a,{\cal M}_b  \}=\imath \delta_{ab}M_b,\qquad \{H_0,I_a\}=\{I_a,I_b\}=0
\label{M-Mcal}
\ee
As we mentioned in Introduction, presented formulae are applicable not only for the nonrelativistic conformal mechanics on $N$-dimensional Euclidean space
defined by the Hamiltonian \eqref{h1} but for the generic finite-dimensional system with conformal symmetry, including relativistic one. Typical example of such a system is a particle moving in the  near-horizon limit of extreme black hole. Several examples of such systems were investigated by A.Galajinsky and his  collaborators (see Refs.~\cite{Galaj0,Galaj1,Galaj2}).

\section{Deformed oscillator and  Coulomb \newline systems}
Let us extend the above consideration  to  the deformed $N$-dimensional oscillator and Coulomb systems defined by
the Hamiltonians
\be
\mathcal{H}_{osc/Coul}=\frac{p^2_r}{2}+\frac{\mathcal{I}}{r^2}+V_{osc/Coul}(r)
=Z\bar Z+ V_{osc/Coul}(r),
\label{flat}\ee
where
\be
V_{osc}=\frac{ \omega^2 r^2}{2}=\omega^2\mathcal{K}=-\frac{{2 \omega^2{\cal I}}}{({\bar Z}- { Z})^2} ,
\qquad
V_{Coul}=-\frac{\gamma}{r}=-\frac{\gamma}{\sqrt{2\mathcal{K}}}=-\gamma\frac{\imath(\bar Z-Z)}{2\sqrt{\mathcal{I}}}.
\ee
Clearly, the action variables of the angular mechanics $I_a$ together with the corresponding
Hamiltonian define  Liouville constants  of motion:
\be
\label{H-I}
\{H_{osc/Coul},I_a\}=\{I_a,I_b\}=0.
\ee
To endow  these systems by superintegrability  property we choose the angular part given by \eqref{angular0} with rational $k_a$, see \cite{rapid}.
Below we construct the additional constants of motion and calculate their algebra for both systems in terms of complex variables \eqref{px2}
introduced in previous section.

\subsection{Oscillator case}
The  $2N-2$  constants of motion of the deformed oscillator $\mathcal{H}_{osc}$ in the coordinates
\eqref{px2} are appeared to look as:
\be
\label{Mosc}
 {\cal M}^{osc}_a=\left(Z^2-\frac{2\omega^2\mathcal{I}}{({\bar Z}- { Z})^2}\right)^{n_a}u_a^{2m_a},
\qquad
|{\cal M}_a^{osc}|^2=\left(\mathcal{H}_{osc}^2-2\omega^2\mathcal{I}\right)^{n_a}I_{a}^{2m_a}.
\ee
The last equation together with \eqref{angular0} means that only the arguments of these complex quantities
give rise to new integrals independent of the Liouville ones.

In fact, they are based on the simpler quantities $A_a$ and $B_a$, which oscillate in time with the same frequency $w$:
\be
\label{Aa}
A_a=\left(Z+\frac{\omega\sqrt{2\mathcal{I}}}{{\bar Z}- { Z}} \right)u_a^{\frac1{k_a}},
\qquad
B_a=\left(Z-\frac{\omega\sqrt{2\mathcal{I}}}{{\bar Z}- { Z}}\right)u_a^{\frac1{k_a}}:
\ee
\be
\{{\cal H}_{osc}, A_a\}=\imath\omega  A_a,
\qquad
\{{\cal H}_{osc}, B_a\}=-\imath\omega  B_a.
\ee
So, the product $A_a B_b$ is preserved,
\be
\label{H-AB}
\{{\cal H}_{osc}, A_a B_b\}=0,
\ee
but is not single valued.
Thus, we have to  take its $n_a$th power to get
a well defined constant of motion, which is precisely \eqref{Mosc}:
\be
\qquad
 {\cal M}^{osc}_a=(A_aB_a)^{n_a}.
\label{H-A}\ee
Note that the reflection $\omega\to -\omega$ in the parameter space maps between $ A_a$ and  $B_a$.
Together with complex conjugate, they are subjected to the following rules:
\be
\label{absAB}
| B_a |^2
 =\frac{{\cal H}_{osc}-\omega\sqrt{2{\cal I}}}{{\cal H}_{osc}+\omega\sqrt{2{\cal I}}} | A_a |^2,
 \qquad
 | A_a|^2=I_a^\frac{1}{k_a}\left({\cal H}_{osc}+\omega\sqrt{2{\cal I}}\right).
\ee
The complex observables  $ A_a$ and  $B_a$ are in involution,
\be
\label{A-A}
\{A_a, A_b\}=\{B_a, B_b\}=\{A_a, B_b\}=0,
\ee
so that the constants  of motion \eqref{Mosc} commute as well:
\be
\label{M-Mosc}
\{{\cal M}^{osc}_a,{\cal M}^{osc}_b  \}=0.
\ee
However, in contrast to the simplicity of the  relations \eqref{M-Mcal},
the Poisson brackets between ${\cal M}^{osc}_a$ and $\overline{\cal M}^{osc}_b$
are more elaborate. They can be derived from the Poisson brackets between
$ A_a$ and  $B_a$ and their conjugates having the following form:
\begin{gather}
\label{A-barB}
\{A_a,\bar B_b\}=-\frac{\imath \delta_{ab}}{k_a^2 I_a}A_a\bar B_a,
\qquad
\{\bar A_a, B_b\}=\frac{\imath \delta_{ab}}{k_a^2 I_a}\bar A_a B_a,
\\
\label{A-barA}
\{A_a,\bar A_b\}=
-\frac{ 2\imath\omega A_a\bar A_b }{\mathcal{H}_{osc}+\omega \sqrt{2{\cal I}}}
-\frac{\imath\delta_{ab}}{k_a^2}I_a^{\frac{1}{k_a}-1}
(\mathcal{H}_{osc}+\omega \sqrt{2{\cal I}}),
\\
\label{B-barB}
\{B_a,\bar B_b\}=
\frac{ 2\imath\omega A_a\bar A_b }{\mathcal{H}_{osc}-\omega \sqrt{2{\cal I}}}
-\frac{\imath\delta_{ab}}{k_a^2}I_a^{\frac{1}{k_a}-1}
(\mathcal{H}_{osc}-\omega \sqrt{2{\cal I}}).
\end{gather}
Hence, we have extended the "holomorphic factorization" formalism to the $N$-oscillator.

\subsection{Coulomb case}
The $2N-2$ locally defined integrals  of the generalized  Coulomb Hamiltonian can be written in the coordinates \eqref{px2}  as follows
 \be
 \label{Mcoul}
  M_a^{Coul}=\left(Z-\frac{\imath\gamma}{2\sqrt{\cal{I}}} \right)u_a^{\frac{1}{k_a}},
  \qquad
 \{ {\cal H}_{Coul},   M_a^{Coul} \}=0.
 \ee
Like in  the previous cases,  only their arguments  produce  conserved quantities
independent from the Liouville integrals  \eqref{H-I}  since
 \be
 \label{absMc}
 \big| M_a^{Coul}\big|^2=\left(\mathcal{ H}_{Coul}+\frac{\gamma^2}{4\cal{I}}\right) I_{a}^\frac{1}{k_a}.
 \ee
They form the following
algebra, which can be verified using the Poisson brackets \eqref{z-z}:
 \be
 \label{M-Mcoul}
\big\{ M_a^{Coul},\overline{M}_b^{Coul}  \big\}
=
\frac{\imath\gamma^2 M_a^{Coul}\overline{M}_b^{Coul} }
    {\sqrt{{2\cal I}}(\gamma^2+4{\cal I}{\cal H}_{Coul})}
- \frac{\imath\delta_{ab}I_a^{\frac{1}{k_a}-1}}{k_a^2}\left({\cal H}_{Coul}+\frac{\gamma^2}{\sqrt{8{\cal I}}}\right),
\nonumber
\ee
\be
  \big\{M_a^{Coul},M_b^{Coul}\big\}=0,
\ee
Let us also present the Poisson brackets of these quantities with Liouville constants of motion
\be
\label{I-Mcoul}\big\{ I_a,M_b^{Coul}\big\}=\frac{\imath \delta_{ab}}{k_b}M_b^{Coul}.
\ee
Similar to the  previous cases, we are forced to take certain powers of the local quantities
\eqref{Mcoul} in order to get
the valid, globally defined  additional constants of motion of the deformed Coulomb problem:
\be
\label{cMcoul}
 {\cal M}_a^{Coul}=\big(M_a^{Coul}\big)^{n_a}=\left(Z-\frac{\imath\gamma}{2\sqrt{\cal{I}}} \right)^{n_a}u_a^{ m_a}.
\ee
Their algebra can be deduced from the Poisson bracket relations \eqref{M-Mcoul} and \eqref{I-Mcoul}.
\\

So, in this Section we extended the method of "holomorphic factorization"  initially developed for the two-dimensional oscillator and Coulomb system, to the superintegrable generalizations of Coulomb and oscillator systems in any dimension.
For this purpose we parameterized the angular parts of these systems by action-angle variables.
To our surprise, we were able to get, in these general terms, the symmetry algebra of these systems.
Notice, that above formulae hold not only on the Euclidean spaces, but for the more general one, if we choose $\mathcal{I}$ be the system with a phase space different from $T_* S^{N-1}$.

\section{Oscillator-Coulomb correspondence}

As is known, the energy surface of the radial oscillator  can  be transformed to the energy surface  of the radial Coulomb problem by
 transformation ${\tilde r}=\lambda r^2$,${\tilde p}_{\tilde r}=p_r/2\lambda r$ where $r, p_r$ are radial coordinate and momentum of oscillator,
  ${\tilde r}, {\tilde p}_{\tilde r}$ are those of Coulomb problem, and $\lambda$ is     an arbitrary positive constant number (see,e.g.\cite{ter,ter1} for the review).
   Extension of oscillator-Coulomb correspondence  from the radial part  to the whole system, as well as to its quantum counterpart yields additional restrictions on the geometry   of configuration spaces.  Namely, only   $N=2,4,8,16$ -dimensional oscillator could be transformed
to the Coulomb system, that is $N=2,3,5,9$ dimensional Coulomb problem. These dimensions are distinguished due to Hopf maps  $S^1/S^0=S^1$,
 $S^3/S^1=S^2$, $S^7/S^3=S^4$, which allow to transform  spherical (angular) part of oscillator to those of  Coulomb problem.
  Indeed,  for the complete correspondence between oscillator end Coulomb system we should be able to transform the
  angular part of oscillator (that is particle on $S^{D-1}$)  to the angular part of Coulomb problem, i.e. to $S^{d-1}$.
  Thus, the only admissible dimensions  are $D=2,4,8,16$ and $d=2,3,5,9$.  In the first three cases we have to reduce
   the initial system by
   $Z_2$, $U(1)$ and $SU(2)$. For the latter case, in spite of many attempts, we do not know rigorous derivation of this correspondence,
 due  to the fact  that $S^7$ sphere has no Lie group structure. Respectively, in the generic case we  get the  extension of two-/three/five- dimensional Coulomb system specified by the presence $Z_2$/Dirac/$SU(2)$ Yang  monopole \cite{hopf}.
 In the  deformed Coulomb and oscillator problems  considered here we do not require  that the angular parts of the systems should be spheres.
 Hence, trying to relate these systems we are not restricted by the  systems of mentioned dimensions.  Instead, we can try to relate the deformed oscillator and Coulomb systems  of the same dimension and find the restrictions to the structure of their angular parts.

Below we  describe this correspondence in terms complex variables introduced in previous Section.
Through this subsection  we will use "untilded" notation  for the  description of oscillator, and  the "tilded" notation for the description of Coulomb system.

The expression of the "Lobachevsky variable" \eqref{px2}
via radial coordinate and momentum  forces to relate the angular parts of oscillator and Coulomb problem by the expression $\tilde{\mathcal{I}}=\mathcal{I}/4$.
The latter  induces the following relations between "angle-like" variables $\Lambda,{\tilde\Lambda}$:  $\tilde{\Lambda}=2\Lambda$.
Altogether  read
\be
{\tilde Z}= \frac{\imath (\bar Z-Z)}{\lambda\sqrt{\mathcal{I}} } Z ,
\quad \tilde{\mathcal{I}}  = \frac{\mathcal{I}}{4},
\quad
\tilde{\Lambda}=2{\Lambda}
\nonumber
\ee
\be
\Updownarrow
\nonumber
\ee
\be
Z=2\sqrt{\lambda} \sqrt[4]{\tilde{\mathcal{I}}} \frac{{\tilde Z}}{\sqrt{\imath({\bar{\tilde Z}}-{\tilde Z})}},
\quad \mathcal{I}=4\tilde{\mathcal{I}},
\quad
 \Lambda=\frac{\tilde{\Lambda}}{2}.
\ee
This transformation is canonical in a sense, that preserve Poisson brackets between  $Z,\bar Z, \Lambda, \mathcal{I}$, and their tilded counterparts.
To make the transformation canonical, we preserve the angular variables unchanged $\tilde{u}_a=u_a$,  which implies to introduce for superintagrable systems the following identification
\be
\tilde{k}_a =\frac{k_a}{2}\qquad\Rightarrow \qquad \tilde{n}_a=n_a,\quad\tilde{m}_a=2m_a.
\ee
Then we can see, that this transformation relates the energy surfaces of oscillator and Coulomb systems:
\be
Z\bar Z+\Omega^2\frac{{2{\cal I}}}{(\imath ({\bar Z}- { Z}))^2}-E_{\rm osc}=0
\qquad \Leftrightarrow \qquad
\frac{2\lambda \sqrt{\mathcal{\tilde{I}} } } {\imath(\bar{\tilde{ Z} }- \tilde{Z})}\left( \tilde{Z}\tilde{\bar Z}-\gamma\frac{\imath(\tilde{\bar Z}-\tilde{Z})}{2\sqrt{\mathcal{I}}}  - {\cal \tilde{E}}_{Coul}\right)=0,
\ee
where
\be
\tilde{\gamma}=\frac{E_{\rm osc}}{\lambda},\qquad \mathcal{\tilde{E}}_{Coul}=-\frac{2\Omega^2}{\lambda^2}.
\ee
The generators of hidden symmetries also transform one into the other on the energy surface
\be
\mathcal{M}_{(a)osc}=\left(\imath\lambda\sqrt[4]{2{\cal \tilde{I}}}\right)^{n_a}\mathcal{M}_{(a)Coul}
\ee
Finally, let us write down the relation between generators of conformal symmetries defined on "tilded" and untilded spaces.
\be
\mathcal{H}_0= \lambda \mathcal{\tilde{H}}_0\sqrt{ 2\mathcal{\tilde{K}}}, \qquad \mathcal{D}=2\mathcal{\tilde{D}},
\qquad \mathcal{K}= \frac{2\sqrt{2\mathcal{\tilde{K}}}}{\lambda}.
\ee

In this Section we  transformed   deformed oscillator into deformed Coulomb problem,  preserving intact angular coordinates.
Performing proper transformations of angular part of oscillator, including its reduction, we can get variety of superintegrable deformations of Coulomb problem. However, they will  belong to the same class of systems  under consideration, since the latter are formulated in most general, action-angle variables, terms.

\section{Spherical and pseudospherical \newline generalizations}
Oscillator and Coulomb systems  admit superintegrable generalizations to $N$-dimensional spheres and two-sheet hyperboloids (pseudospheres),
 which are given by the Hamiltonians \cite{higgs}
\be
\label{Hv}
\mathbb{S}^{N}:
\qquad {\cal H}_V=\frac{p_{\chi}^2}{2 r_0^2}+\frac{{\cal I}}{ r_0^2 \sin^2\chi}+V(\tan\chi),
\nonumber
\ee
\be
\mathbb{H}^{N}:\quad {\cal H}_V=\frac{p_{\chi}^2}{2 r_0^2}+\frac{{\cal I}}{ r_0^2\sinh^2\chi}+V(\tanh\chi)
\ee
with  the  potentials
\begin{align}
\mathbb{S}^{N}:
&\qquad  V_{osc}(\tan\chi)=\frac{r^2_0\omega^2\tan^2\chi}{2},
&& V_{Coul}(\tan\chi)=-\frac{\gamma}{r_0}\cot\chi,
\label{Vsph}
\\
\mathbb{H}^{N}:
&  \qquad  V_{osc}(\tanh\chi)=\frac{r^2_0\omega^2\tanh^2\chi}{2},
&&  V_{Coul}(\tanh\chi)=-\frac{\gamma}{r_0}\coth\chi.
\label{Vhyp}
\end{align}
 Here ${\cal I}$ is a quadratic Casimir element of the orthogonal algebra $so(N)$.  To get  integrable deformations of these systems,  we  replace it, as in Euclidean case,  by some
integrable (angular) Hamiltonian  depending on the action variables \cite{hlnsy}.
The  particular angular Hamiltonian \eqref{angular0} defines superintegrable systems
as in the flat case.
About decade ago the so-called $\kappa$-dependent formalism was developed \cite{kappa,kappa1,kappa2} where the oscillator and Coulomb  systems on plane and on the
  two-dimensional
sphere and  hyperboloid  were described in the unified way.

Introduce, following that papers,
\be
\label{T}
T_{\kappa}=\frac{S_{\kappa}}{C_{\kappa}}
\qquad\text{with}\qquad
C_{\kappa}(x)=
\begin{cases}
\cos{\sqrt{\kappa}x}  &   \kappa>0,  \\
 1  & \kappa=0,\\
 \cosh{\sqrt{-\kappa}x}  &  \kappa<0,
\end{cases}
\nonumber
\ee
\be
\label{T}
S_{\kappa}(x)=
\begin{cases}
\displaystyle
\frac{\sin{\sqrt{\kappa}x}}{{\sqrt{\kappa}}}   & \kappa>0,  \\
 x  & \kappa=0,\\
 \displaystyle
 \frac{\sinh{\sqrt{-\kappa}x}}{\sqrt{-\kappa}}  & \kappa<0,
\end{cases}
\ee
where the parameter $\kappa$  in two-dimensional  case
coincides with the curvature of (pseudo)sphere,
\be
\label{kappa}
\mathbb{S}^{N}: \quad\kappa=\frac{1}{r_0^2},
\qquad\qquad
\mathbb{H}^{N}: \quad\kappa=-\frac{1}{r_0^2}.
\ee
The case  $\kappa=\pm 1$ corresponds to a unit sphere/pseudosphere. 
 For $\kappa\neq 0$ we  identify
 \be
 \label{x-kappa}
 x=r_0\chi=\frac{\chi}{\sqrt{\kappa}},
 \qquad
  p_x= \frac{p_\chi}{r_0} = \sqrt{\kappa} p_\chi.
  \ee
The  "holomorphic factorization" approach to two-dimensional systems  was combined with $\kappa$-dependent formalism by Ranada.
Let  us show that it can be straightly extended to any dimension.
For this purpose introduce   a (pseudo)spherical analog of  $Z$, $\bar Z$ coordinates  and obtain their Poisson bracket:
\be
\label{z-kappa}
Z=\sqrt{|\kappa|}\frac{{p_\chi}}{\sqrt2}+\frac{\imath\sqrt{\cal I}}{T_\kappa},
\qquad
\{\bar{Z},Z\}=\frac{\imath(Z-\bar{Z})^2}{2\sqrt{2\cal I}}-\imath\kappa {\sqrt{{2\cal I}}}.
\ee
The Poisson brackets between $Z$,  $u_a$ and ${\bar u}_a$ remain unchanged [see relations \eqref{z-z}].

In these  terms the $\kappa$-deformed Hamiltonian reads
\be
\label{Hkap}
\mathcal{H}_{osc/Coul}=\mathcal{H}_0+V_{osc/Coul},
\qquad
\mathcal{H}_{0}=\frac{p_{r}^2}{2}+\frac{{\cal I}}{S_{\kappa}^2}+\kappa  \mathcal{ I}
= Z \bar Z+\kappa \mathcal{I},
\ee
where using \eqref{T}, \eqref{kappa},
\eqref{x-kappa}, \eqref{z-kappa},   the oscillator and Coulomb potentials on sphere \eqref{Vsph} can be expressed
as follows:
\be
\label{Vkap}
V_{osc}=\frac{\omega^2T^2_{\kappa}}{2}=-\frac{{2\omega^2{\cal I}}}{ ({\bar Z}- { Z})^2} ,
\qquad
V_{Coul}=-\frac{\gamma}{T_{\kappa}}=-\imath\gamma\frac{\bar Z-Z}{2\sqrt{\mathcal{I}}}.
\ee
The (local  and global) constants of motion and related quantities have the same expressions  in terms of  $Z$, $\bar Z$ as in the flat case, with the Hamiltonians shifted  in agreement with \eqref{Hkap}
\be
\label{shift}
{\cal H}\to {\cal H}-\kappa{\cal I}.
\ee
%

For the free system on sphere, ${\cal H}_0$, the most of  Poisson brackets among the integrals
survive from the flat case [see relations \eqref{M-M}, \eqref{M-bMcal} and \eqref{M-Mcal}]. The only brackets, which acquire
 extra $\kappa$-dependent terms, are:
\begin{gather}
\label{M-Mkap}
\{M_a,\overline{M}_b  \}
=\left(\frac{ \imath\kappa \sqrt{ 2{\cal I}} } { {\cal H}_0-\kappa{\cal I} }
-\frac{\imath\delta_{ab}}{k_a^2I_a}\right)
M_a\overline{M}_b
=-\frac{\imath\delta_{ab}}{k_a^2} I_a^{\frac{1}{k_a}-1}({\cal H}_0-\kappa{\cal I})
+\frac{ \imath\kappa \sqrt{ 2{\cal I}} } { {\cal H}_0-\kappa{\cal I} } M_a\overline{M}_b,
\\
\{{\cal M}_a,\overline{{\cal M}}_b  \}
=\imath\left(\frac{ \kappa n_an_b\sqrt{ 2{\cal I}} } { {\cal H}_0-\kappa{\cal I} }
-\frac{ m_a^2\delta_{ab}}{I_a}\right){\cal M}_a\overline{{\cal M}}_b.
\end{gather}

Let us write down also the  deformation of conformal algebra \eqref{ca}
\be
\{\mathcal{H}_0,\mathcal{D}\}=2(\mathcal{H}_0-\kappa { \cal I})(1+2\kappa \mathcal{K}) ,
\qquad
\{\mathcal{H}_0, \mathcal{K} \}=\mathcal{D}(1+2\kappa \mathcal{K}) ,
\qquad
\{\mathcal{D},\mathcal{K}\}=2\mathcal{K}(1+2\kappa \mathcal{K} ).
\ee

\smallskip

For the Coulomb problem on sphere, the Poisson brackets between the local integrals \eqref{I-Mcoul} remain unaffected, while
the relations \eqref{M-Mcoul} undergo a similar modification:
\be
\label{M-Mcoulkap}
\begin{aligned}
\big\{ M_a^{Coul},\overline{M}_b^{Coul}  \big\}
&=\left[\frac{\imath\sqrt{2{\cal I}}\left(\frac{\gamma^2}{4{\cal I}^2}+\kappa\right)}
    {{\cal H}_{coul}-\kappa{\cal I}+\frac{\gamma^2}{4{\cal I}^2}}
    -\frac{\imath\delta_{ab}}{k_a^2I_a}\right] M_a^{Coul}\overline{M}_b^{Coul}
\\
&=
\imath\sqrt{2{\cal I}}\left(\frac{\gamma^2}{4{\cal I}^2}+\kappa\right)\frac{ M_a^{Coul}\overline{M}_b^{Coul} }
    {{\cal H}_{coul}-\kappa{\cal I}+\frac{\gamma^2}{4{\cal I}^2}}
- \frac{\imath\delta_{ab}}{k_a^2}I_a^{\frac{1}{k_a}-1}\left({\cal H}_{Coul}-\kappa{\cal I}+\frac{\gamma^2}{4{\cal I}^2}\right).
\end{aligned}
\ee

\smallskip

Consider now the spherical system \eqref{Hv} with the oscillator potential.
Line for the flat case, the integrals of motion are based on  the simpler local
quantities $A$ and $B$,
\be
\label{Aa-sp}
A_a=(z+\frac{\imath\omega T_\kappa}{\sqrt2})u_a^{\frac1{k_a}},
\qquad
B_a=(z-\frac{\imath\omega T_\kappa}{\sqrt2})u_a^{\frac1{k_a}},
\qquad
 {\cal M}^{osc}_a=(A_aB_a)^{n_a},
\ee
which evolve in time under  the following rule:
\be
\label{H-A-sp}
\{{\cal H}_{osc}, A_a\}=\imath\omega(1+\kappa T_\kappa^2)  A_a,
\qquad
\{{\cal H}_{osc}, B_a\}=-\imath\omega(1+\kappa T_\kappa^2)   B_a.
\ee
They are $\kappa$-deformations of the harmonic oscillating quantities
\eqref{Aa}, \eqref{H-A} in the flat case. Unlike them, they do not oscillate
harmonically, but
the product $A_a B_b$ is still preserved.

The Poisson brackets between local quantities can be  calculated explicitly
giving rise to $\kappa$-deformations of the relations \eqref{A-barB},
\eqref{A-barA}, \eqref{B-barB}:
\begin{gather}
\label{A-Bkap}
\{A_a, B_b\}=-\frac{\imath \kappa\omega  T_\kappa^2}{z^2 + \frac{\omega ^2T_\kappa^2}{2}}A_a  B_b,
\qquad
\{A_a,\bar B_b\}=-\frac{\imath \delta_{ab}}{k_a^2 I_a}A_a\bar B_a
+\frac{ \imath\kappa  \sqrt{ 2{\cal I}}  A_a\bar A_b} { {\cal H}_{osc}-\kappa{\cal I} +\omega  \sqrt{2{\cal I}}},
\\
\label{A-barAkap}
\{A_a,\bar A_b\}=\imath
\frac{\kappa(\sqrt{2{\cal I}} -2\omega T_\kappa) -2\omega  }{\mathcal{H}_{osc}-\kappa {\cal I}+\omega \sqrt{2{\cal I}}}A_a\bar A_b
-\frac{\imath\delta_{ab}}{k_a^2}I_a^{\frac{1}{k_a}-1}
(\mathcal{H}_{osc}-\kappa {\cal I}+\omega \sqrt{2{\cal I}}),
\\
\label{B-barBkap}
\{B_a,\bar B_b\}=\imath
\frac{\kappa(\sqrt{2{\cal I}} +2\omega T_\kappa) +2\omega  }{\mathcal{H}_{osc}-\kappa {\cal I}-\omega \sqrt{2{\cal I}}}A_a\bar A_b
-\frac{\imath\delta_{ab}}{k_a^2}I_a^{\frac{1}{k_a}-1}
(\mathcal{H}_{osc}-\kappa {\cal I}-\omega \sqrt{2{\cal I}}).
\end{gather}
The Poisson brackets between the true integrals of motion ${\cal M}^{osc}_a$, ${\cal M}^{Coul}_a$ and their conjugate
are based on the local brackets
\eqref{M-Mcoulkap}, \eqref{A-Bkap}, \eqref{A-barAkap}, \eqref{B-barBkap}
and can be easily obtained.

\section{Examples of spherical part}
In previous Sections we extended "holomorphic factorization approach" to higher-dimensional superintegrable systems with oscillator and Coulomb potentials, including those on spheres and hyperboloids.
For this purpose we separated the "radial" and "angular" variables in these systems. Then we combined the radial coordinate and momentum in single complex coordinate parameterizing Klein model of Lobachevsky space, and combined  "angular" coordinates  and their conjugated momenta in complex coordinates
 by the use of action-angle variables.
However, action-angle variables are not in common use  in present math-physical society, and their explicit expressions are  not common even for the
such  textbook models like oscillator and Coulomb problems.

For clarifying  the relation of the above formulations of constants of motion with their conventional representations  first present the
action-angle variables of the angular part(s) of non-deformed, oscillator and Coulomb systems (on Euclidean space, sphere and hyperboloids).
 Its  Hamiltonian is given by  the quadratic Casimir element of $so(N)$ algebra on $(N-1)$-sphere, ${\cal I}=L_{N}^2/2$. It can be decomposed by the eigenvalues of the embedded $SO(a)$ angular momenta  defining the action variables
 $I_a $. For the details of derivation  of  their explicit expressions, for those of  conjugated angle variables we refer to   Appendix in Ref.~\cite{hlnsy}.
  The action variables are given by the expressions
  \be
  I_a=\sqrt{j_{a+1}}-\sqrt{j_{a}},\quad {\rm where}\quad j_{a+1}=p^{2}_{a}+\frac{j_{a}}{\sin^2 \theta_{a}},\qquad j_0=0,\qquad a=1,\ldots N-1.
  \ee
 This gives rise   the angular Hamiltonian which belongs to the family \eqref{angular0}
\be
{\cal I}=\frac12\left(\sum_{a=1}^{N-1}I_a \right)^2.
\ee
Its substitution to the Hamiltonians \eqref{flat},\eqref{Hv} leads to well-known oscillator and Coulomb systems on the Euclidean spaces, spheres and hyperboloids.

The expressions for angle variables are more complicated,
\be\Phi_a=\sum_{l=a}^{N-1}a_l+ \sum_{l=a+1}^{N-1}b_l,
\ee
where
\be
a_l=\arcsin\sqrt{\frac{j_{l+1}}{j_{l+1}-j_{l}}} \cos{\theta_l},\quad b_l=\arctan \frac{\sqrt{j_l}\cos\theta_l}{p_l\sin\theta_l}.
\ee
Direct transformations give the following
expressions for $u_a $  coordinates:
\be
u_a=\sqrt{\sqrt{j_{a+1}}-\sqrt{j_a}}\;{\rm e}^{\imath a_a} \prod_{l=a+1}^{N-1}{\rm e}^{\imath(a_l+b_l)},
\ee
with
\be
{\rm e}^{\imath a_l}=
\frac{p_l\sin\theta_l +\imath\sqrt{j_{l+1}}\cos\theta_l} {\sqrt{j_{l+1}-j_{l}}  },\qquad {\rm e}^{\imath b_l}=\frac{p_l\sin\theta_l +\imath\sqrt{j_{l}}\cos\theta_l} {\sqrt{j_{l+1}-j_{l}} \sin\theta_l }
\ee
With these expressions at hand we can express  ``holomorphic representation" of constants of motion via initial coordinates.
In two-dimensional  case it has transparent  relation with conventional representations of hidden constants of motion, like Fradkin tensor (for the oscillator) and Runge-Lenz vector (for Coulomb problem). In the higher dimensional cases the relation of these two representations is more complicated.

\smallskip

This construction could easily be modified to the system whose Hamiltonian is given in the angle variables by the
 generic expression \eqref{angular0}. We define it
by the recurrence relation
\be
{\cal I}\equiv \frac12 {j_N}, \qquad j_{a}=p^2_{a-1}+\frac{j_{a-1}}{\sin^2 k_{a-1}\theta_{a-1}},\qquad a=1,\ldots N-1, \qquad j_0=c_0.
\ee
It describes  particle moving on the space (spherical segment)   equipped with the diagonal metric
\be
ds^2=g_{ll}(d\theta_l)^2, \qquad g_{N-1.N-1}=1,\qquad  g_{ll}=\prod_{m=l}^{N-1}\sin^2 k_{m}\theta_{m}
\ee
and interacting with the  potential field
\be
U=\frac{c_0}
{ \prod_{l=1}^{N-1} \sin^2 k_l \theta_l}.
\ee
Redefining the angles, $\theta_a\to \theta_{a}/k_a$, we can represent the above metric  in the form
\be
ds^2=\frac{1}{k_{a}^2}\prod_{a=1}^{N-1}\sin^2\theta_{a}(d\theta_a)^2.
\ee

It is obvious, that  the functions $j_k(\theta_a , p_a)$ define commuting constants of motions of the system.
Similar to derivation given in Appendix of Ref.~\cite{hlnsy} we can use  action-angle variable formulation, and find that the Hamiltonian is given by the expression \eqref{angular0}. The
 action variables are related  with the initial ones by the expressions
\be
{I}_a=\frac{1}{2\pi}\int^{\theta_{min}}_{\theta_{max}}\sqrt{j_{a+1}-\frac{j_{a}}{\sin^2k_a \theta_a}}\,d\theta_a =\frac{\sqrt{j_{a+1}}-\sqrt{j_{a}} }{k_a}  \quad\Rightarrow\quad j_a=\Bigg( \sum^{N-1}_{a=1}k_a I_a+ c_0\Bigg)^2.
  \ee
  The angle variables read
  \be
  \Phi_a=\sum^{N-1}_{l=a} \frac{k_a}{k_l} a_l + \sum_{l=a+1}^{N-1} \frac{k_a}{k_l}b_l,
\nonumber
\ee
\be
  a_l=\arcsin \sqrt{\frac{j_{l+1}}{j_{l+1}-j_{l}}}\cos{k_l\theta_l},
  \qquad  b_l=\arctan\frac{\sqrt{j_l}\cos k_l\theta_l}{p_l\sin k_l\theta_l}.
  \ee
Thus,
\be
u_a= \frac{1}{k_a}\sqrt{\sqrt{j_{a+1}}-\sqrt{j_a}}  \prod_{l=a}^{N-1} \left(\frac{p_l\sin k_l\theta_l +\imath\sqrt{j_{l+1}}\cos k_l\theta_l} {\sqrt{j_{l+1}-j_{l}}  }\right)^{\frac{k_a}{k_l}}\times
\nonumber\ee
\be
\times\prod_{l=a+1}^{N-1}\left(\frac{p_l\sin k_l \theta_l +\imath\sqrt{j_{l}}\cos k_l \theta_l} {\sqrt{j_{l+1}-j_{l}} \sin\theta_l }\right)^{\frac{k_a}{k_l}}.
 \ee
Hence, we constructed the superintegrable system with higher order constants of motion, which admits separation of variables.
Since the classical spectrum of its angular part is isospectral with the   "angular  Calogero model", we can state that they become, under appropriate choice of constants $k_i$, $c_0$,  canonically equivalent with angular part of rational Calogero model \cite{flp}.
In fact this means equivalence of these two systems.  However, we can't present explicit mapping of one system to other.

Now consider the spherical Hamiltonian of the particle moving near horizon of the external Myers -Perry   black hole in odd dimensions ($2n+1$)\cite{Galaj1}. Although one deals with the relativistic system , the initial Hamiltonian can be brought to $n$-dimensional non-relativistic form.
Angular part of it in terms of spherical variables will have the following form.
\be
{\cal I}=\frac{1}{2}F_{n-1}, \quad
F_a=P_{\theta_a}^2+\frac{g_{a+1}^2}{\cos^2{\theta_a}}+\frac{F_{a-1}}{\sin^2 {\theta_{a}}}
\ee
As was mentioned ${\cal I}$ is identified with the Casimir element of the conformal group.
 The aim is to describe this system in terms of complex variables ($u_a$). Firstly the introduction of action-angle variables is needed. Action variables can be computed.
\be
I_a=\frac{1}{2\pi}\int d\theta_a P_{\theta_a}=\frac{1}{2}(\sqrt{F_a}+\sqrt{F_{a-1}}-|g_{a+1}|)
\ee
Inverting this relation one finds.
\be
{\cal I}=\frac{1}{2}\left(2\sum_{a=1}^{N-1} I_a + \sum_{a=1}^{N}|g_a| \right)^2
\ee
Since action and angle variables are canonically conjugated
corresponding angle variables can be found via taking derivative of an action.
\be
\Phi_a=\frac{\partial S}{\partial I_a}=\sum_{l=a}^{n-1}\arcsin X_l+2\sum_{l=a+1}^{n-1}\arctan Y_{l}
\ee
where
\be
X_l=\frac{(F_l+F_{l-1}-g_{l+1}^2)-2F_{l}\sin^2{\theta_l}}{\sqrt{(F_{l-1}-F_l-g_{l+1}^2)^2-4F_l g_{l+1}^2}}
\ee
\be
Y_l=2\frac{(F_l+F_{l-1}-g_{l+1}^2)P_{\theta_l}\sin \theta_l \cos \theta_l-\sin^2 \theta_l\sqrt{F_l(F_l+F_{l-1}-g_{l+1}^2)^2-F_{l}^2F_{l-1}} }{\sqrt{F_{l-1}}(F_l+F_{l-1}-g_{l+1}^2-2F_{l}\sin^2{\theta_l})}
\ee
 $u_a$ variable contains exponents of angle variables and it is useful to give the expressions of these exponents.
\be
e^{i\arcsin X_l}=\sqrt{1-X^2_l}+\imath X_l=\frac{\sqrt{F_l}P_{\theta_l}\sin2\theta_l-\imath(F_l\cos 2\theta_l+F_{l-1}-g_{l+1}^2)}{\sqrt{(F_{l-1}-F_l-g_{l+1}^2)^2-4F_l g_{l+1}^2}}
\label{exp1}
\ee
\be
e^{2 \imath \arctan Y_l}=\frac{1+\imath Y_l}{1- \imath Y_l}=
\nonumber
\ee
\be
\scriptstyle
=\frac{\imath \sqrt{F_{l-1}}(F_{l-1}+F_l\cos^2 2\theta_l-g_{l+1}^2)
-P_{\theta_l}\sin 2\theta_l(F_{l-1}+F_l-g_{l+1}^2)
+2\sin^2 \theta_l\sqrt{F_l(F_{l-1}^2+F_{l-1}(F_l-2g_{l+1}^2)+(F_l-g_{l+1}^2)^2)}}
{\imath \sqrt{F_{l-1}}(F_{l-1}+F_l\cos^2 2\theta_l-g_{l+1}^2)+P_{\theta_l}\sin 2\theta_l(F_{l-1}+F_l-g_{l+1}^2)
-2\sin^2 \theta_l\sqrt{F_l(F_{l-1}^2+F_{l-1}(F_l-2g_{l+1}^2)+(F_l-g_{l+1}^2)^2)}}
\label{exp2}
\ee
And finally the expression of $u_a$ can be written.
\be
\begin{split}
u_a=
\sqrt{\frac{1}{2}(\sqrt{F_a}+\sqrt{F_{a-1}}-|g_{a+1}|)}\prod_{l=a}^{n-1}\left(\frac{\sqrt{F_l}P_{\theta_l}\sin2\theta_l-\imath(F_l\cos 2\theta_l+F_{l-1}-g_{l+1}^2)}{\sqrt{(F_{l-1}-F_a-g_{l+1}^2)^2-4F_l g_{l+1}^2}}\right)\times
\\
\scriptstyle
\times\prod_{l=a+1}^{n-1}
\left(
\frac{\imath \sqrt{F_{l-1}}(F_{l-1}+F_l\cos^2 2\theta_l-g_{l+1}^2)
-P_{\theta_l}\sin 2\theta_l(F_{l-1}+F_l-g_{l+1}^2)
+2\sin^2 \theta_l\sqrt{F_l(F_{l-1}^2+F_{l-1}(F_l-2g_{l+1}^2)+(F_l-g_{l+1}^2)^2)}}
{\imath \sqrt{F_{l-1}}(F_{l-1}+F_l\cos^2 2\theta_l-g_{l+1}^2)+P_{\theta_l}\sin 2\theta_l(F_{l-1}+F_l-g_{l+1}^2)
-2\sin^2 \theta_l\sqrt{F_l(F_{l-1}^2+F_{l-1}(F_l-2g_{l+1}^2)+(F_l-g_{l+1}^2)^2)}}\right)
\end{split}
\ee

\section{Concluding remarks}
In this chapter we discuss   Tremblay-Turbiner-Winternitz and Post-Wintenitz systems
and their relation  with $N$-dimensional rational Calogero model with oscillator and Coulomb potentials. We write the hidden symmetries of this systems using complex variables.
Then we investigated  superintegrable deformations of oscillator and Coulomb problems separating their "radial" and "angular" parts, where the latter was described in terms of action-angle variables. We encoded phase space coordinates in the complex ones: the complex coordinate $z$ involved radial variables parameterizing Klein model of Lobachevsky plane, and complex coordinates $u_a$ encoding  action-angle variables of the angular part.
Then we combined the whole set of constants of motion (independent from Hamiltonian)  in  $N-1$ holomorphic  functions $\mathcal{M}_a$, generalizing the so-called "Holomorphic factorization" earlier developed for two-dimensional generalized oscillator/Coulomb systems. Then we presented their algebra, which among nontrivial relations possesses chirality property $\{\mathcal{M}_a, \mathcal{M}_a\}=0$. Hence, presented representation can obviously considered as a classical trace of ''quantum factorization" of respective Hamiltoinian. Seems that  it could be used for the construction of  supersymmetric extensions of these systems. The lack of given representation is the use of  the action-angle formulation of the angular parts of the original systems.

In this context  one should mention the earlier work \cite{hkl}, where symmetries of the angular parts of conformal mechanics (and those with additional oscillator potential) were related with the symmetries of the whole system by the use of  coordinate $z$ and conformal algebra generators \eqref{uk2}.
That study was done in most general terms, without referring to action-angle variables and to specific form of angular part. Quantum mechanical aspects were also considered there.
Hence, it seems  to be natural to combine these two approaches for and at first,  exclude the action-angle argument from present formulations, and at second, use presented constructions for the quantum considerations of systems, in particular, for construction of spectrum and wavefunctions within operator approach. We are planning to present this elsewhere.

\newpage
\chapter{$\mathbb{C}^N$-Smorodinsky-Winternitz system}
\setcounter{equation}{0}
\setcounter{section}{0}
\section{Introduction}

Current chapter is based on my single-authored paper \cite{shmsw}.

 The one-dimensional singular oscillator is a textbook example of a system which is exactly solvable both on classical and quantum levels.The sum of its $N$ copies, i.e. $N$-dimensional singular isotropic oscillator is, obviously, exactly solvable as well.
It is given by the Hamiltonian
\be\label{SWR0}
H=\sum_{i=1}^{N}I_i,\qquad{\rm with}\quad  I_i=\frac{p_i^2}{2}+\frac{g_i^2}{2x_i^2}+\frac{\omega^2x_i^2}{2},\qquad \{p_i,x_j\}=\delta_{ij},\quad \{p_i,p_j\}=\{x_i,x_j\}=0
\ee
  It is not obvious that in addition to Liouville Integrals $I_i$ this system possesses  supplementary series of constants of motion, and is respectively, {\sl maximally superintegrable}, i.e. possesses $2N-1$ functionally independent constants of motion.
All these constants of motion are of the second order on momenta.
 It seems that this  was first noticed  by Smorodinsky and Winternitz, who then investigated superintegrability properties of this system in great detail \cite{sw1,sw2,sw3}. For this reason this model is sometimes called Smorodinsky-Winternitz system and we will use this name as well.  For sure, such a simple and internally rich system would attract wide attention in the community of mathematical and theoretical physics, and that is one of the main reasons why  there are so many publications devoted to its study and further generalizations. Besides the above-mentioned publications, we should as well mention the  references \cite{Evans,evans,evans2,heinzl,pogosyan,Hoque,Hoque2,miller}(see   the recent  PhD thesis on this subject  with expanded list of references  \cite{PhD}). Notice also  that  Smorodinsky-Winternitz system  is a simplest case of the generalized Calogero model(with oscillator potential) associated with an arbitrary Coxeter root system \cite{algebra1}. Thus, one hopes that observations done in this simple model could be somehow extended to the Calogero models.
 There is a  well-known superintegrable generalization of the oscillator to sphere, which is known as Higgs oscillator\cite{higgs,leemon}
 Smorodinsky-Winternitz  model  admits superintegrable generalization  of the  sphere as well \cite{pogosyanSph}.  Though it was first suggested  by Rosochatius in XIX century (without noticing its superintegrability) \cite{rosochatius}, it  was later rediscovered  by many  other authors  as well (e.g. \cite{oksana,Galaj1}) . Superintegrable generalization of Calogero model on the sphere also exists \cite{rapid,rapid1,rapid2}.

  In this chapter we consider simple generalization of the Smorodinsky-Winternitz system {\sl interacting with constant magnetic field}.
  It is defined on the $N$-dimensional complex Euclidian space parameterized by the coordinates $z^a$ by the Hamiltonian
  \be\label{Hamplan0}
\mathcal{H}=\sum_{a=1}^N\left(\pi_a\bar{\pi}_a+\frac{g_a^2}{z^a\bar{z}^a}+\omega^{2}z^a\bar{z}^a\right),\qquad{\rm with}\quad  \{\pi_a,z^b\}=\delta_{ab},\quad \{{\pi}_a,\bar{\pi}_b\}=\imath B \delta_{ab}
\ee
The (complex) momenta  $\pi_a$  have nonzero Poisson brackets due to the presence of magnetic field  with   magnitude $B$ \cite{SupHam,march}. We will refer this model as $\mathbb{C}^N$-Smorodinsky-Winternitz system. For sure, in the absence of magnetic field this model could be easily reduced to the conventional Smorodinsky-Winternitz model, but the presence of magnetic field could have  nontrivial impact which will be studied in this chapter. So, {\sl our main goal is to investigate the whole symmetry algebra of this system}.
Notice that this is not only for academic interest: the matter is that  $\mathbb{C}^1$-Smorodinsky-Winternitz system is a popular model for the qualitative study of the so-called quantum ring \cite{chp,chp1,chp2}, and the study of its behaviour in external magnetic field is quite a natural task. Respectively, $\mathbb{C}^N$-Smorodinsky-Winternitz  could be viewed as an ensemble of $N$ quantum rings interacting with external magnetic field. So  investigation of its symmetry algebra is of the physical importance.

 Since $\mathbb{C}^2$-Smorodinsky-Winternitz system is manifestly invariant with respect to $U(1)$ group action, we can  perform its Kustaanheimo-Stiefel transformation, in order to obtain three-dimensional Coulomb-like system. It  was done about ten years ago \cite{mardoyan0}, but in the absence of magnetic field in initial system. Repeating this transformation for the system with constant magnetic field we get unexpected result:  it has no qualitative impact in the resulting system, which was referred in \cite{mardoyan1} as "generalized MICZ-Kepler system"\cite{mic,zw,kep}. In addition, we obtain, in this way, the explicit expression of its symmetry generators and their  symmetry algebra, which as far as we know was not constructed before.

We already mentioned that both oscillator and Smorodinsky-Winternitz system admit superintegrable generalizations to the spheres.
On the other hand the isotropic oscillator on $\mathbb{C}^N$ admits the superintegrable generalization on the complex projective space, moreover, the inclusion of constant magnetic field preserves all symmetries of that system \cite{cpn,qcpn}.
It will be shown that introduction of a constant magnetic field doesn't change these properties of the $\mathbb{C}^N$-Smorodinsky-Winternitz system.
Thus, presented model could be viewed as a  first step towards the construction of the analog of Smorodinsky-Winternitz system on $\mathbb{CP}^N$.

The chapter  is organized as follows.

In the {\sl Section 3.2} we review the main properties of the conventional ($\mathbb{R}^N$-)Smorodinsky-Winternitz system, presenting explicit expressions of its symmetry generators, as well as wavefunctions and Energy spectrum.
We also present  symmetry algebra in a very simple, and seemingly new form via  redefinition of  symmetry generators.

In the  {\sl Section 3.3} we present  $\mathbb{C}^N$-Smorodinsky-Winternitz system in a constant magnetic field, find  the explicit expressions of its  constants of motion. We compute their algebra and find that it is independent from the magnitude of constant magnetic field. Then we quantize a system and obtain wavefunctions and energy spectrum. We notice that the $\mathbb{C}^N$-Smorodinsky-Winternitz system has the same degree of degeneracy as $\mathbb{R}^N$- one, due to the lost part of additional symmetry.

In the  {\sl Section 3.4} we perform Kustaanheimo-Stiefel transformation of the $\mathbb{C}^2$-Smorodinsky-Winternitz system in constant magnetic field and obtain, in this way, the so-called ``generalized MICZ-Kepler system". We find  that  constant magnetic field appearing in the initial system, does not lead to any changes in the resulting one.

In the {\sl Section 3.5} we discuss the obtained results and possibilities of further generalizations. Possible extensions of discussed system include supersymmetrization and quaternionic generalization as well as generalization of  these systems in curved background.

\section{Smorodinsky-Winternitz system on $\mathbb{R}^N$}

Smorodinsky-Winternitz system is defined  as a sum of $N$ copies of one-dimensional singular oscillators \eqref{SWR0},
%
each of them defined by  generators $I_i$ which  obviously form its  Liouville integrals
$\{I_i,I_j\}=0$.
About fifty years ago it was  noticed that this system possesses additional set of constants of motion given by the expressions \cite{sw1}
\be
I_{ij}=L_{ij}L_{ji}-\frac{g_i^2x_j^2}{x_i^2}-\frac{g_j^2x_i^2}{x_j^2},\qquad \{ I_{ij}, H\}=0,
\ee
where $L_{ij}$ are the generators of $SO(N)$ algebra,
\be
 L_{ij}=p_i x_j-p_j x_i\; :\quad \{L_{ij},L_{kl}\}=\delta_{ik}L_{jl}+\delta_{jl}L_{ik}-\delta_{il}L_{jk}-\delta_{jk}L_{il}.
\ee
The generators $I_{ij}$ provides additional $N-1$ functionally independent constants of motions and so this system is maximally superintegrable.
These generators define highly nonlinear symmetry algebra,
\be
\{I_i,I_{jk}\}=\delta_{ij}S_{ik}-\delta_{ik}S_{ij},
\quad
\{I_{ij},I_{kl}\}=\delta_{jk}T_{ijl}+\delta_{ik}T_{jkl}-\delta_{jl}T_{ikl}-\delta_{il}T_{ijk}
\ee
where
\be
S^2_{ij}=-16(I_iI_jI_{ij}+I_i^2g_j^2-I_j^2g_i^2+\frac{\omega^2}{4}I_{ij}^2-g_i^2g_j^2\omega^2)
\ee
\be
T_{ijk}^2=-16(I_{ij}I_{jk}I_{ik}+g_k^2I_{ij}^2+g_j^2I_{ik}^2+g_i^2I_{jk}^2-4g_i^2g_j^2g_k^2).
\ee
The  generators $S^2_{ij}$ and $T^2_{ijk}$ are of the sixth-order in momenta and antisymmetric over $i,j,k$ indices.
The above symmetry algebra could be written in a compact form if we introduce the notation
\be
M_{ij}=I_{ij},\quad M_{0i}=I_i,\quad M_{ii}=g_i^2,\quad M_{00}=\frac{\omega^2}{4}, \quad R_{ijk}=T_{ijk},\quad R_{ij0}=S_{ij}.
\ee
Then one can introduce capital letters   which will take values from $0$ to $N$. It is worth to mention that $M_{IJ}$ is  symmetric, whereas $R_{IJK}$ is antisymmetric with respect to all indices.
In this terms  the whole symmetry algebra of Smorodinsky-Winternitz system  reads
\be
\{M_{IJ},M_{KL}\}=\delta_{JK}R_{IJL}+\delta_{IK}R_{JKL}-\delta_{JL}R_{IKL}-\delta_{IL}R_{IJK}
\ee
where
\be
R_{IJK}^2=-16(M_{IJ}M_{JK}M_{IK}+M_{IJ}^2M_{KK}+M_{IK}^2M_{JJ}+M_{KL}^2M_{II}-4M_{II}M_{JJ}M_{KK})
\ee
One important fact should be mentioned, although in this algebra on the right side we have sum of many terms (square roots), only one term always survives, since in case of three indices are equal, the result is automatically $0$. Consequently in this algebra we always have one square root on the right hand side.
 Quantum-mechanically the maximal superintegrability is reflected in the dependence of its energy spectrum  on the single,``principal" quantum number only.
Having in mind that in Cartesian coordinates the system decouples to the set of one-dimensional singular oscillators,
we can immediately extract the expressions for its wavefunctions and spectrum from the standard textbooks on quantum mechanics, e.g. \cite{landau},
\be
E_{n|\omega}=\hbar\omega\Big(2n+N+\sum_{i=1}^N\sqrt{\frac14+\frac{g_i^2}{\hbar^2}}\Big),\qquad \Psi=\prod_{i=1}^N\psi(x_i,n_i),
\quad n=\sum_{i=1}^N n_i
\ee
where
\be
\psi(x_{i},n_i)=F\Big(-n_i,1+\sqrt{\frac14+\frac{g_i^2}{\hbar^2}},\frac{\omega x_i^2}{\hbar}\Big)\Big(\frac{\omega x_i^2}{\hbar}\Big)^{\frac{1+\sqrt{1+4g_i^2/\hbar^2}}{4}}e^{-\frac{\omega x_i^2}{2\hbar}}
\ee
Here $F$ is the confluent hypergeometric function.
With these expressions at hands we are ready to study Smorodinsky-Winternitz system on complex Euclidean space in the presence of constant magnetic field.

\section{$\mathbb{C}^N$-Smorodinsky-Winternitz system}
Now let us study $2N$-dimensional  analog of Smorodinsky-Winternitz system interacting with constant magnetic field.
It is defined by \eqref{Hamplan0} and could be viewed as an analog of Smorodinsky-Winternitz system on complex Euclidian space $\left (\mathbb{C}^N, ds^2=\sum_{a=1}^N dz^ad{\bar z}^a \right)$. Thus, we will refer it as $\mathbb{C}^N$-Smorodinsky-Winternitz system.
The analog of SW-system which respects the inclusion of constant magnetic field is defined as follows,
\be\label{Hamplan}
\mathcal{H}=\sum_{a}I_a,\qquad I_{a}=\pi_a\bar{\pi}_a+\frac{g_a^2}{z^a\bar{z}^a}+\omega^{2}z^a\bar{z}^a\;,
\ee
where $z^a, \pi_a$  are complex (phase space) variables with the following non-zero Poisson bracket relations
\be
\{\pi_a,z^b\}=\delta_{ab},\qquad \{\bar{\pi}_a,\bar{z}^b\}=\delta_{ab}, \quad \{{\pi}_a,\bar{\pi}_b\}=\imath B \delta_{ab}.
\ee

For sure, it can be interpreted  as a sum of  $N$ two-dimensional singular oscillators interacting with constant magnetic field perpendicular to the plane. It is obvious that  in addition to $N$ commuting  constants of motion $I_a$
this system has another set of $N$ constants of motion defining
 manifest  $(U(1))^N$ symmetries of the system
\be
L_{a\bar{a}}=\imath(\pi_a z^a-\bar{\pi}_a\bar{z}^a)-{B}z^a{\bar z}^a\; :\{L_{a\bar a}, \mathcal{H}\}=0
\ee
and supplementary, non-obvious,   set of constants of motion defined in complete analogy with those of conventional Smorodinsky-Winternitz system:
\be
I_{ab}=L_{a\bar b}L_{b\bar a}+\Big(\frac{g_a^2z^b\bar{z}^b}{z^a\bar{z}^a}+\frac{g_b^2z^a\bar{z}^a}{z^b\bar{z}^b}\Big),\quad
\{I_{ab},\mathcal{H}\}=0,\qquad
 a\neq b
\ee
with $L_{a\bar b}$ being generators of $SU(N)$ algebra
\be\label{SUN}
L_{a\bar b}=\imath(\pi_az^b-\bar{\pi}_b\bar{z}^a)-{B}\bar{z}^a z^b\;:\quad \{L_{a\bar b},L_{c\bar d}\}=i\delta_{a\bar d}L_{c\bar{b}}-i\delta_{c\bar b}L_{c\bar d}.
\ee
These symmetry generators, and $I_a$ obviously commute with  $L_{a\bar a}$ due to manifest $U(1)^N$ symmetry
\be
\{L_{a\bar a}, I_b\}=\{L_{a\bar a}, I_{bc}\}= \{L_{a\bar a}, L_{b\bar b}\}=\{I_{a}, I_b\}=0
\ee
The rest Poisson brackets between them are highly nontrivial
\be
\{I_{a},I_{bc}\}= \delta_{ab}S_{ac}-\delta_{ac}S_{ab},\qquad \{I_{ab},I_{cd}\}=\delta_{bc}T_{abd}+\delta_{ac}T_{bcd}-\delta_{bd}T_{acd}-\delta_{ad}T_{abc} ,
\ee
where
\be
S_{ab}^2=4I_{ab}I_aI_b-(L_{a\bar a}I_b+L_{b \bar b}I_a)^2-4g_a^2I_b^2-4g_b^2I_a^2-4\omega^2I_{ab}(I_{ab}-L_{a\bar a}L_{b\bar b})
\nonumber\ee
\be
+4\omega^2g_b^2L_{a\bar a}^2
+4g_a^2\omega^2L_{b \bar b}^2
+16g_a^2g_b^2\omega^2
-2B(I_{ab}-L_{a\bar a}L_{b \bar b})(L_{a\bar a}I_b+L_{b \bar b}I_a)
\nonumber\ee
\be
-B^2(I_{ab}-L_{a \bar a}L_{b \bar b})^2
+4B(g_b^2I_aL_{a\bar a}+g_a^2I_bL_{b\bar b})+4B^2g_a^2g_b^2
\ee
\be
T_{abc}^2=2(I_{ab}-L_{a\bar a}L_{b \bar b})(I_{bc}-L_{b\bar b}L_{c \bar c})(I_{ac}-L_{a\bar a}L_{c\bar c})+2I_{ab}I_{ac}I_{bc}+L_{a\bar a}^2L_{b \bar b}^2L_{c\bar c}^2
\nonumber\ee
\be
-4(g_c^2I_{ab}(I_{ab} - L_{a\bar a}L_{b\bar b}) + g_a^2I_{bc}(I_{bc} - L_{b \bar b}L_{c \bar c}) +
    g_b^2I_{ac}(I_{ac} - L_{a\bar a}L_{c \bar c}))
\nonumber\ee
\be
-(I_{bc}^2L_{a \bar a}^2+I_{ab}^2L_{c \bar c}^2+I_{ac}^2L_{b\bar b}^2)
+ 4g_b^2 g_c^2L_{a\bar a}^2 +
 4g_a^2 g_c^2L_{b \bar b}^2 + 4g_a^2 g_b^2L_{c\bar c}^2 + 16g_a^2 g_b^2 g_c^2
\ee
To write the symmetry algebra in a simpler form we can redefine the generators
\be
M_{aa}=L_{a\bar a}^2+4g_a^2,\quad M_{ab}=I_{ab}-\frac{1}{2}L_{a\bar a}L_{b\bar b}, \quad
M_{a0}=I_a-\frac{B}{2}L_{a\bar a},\quad
M_{00}=4\omega^2+B^2.
\ee
Since $L_{a\bar a}$ commute with all other generators Poisson brackets of $M$ will exactly coincide with the Poisson brackets of $I_{ab}$ and $I_a$. Similarly the $R$ tensor is defined as in the real case. So the algebra will have the following form
\be
\{M_{ab},M_{cd}\}=\delta_{bc}T_{abd}+\delta_{ac}T_{bcd}-\delta_{bd}T_{acd}-\delta_{ad}T_{abc} ,\quad  \{M_{a0},M_{ab}\}=\delta_{ab}S_{ac}-\delta_{ac}S_{ab}.
\ee
where
\be
S_{ab}^2=4M_{ab}M_{a0}M_{b0}+\Big(\omega^2+\frac{B^2}{4}\Big)(M_{aa}M_{bb}-4M_{ab}^2)-M_{b0}^2 M_{aa}-M_{a0}^2 M_{bb}
\ee
\be
T_{abc}^2=4M_{ab}M_{bc}M_{ac}-M_{ab}^2M_{cc}-M_{ac}^2M_{bb}-M_{bc}^2M_{aa}+\frac{1}{4}M_{aa}M_{bb}M_{cc}
\ee
Needless to say that $L_{a\bar a}$ commute with all the other constants of motion. Finally
the full symmetry algebra then reads
\be
\{M_{AB},M_{CD}\}=\delta_{BC}R_{ABD}+\delta_{AC}R_{BCD}-\delta_{BD}R_{ACD}-\delta_{AD}R_{ABC}
\ee
where
\be
R_{ABC}^2=4M_{AB}M_{BC}M_{AC}-M_{AB}^2M_{CC}-M_{AC}^2M_{BB}-M_{BC}^2M_{AA}+\frac{1}{4}M_{AA}M_{BB}M_{CC}
\ee
Again capital letters take values from $0$ to $N$. In the complex case $R_{ABC}$ and $M_{AB}$ are again respectively antisymmetric and symmetric as in the real case.
Up to multiplication by a constant this has the same form as the symmetry algebra for the real case.

 Let us briefly discuss the number of conserved quantities. We have $N$ real functionally independent  constants of motion ($I_a$). Moreover  let us mention that $I_{ab}$  is also real, and although it has $N(N-1)/2$ components, the number of functionally independent constants of motion is $N-1$.
 In addition to this, the complex system has $N$ real conserved quantities  ($L_{a\bar a}$). So the total number of constants of motion is $3N-1$ and it is superintegrable (but not maximally superintegrable). Especially if $N=1$ the system is integrable. For $N=2$ the system is superintegrable, but it has only one additional constant of motion (minimally superintegrable).

\subsection{Quantization}

Quantization will be done using the fact that $\mathbb{C}^N$-Smorodinsky-Winternitz system is a sum of two dimensional singular oscillators. This allows to write the wave function as a product of $N$ wave functions and total energy of the system as a sum of the energies of its subsystems. So the initial problem reduces to two-dimensional one.

\be
\hat I_{a}\Psi_{a}(z_a,\bar z_a)= E_a \Psi_{a}(z_a,\bar z_a) , \quad
\hat{H}\Psi_{tot}=E_{tot}\Psi_{tot},
\nonumber\ee
\be
\Psi_{tot}=\prod_{a=1}^N\Psi_{a}(z_a,\bar z_{a}), \quad E_{tot}=\sum_{a}^N E_{a}.
\ee
After this reduction, complex indices can be temporarily dropped. Now it is obvious to introduce the momenta operators and commutation relations, which will have the following form in the presence of constant magnetic field.
\be
\hat{\pi}=-\imath(\hbar\partial+\frac{B}{2}\bar{z}),\quad
\hat{\bar\pi}=-\imath(\hbar\bar{\partial}-\frac{B}{2}z)
\quad
[\pi,\bar{\pi}]=\hbar B,\quad [\pi,z]=-\imath\hbar
\ee
Schr{\"o}dinger equation can be written down
\be
\Big[-\hbar^2\partial\bar{\partial}+\Big(\omega^2+\frac{B^2}{4}\Big)z\bar{z}-\hbar\frac{B}{2}(\bar z \bar{\partial}-\partial z)+\frac{g^2}{z\bar z}\Big]\Psi(z,\bar z)=E\Psi(z,\bar z).
\ee
Even in this two-dimensional system additional separation of variables can be done if one writes this system in a polar coordinates using the fact that $z=\frac{r}{\sqrt2}e^{i\phi}$.

\be
\Big[\frac{\partial^2}{\partial r^2}+\frac{1}{r}\frac{\partial}{\partial r}+\frac{2}{\hbar^2}\Big(E+\frac{\hbar^2}{2r^2}\frac{\partial^2}{\partial \phi^2}-\frac{2g^2}{r^2}-\frac12\Big(\omega^2+\frac{B^2}{4}\Big)r^2+\frac{\imath B\hbar}{2}\frac{\partial}{\partial \phi}\Big)\Big]\Psi(r,\phi)=0.
\label{schrodinger}
\ee
Further separation of variables can be done and one can use the fact that
L is a constant of motion.
\be
\Psi(r,\phi)=R(r)\Phi(\phi),\quad \hat{L}\Phi=\hbar m\Phi .
\ee
Using the explicit form of the $U(1)$ generator, normalized solution can be written
\be
\hat L=-\imath\hbar\frac{\partial}{\partial \phi},
\quad
\Phi(\phi)=\frac{1}{\sqrt{2\pi}}e^{\imath m\phi}.
\ee
This result allows to write the equation \eqref{schrodinger} in the following form
\be
\Big[\frac{d^2}{d r^2}+\frac{1}{r}\frac{d}{d r}+\frac{2}{\hbar^2}\Big(E-\frac{\hbar^2 m^2}{2r^2}-\frac{2g^2}{r^2}-\frac12\Big(\omega^2+\frac{B^2}{4}\Big)r^2-\frac{ B\hbar m}{2}\Big)\Big]R(r)=0.
\ee
Solution of this kind of Schr{\"o}dinger equation can be
 written down. The final result for the wave functions of two-dimensional system and the energy spectrum are as follows
\be
\psi(z,\bar z,n,m)=\frac{C_{n,m}}{\sqrt{2\pi}}(\sqrt{z/\bar z})^mF\Big(-n,\sqrt{m^2+\frac{4g^2}{\hbar^2}}+1,\frac{2\sqrt{\omega^2+\frac{B^2}{4}}}{\hbar}z\bar z\Big)\times
\ee
\be
\times\Big(\frac{2\sqrt{\omega^2+\frac{B^2}{4}}}{\hbar}z\bar z\Big)^{1/2\sqrt{m^2+\frac{4g^2}{\hbar^2}}}e^{-\frac{2\sqrt{\omega^2+\frac{B^2}{4}}}{\hbar}z\bar z}
\nonumber
\ee
\be
E=\hbar\sqrt{\omega^2+\frac{B^2}{4}}\Big(2n+1+\sqrt{m^2+\frac{4g^2}{\hbar^2}}\Big)+\frac{B\hbar m}{2}
\ee
Finally the indices of $\mathbb{C}^N$ can be recovered. The total wave function is a product of the wavefunctions and the total energy is the sum of the energies of two-dimensional subsystems
\be
\Psi(z,\bar z)=\prod_{a=1}^N\psi(z_a,\bar z_a,n_a,m_a)
\ee
\be
E_{tot}=\sum_{a=1}^NE_{n_a,m_a}=\hbar\sqrt{\omega^2+\frac{B^2}{4}}\Big(2n+N+\sum_{a=1}^N\sqrt{m_a^2+\frac{4g_a^2}{\hbar^2}}\Big)+\frac{B\hbar }{2}\sum_{a=1}^N m_a,
\ee
\be
n=\sum_{a=1}^N n_a, \quad n=0,1,2... \qquad m_a=0,\pm 1,\pm 2,...
\ee
In contrast to the real case  the energy spectrum of the $\mathbb{C}^N$-Smorodinky-Winternitz system depends on $N+1$ quantum numbers, namely $n$ and $m_a$ .

\section{Kustaanheimo-Stiefel transformation}
Since $\mathbb{C}^N$-Smorodinsky-Winternitz system has manifest $U(1)$ invariance, we could apply its respective reduction procedure related with first Hopf map $S^3/S^1=S^2$, which is known as Kustaanheimo-Stiefel transformation, for the particular case of $N=2$.
Such a reduction was performed  decade ago  \cite{mardoyan0} and was found to be resulted in the so-called
``generalized MICZ-Kepler problem" suggested by  Mardoyan a bit earlier \cite{mardoyan1,mardoyan2}. However the initial system was considered, it was not specified by the presence of constant magnetic field, furthermore, the symmetry algebra of the reduced system was not obtained there. Hence, it is at least deductive to perform Kustaanheimo-Stiefel transformation to the $\mathbb{C}^2$-Smorodinsky-Winternitz  system with constant magnetic field in order to find its impact  (appearing in the initial system) in the resulting one. Furthermore, it is natural way to find  the constants of motion of the ``generalized MICZ-Kepler system" and construct their algebra.

So,  let us perform the reduction of $\mathbb{C}^2$-Smorodinsky-Winternitz system by the $U(1)$-group action given by the generator
\be
J_0=L_{11}+L_{22}=\imath(z\pi-\bar z \bar \pi)-B z\bar z
\ee
For this purpose we have to choose six  independent functions of initial phase space variables which commute with that generators,
\be
q_k=z\sigma_k\bar z,\quad p_k=\frac{z \sigma_k \pi+\bar \pi\sigma_k \bar z }{2z\bar z},\quad k=1,2,3
\ee
where $\sigma_k$ are standard $2\times 2$ Pauli matrices. Matrix indices are dropped here. This transformation is called Kustaanheimo-Stiefel transformation.
Then we calculate their Poisson brackets and fix the value of $U(1)$- generator $J_0=2s$.
As a result, we get the reduced Poisson brackets
\be
\{q_k,q_l\}=0,\quad \{p_k,q_l\}=\delta_{kl},\quad \{p_k,p_l\}=s\epsilon_{klm}\frac{q_m}{|q|^3}
\ee
Expressing the Hamiltonian via $q_i, p_i, J_0$ and fixing the value of the latter one, we get
\be
H_{SW}=2|q|\Big[\frac{p^2}{2}+\frac{s^2}{2|q|^2}+\frac{Bs}{2|q|}+\frac{1}{2}\Big(\frac{B^2}{4}+\omega^2\Big)+\frac{g_1^2}{|q|(|q|+q_3)}+\frac{g_2^2}{|q|(|q|-q_3)}\Big]
\ee
So, we reduced the $\mathbb{C}^2$-Smorodinsky-Winternitz Hamiltonian to the three-dimensional system.
To get the Coulomb-like system we fix the  energy surface or reduced Hamiltonian, $H_{SW}-E_{SW}=0$ and divide it on $2|q|$.
This yields the equation
\be
{\cal H}_{gMICZ}-\mathcal{E}=0,\qquad {\rm with}\quad \mathcal{E}\equiv -\frac{\omega^2+{B^2}/{4} }{2}
\ee
and
\be
{\cal H}_{gMICZ}=\frac{p^2}{2}+\frac{s^2}{2|q|^2}+\frac{g_1^2}{|q|(|q|+q_3)}+\frac{g_2^2}{|q|(|q|-q_3)}-\frac{\gamma}{|q|} \qquad {\rm with}\quad \gamma\equiv\frac{E_{SW}-Bs}{2}.
\ee
The latter expression defines the Hamiltonian of ``generalized MICZ-Kepler problem". Hence, we transformed the energy surface of the reduced $\mathbb{C}^2$-Smorodinsky-Winternitz Hamiltonian to those of (three-dimensional) ``Generalized MICZ-Kepler system".
Additionally it has an inverse square potential and this system has an interaction with a Dirac monopole magnetic field which affects the symplectic structure.

{\sl Surprisingly, the reduced system contains interaction with Dirac monopole field only, i.e. the constant magnetic field in the original system does not contribute in the reduced one. All dependence on $B$ is hidden in $s$ and $\gamma$, which are fixed,  so the reduced system does not depend on $B$ explicitly.}

Now this reduction can be done for constants of motion. Before doing that it is convenient to present the
initial generators of $u(2)$ algebra  given by \eqref{SUN} in the form
\be
J_{0}=i(z\pi-\bar z \bar \pi)-{Bz\bar z}, \quad J_k=\frac{i}{2}(z\sigma_k\pi- \bar \pi \sigma_k\bar z)-\frac{Bz\sigma_k\bar z}{2},
\nonumber\ee
\be
\{J_0, J_i\}=0,\quad \{J_i,J_j\}=\varepsilon_{ijk}J_k.
\ee
After reduction  we get   $J_{0}
=2s$.   After the reduction, the rest $su(2)$ generators result in the  generators of the $so(3)$ rotations of three-dimensional Euclidian space
with  the Dirac monopole placed in the beginning of Cartesian coordinate frame,
\be
J_k=\epsilon_{klm}p_{l}q_{m}-s\frac{q_k}{|q|}
\ee
Then the symmetry generators for the ``generalized MICZ-Kepler system" can be written down,
\be
{\cal I}=\frac{I_1-I_2}{2}+\frac{B}{4}(L_{22}-L_{11})=p_1J_2-p_2J_1+\frac{x_3\gamma}{r}+\frac{g_1^2(r-x_3)}{r(r+x_{3})}-\frac{g_2^2(r+x_3)}{r(r-x_3)}
\ee
\be
{\cal L}=\frac{1}{2}(L_{22}-L_{11})=J_{3}=p_{1}q_{2}-q_{1}p_{2}-\frac{sq_{3}}{|q|},
\nonumber\ee
\be
{\cal J}=I_{12}=J_1^2+J_2^2+\frac{g_1^2(r-q_3)}{r+q_3}+\frac{g_2^2(r+q_3)}{r-q_3}.
\ee
It is important to notice that ${\cal I}$ is a generalization of the $z$-component of the Runge-Lenz vector.

The  relation of the initial system and the reduced one will allow to find the symmetry algebra of the final system
  using the previously obtained result for the complex Smorodinsky-Winternitz system. First of all the constants of motion in the initial system will also commute with the reduced Hamiltonian.
\be
\{{\cal H}_{gMICZ},{\cal I}\}=\{{\cal H}_{gMICZ},{\cal J}\}=\{{\cal H}_{gMICZ},{\cal L}\}=0
\ee
Moreover, since in the initial system $L_{a\bar a}$ generators commute with all the other constants of motion one can write.
\be
\{{\cal L},{\cal J}\}=\{{\cal L},{\cal I}\}=0
\ee
There is only one non-trivial commutator
\be
\{{\cal I},{\cal J}\}=S
\ee
$S$ here coincides with $S_{12}$ of $\mathbb{C}^{2}$-Smorodinsky-Winternitz system and can be written using the generators of the reduced system.
\be
S^2=2{\cal H}_{gMICZ}\Big[4\Big({\cal J}+\frac{1}{2}\Big({\cal L}^2-s^2\Big)\Big)^2-\Big(4g_2^2+({\cal L}+s)^2\Big)\Big(4g_1^2+({\cal L}-s)^2\Big)\Big]
-\Big(4g_2^2+({\cal L}+s)^2\Big)\Big({\cal I}+\gamma\Big)^2
\nonumber
\ee
\be
-\Big(4g_1^2+({\cal L}-s)^2\Big)\Big({\cal I}-\gamma\Big)^2-4\Big({\cal J}+\frac{1}{2}({\cal L}^2-s^2)\Big)\Big({\cal I}-\gamma\Big)\Big({\cal I}+\gamma\Big)
\ee
There is a crucial fact that should be mentioned. Although the initial system had an interaction with magnetic field, after reduction we don't have any dependence on $B$ both in symplectic structure and in generators of the symmetry algebra, at least in classical level. In other words, the reduced system does not feel the  magnetic field of the initial system.


\section{Discussion and outlook}

In this chapter we formulated  the analog of the Smorodinksy-Winternitz system  interacting with a constant magnetic field on the $N$-dimensional complex Euclidian space $\mathbb{C}^N$. We found out it has $3N-1$ functionally independent  constants of motion and derived
the symmetry algebra of this system.
Quantization of these systems is also discussed. While  for the real Smorodinsky-Winternitz system energy spectrum is totally degenerate and depends on single ("principal") quantum number, the  $\mathbb{C}^N$-Smorodinsky-Winternitz  energy spectrum depends on $N+1$ quantum numbers. Then we performed Kustaanheimo-Stiefel transformation of the $\mathbb{C}^2$-Smorodinsky-Winternitz system and reduced it to the  so-called "generalized  MICZ-Kepler problem". We   obtained the symmetry algebra of the latter system using the result obtained for the initial ones. Moreover, we have shown that the presence of constant magnetic field in the initial problem does not affect the reduced system.

There are several generalizations one can perform for this system. Straightforward   task is the construction of a quaternionic ($\mathbb{H}^N$-)  analog of this system. While complex structure allows to introduce constant magnetic field without violating the superintegrability, quaternionic structure should  allow to introduce interaction with  $SU(2)$ instanton.
It seems that one can also introduce the superintegrable analogs of the $\mathbb{C}^N$-/$\mathbb{H}^N$-Smorodinsky-Winternitz  systems on
the complex/quaternionic   projective space $\mathbb{CP}^N$/$\mathbb{HP}^N$, having in mind the existence of such generalization for the $\mathbb{C}^N$-/($\mathbb{H}^N$-) oscillator \cite{cpn,hpn}. We expect that the  inclusion of a constant magnetic/instanton field does not cause any qualitative changes for this system.
 These generalizations will be discussed later on.
 %

\newpage

\chapter{$\mathbb{CP}^N$-Rosochatius system}
\setcounter{equation}{0}
\setcounter{section}{0}

\section{Introduction}

This chapter is based on the article written with Armen Nersessian and Evgeny Ivanov\cite{shmros}.

The ($D$-dimensional) isotropic oscillator and the relevant Coulomb problem play a pivotal role among other textbook examples of $D$-dimensional integrable systems.
  They are distinguished by the ``maximal superintegrability" property, which is the existence of $2D-1$ functionally independent constants of motion \cite{perelomov}.
The rational Calogero model with oscillator potential \cite{calogero1,calogero2},  being a nontrivial generalization  of isotropic oscillator,  is also  maximally superintegrable \cite{woj83}.
Moreover, Calogero model with Coulomb potential is superintegrable too \cite{rapid,rapid1,rapid2}.
All these systems, being originally defined on a plane, admit the maximally superintegrable deformations to the spheres (see Ref. \cite{higgs} for the spherical generalizations
 of the oscillator and Coulomb problem, and Ref. \cite{rapid} for the  Calogero-oscillator and Calogero-Coulomb ones).
The integrable spherical generalizations  of anisotropic oscillator \cite{Vahagn,Vahagn1}, Stark-Coulomb and two-center Coulomb problems \cite{Calogero-Stark}
are  also known.

In contrast to the spherical extensions, the generalizations to other curved spaces have not  attracted much attention so far.
The only exception  seems to be  the isotropic oscillator on  the complex/quaternionic spaces considered in Ref. \cite{cpn,hpn}.
These systems reveal an important feature: they remain superintegrable after coupling to a constant magnetic/BPST instanton field,
though cease to be maximally superintegrable. One may pose a question:\\

{\sl How to construct the {\it superintegrable} generalizations of Calogero-oscillator and Calogero-Coulomb models on complex and quaternionic projective spaces?}\\

In this chapter we make first steps toward the answer. Due to the complexity of the problem
 we restrict our attention to the simplest particular case. Namely,
 we construct  the superintegrable $\mathbb{CP}^N$-generalization of
 the $N$-dimensional singular oscillator (the simplest rational Calogero-oscillator model) which is defined by the Hamiltonian
\be\label{SWR}
H_{SW}=\sum_{a=1}^{N}\Big(\frac{p_a^2}{2}+\frac{g_a^2}{2x_a^2}+\frac{\omega^2x_a^2}{2}\Big),\qquad
\{p_a,x_b\}=\delta_{ab},\quad \{p_a,p_b\}=\{x_a,x_b\}=0. \ee This
model is less trivial than it looks at first sight: it has  a
variety of hidden constants of motion  which form a nonlinear
symmetry algebra and endow the system with the maximal
superintegrability property, as was mentioned in the previous chapter.

The maximally superintegrable spherical counterpart  of the Smorodinsky-Winternitz system is defined by the Hamiltonian  suggested  by Rosochatius in 1877 \cite{rosochatius}
\be
H_{Ros}=\frac{p^2}{2}-\frac{(xp)^2}{2r^2_0}+ \sum_{a=1}^N\frac{\omega^2_ar^2_0}{x^2_a}+\frac{\omega^2r^2_0 x^2}{2x^2_0}
,\qquad x^2_a+x^2_0=r^2_0\,.
\label{Rosin}\ee
It is a particular case of the integrable systems  obtained by restricting the free particle and oscillator  systems to a sphere.
It was studied  by many authors from different viewpoints, including its re-invention  as a superintegrable spherical generalization of Smorodinsky-Winternitz system \cite{mack,mack2,pogosyanSph,oksana,Galaj1}.
Rosochatius  model, as well as its hybrid with the Neumann model suggested in 1859 \cite{neumann}, attract a stable interest for years due to their relevance to
a wide circle of physical and mathematical problems. Recently, the Rosochatius-Neumann system was encountered, while studying strings \cite{arut,arut1,arut2},
extreme black hole geodesics \cite{Galaj1,sheikh,sheikh1} and  Klein-Gordon equation in curved backgrounds \cite{evnin}.\\
In this chapter we  propose a superintegrable generalization of Rosochatius (and Smorodinsky-Winternitz) system on the complex projective space $\mathbb{CP}^N$.
It is  defined by  the Hamiltonian
 \be\label{Hamr0}
{\cal H}_{Ros}=(1+z\bar z)\frac{(\pi\bar\pi) +(z\pi)(\bar z\bar\pi)}{r^2_0} + r^2_0(1+z\bar z)(\omega^2_0+ \sum_{a=1}^N\frac{\omega^2_a}{z^a{\bar z}^a})-r^2_0\sum_{i=0}^N\omega^2_i,
\ee
and  by the Poisson brackets providing the interaction with a constant magnetic field of the magnitude $B$
 \be \label{com}
 \{\pi_a, z^b\}=\delta_a^b,\quad \{\bar\pi_a, \bar z^b\}=\delta_{\bar a}^{\bar b},\quad \{\pi_a,\bar\pi_b\}=\imath B{r^2_0}\left(\frac{\delta_{a\bar b}}{1+z\bar z}-\frac{\bar z^a z^b}{(1+z\bar z)^2}\right).
 \ee
We will call it {\bf \sl $\mathbb{CP}^N$-Rosochatius system}.

Reducing  this $2N$-dimensional system by the action of  $N$ manifest $U(1)$ symmetries, $z^a\to{\rm e}^{\imath\kappa_a}z^a,\pi_a\to{\rm e}^{-\imath\kappa_a}\pi_a $,
we recover the $N$-dimensional Rosochatius system \eqref{Rosin} (see {\sl Section 3}).

 On the other hand, rescaling the coordinates and momenta as $r_0z^a\to z^a, \pi_a/r_0\to\pi_a$ and taking the limit $r_0\to \infty, \omega_a\to 0$ with
 $r^2_0\omega_a=g_a$ kept finite, we arrive  at the    $\mathbb{C}^N$-Smorodinsky-Winternitz system discussed in the previous chapter.
\be
\mathcal{H}_{SW}=\sum_{a=1}^N\left(\pi_a\bar\pi_a +\omega^2_0 z^a\bar z^a+\frac{g^2_a}{z^a\bar z^a}\right),
\nonumber
\ee
\be
 \{\pi_a, z^b\}=\delta_a^b,\quad \{\bar\pi_a, \bar z^b\}=\delta_{\bar a}^{\bar b},\quad \{\pi_a,\bar\pi_b\}=\imath B\delta_{a\bar b}\,.
\label{SW0}\ee
Since the reductions of $\mathbb{CP}^N$-Rosochatius system yield superintegrable systems, it is quite natural that it proves to be superintegrable on its own.

Finally, note that  $\mathbb{C}^N$-Smorodinsky-Winternitz system \eqref{SW0} can be interpreted as
a set of $N$ two-dimensional ring-shaped oscillators interacting with a constant magnetic field orthogonal to the plane.
As opposed to  \eqref{SW0},  the $\mathbb{CP}^N$-Rosochatius system does not split into a set of $N$ two-dimensional decoupled systems.
Instead, it can be interpreted as describing  {\sl interacting} particles with a position-dependent mass in the two-dimensional quantum rings.

To summarize, the {\sl $\mathbb{CP}^N$-Rosochatius system} suggested is of interest from many points of view.
Its study is the subject of the remainder of this chapter. It is organized as follows.\\

In {\sl Section 4.2} we discuss the simplest systems  on  $\mathbb{CP}^N$, namely  $\mathbb{CP}^N$-Landau problem and the $\mathbb{CP}^N$-oscillator.Then we  derive the potential specifying the $\mathbb{CP^N}$-Rosochatius system.

In  {\sl Section 4.3} we present  classical $\mathbb{CP}^N$-Rosochatius model in a constant magnetic field and find
that, in addition to  $N$ manifest $U(1)$ symmetries,  this system possesses additional $2N-1$ functionally-independent second-order
constants of motion. The latter property  implies the (non-maximal) superintegrability of the model considered.
We present the explicit expressions of  the constants of motion and calculate their algebra. We also show that
the reduction of  $\mathbb{CP}^N$-Rosochatius model by manifest $U(1)$ symmetries reproduces the original $N$-dimensional ($\mathbb{S}^N$-) Rosochatius system.

In   {\sl Section 4.4} we   separate the variables and find  classical solutions of $\mathbb{CP}^N$-Rosochatius model.

In  {\sl Section 4.5} we  study quantum $\mathbb{CP}^N$-Rosochatius system and find its spectrum which depends on $N+1$ quantum numbers,
as well as the relevant  wavefunctions.

In  {\sl Section 4.6} we give an account of  open problems and possible  generalizations.

In the subsequent consideration we  put,  for simplicity,  $r_0=1$.

\section{Models on complex projective spaces}
In this Section we
briefly describe the Landau problem and the oscillator  on a complex projective space, and construct
$\mathbb{CP}^N$-analog of Rosochatius system.

Let us introduce, on the cotangent bundle of $\mathbb{C}^{N+1}$, the canonical Poisson brackets $\{p_i, u^j\}=\delta_{ij}$,
and  define the  $su(N+1)$ algebra with the generators
\be
L_{i\bar j}=\imath ( p_i  u^j-\bar p_j \bar u^i)-\frac{\delta_{i\bar j}}{N}L_0 ,\qquad {\rm where}\quad L_0=\imath \sum_{i=0}^N( p_i  u^i-\bar p_i\bar u^i).
\label{Lij}\ee
 Reducing this phase space  by the action of  generators $L_0, h_0=\sum_i u^i\bar u^i$, and finally fixing their values as $L_0=2B, h_0=1$, we arrive
 at the Poisson brackets \eqref{com} (with $r_0=1$). They describe an electrically charged particle on $\mathbb{CP}^N$
 interacting with  a constant magnetic field of the magnitude $B$ and set the corresponding  twisted symplectic structure
\be
\Omega_0=
dz^a\wedge d\pi_a +d{\bar z}^{a}\wedge d{\bar\pi}_a + \imath Bg_{a\bar b}dz^a\wedge
d{\bar z}^b,
\label{hn}\ee
with $g_{a\bar b}$ being defined in \eqref{FS}.

 The inhomogeneous coordinates and momenta  $z^a, \pi_a$ are related to the homogeneous ones $p_i, u^i$ as  \cite{SupHam}
\be
z^a=\frac{u^a}{u^0},\quad
\pi_a=g_{a\bar b}\Big(\frac{p_b}{\bar u^0}-\bar{z}^b\frac{ p_0}{\bar u^0}\Big).
\ee
The  $su(N+1)$ generators \eqref{Lij} are reduced to the following ones
\bea\label{suN1}
&J_{a\bar b}=\imath(z^b\pi_a-\bar\pi_b\bar z^a)- B\frac{\bar z^a z^b}{1+z\bar z},\quad
J_{a}=\pi_a+\bar z^a(\bar z\bar\pi)+\imath B\frac{{\bar z}^a}{1+z\bar z}
:&
\label{su}\\
&\{J_{{\bar a} b}, J_{\bar c d}\}=
i\delta_{\bar a d}J_{\bar b c}
-i\delta_{\bar c b}J_{\bar a d},\quad \{J_{a}, {\bar J}_{b}\}=-i (J_{a\bar b}+J_0 \delta_{a\bar b}),\quad \{J_a, { J}_{b\bar c}\}=i J_b\delta_{a\bar c},&
\eea
where $J_0\equiv \sum_{a=1}^N J_{a\bar a}+B$.

With these expressions at hand we can now consider some superintegrable systems on $\mathbb{CP}^N$.
\\

{\bf $\mathbb{CP}^N$-Landau problem.}
 The $\mathbb{CP}^N$-Landau problem is defined by the symplectic structure \eqref{hn}  and the free-particle Hamiltonian identified with a  Casimir  of $su(N+1)$ algebra
\be
\mathcal{H}_{0}=(1+z\bar z)\Big((\pi\bar\pi) +(z\pi)(\bar z\bar\pi)\Big)=\frac12\sum_{i,j=0}^{N}L_{i\bar j}L_{j\bar i}-\frac{B^2}{2}=\sum_{a=1}^NJ_a\bar J_a+\frac{\sum_{a,b=1}^NJ_{a\bar b}J_{b \bar a}+J_0^2-B^2}{2}
\nonumber\ee
\be
\{\mathcal{H}_0, L_{ij}\}=0.
\ee
Its   quantization  was done, e.g., in  \cite{nair}.\\

{\bf $\mathbb{CP}^N$-oscillator.}
The $\mathbb{CP}^N$-oscillator is defined by the symplectic structure \eqref{hn} and the Hamiltonian \cite{cpn}

\be
\mathcal{H}_{osc}=\mathcal{H}_{0} +
\omega^2 z\bar z\,.
\label{cpnosc}\ee
It respects manifest $U(N)$ symmetry with the generators $J_{a\bar b}$  \eqref{suN1}, and additional hidden symmetries  given by the proper analog of ``Fradkin tensor",
\be\label{fradkinosc}
I_{a\bar b}={J_a {\bar J}_b} +\omega^2 {\bar z}^a z^b\,.
\ee
The full  symmetry algebra of this system  reads
\bea
&
\{J_{{\bar a} b}, J_{\bar c d}\}=
\imath\delta_{\bar a d}J_{\bar b c}
-\imath\delta_{\bar c b}J_{\bar a d},\quad
\{I_{a\bar b}, J_{c\bar d}\}=\imath\delta_{a\bar d}I_{c\bar b}-
\imath\delta_{c\bar b}I_{a\bar d}& \\
 &\{I_{a \bar b}, I_{c\bar d}\}=\imath\omega^2
 \delta_{a\bar d}J_{c\bar b}-
\imath\omega^2 \delta_{c\bar b} J_{a\bar d}
-\imath I_{c\bar b}(J_{a\bar d}+J_0\delta_{a\bar d})
+\imath I_{a\bar d}(J_{c\bar b}+J_0\delta_{c\bar b})\,,&
\label{constraint}
\eea
where
$
J_0=i(z\pi-\bar\pi\bar z)+B\frac{1}{1+z\bar z}
$.
The Hamiltonian \eqref{cpnosc} is expressed via the symmetry generators as follows
\be
{\cal H}_{osc}=\sum_{a=1}^NI_{a\bar a}+\frac 12\sum_{a,b=1}^N J_{a\bar b}J_{b \bar a}+ \frac{J_0^2-B^2}{2}.
\label{cpnoscconstants}\ee
The quantum mechanics associated with this Hamiltonian was considered in \cite{qcpn}.
In the flat limit, the $\mathbb{CP}^N$-oscillator  goes over to the $\mathbb{C}^N$-oscillator interacting with a constant magnetic field.\\

{\bf $\mathbb{CP}^N$-Rosochatius system.}
The $\mathbb{CP}^N$-oscillator, being   superintegrable system (for $N>1$), has an obvious drawback: it lacks covariance  under  transition from one chart to another.
This non-covariance becomes manifest after expressing  the Hamiltonian \eqref{cpnosc}  via the $SU(N+1)$ symmetry generators and the homogeneous coordinates $u^i$,
\be
{\cal H}_{osc}=\frac{\sum_{i,j=0}^{N} L_{i\bar j}L_{j\bar i}-B^2}{2}+\frac{\omega^2}{u^0{\bar u}^0}-\omega^2.
\label{cpnosc1}\ee
This expression  allows one to immediately construct  $(N+1)$-parameter deformation of the $\mathbb{CP}^N$-oscillator, such that it is manifestly form-invariant
under  passing from one chart to another accompanied by the appropriate change of the parameters $\omega_i$. The relevant potential is
 \be
V_{Ros}=\sum_{i=0}^N\left(\frac{\omega^2_i}{u^i{\bar u}^i}-\omega^2_i\right),\qquad {\rm with}\quad \sum_{i=0}^N u^i{\bar u}^i=1.
\label{ros}\ee
In the case when all parameters $\omega_i$ are equal, the system is  globally  defined on the complex projective space with the punctured points $u^i=0\,$.

The system with the potential \eqref{ros}  is just  the $\mathbb{CP}^N$-Rosochatius system mentioned in Introduction.
Now we turn to its investigation as the main subject of the present chapter.

\section{$\mathbb{CP}^N$-Rosochatius system}
We consider   the  $N$-parameter deformation of the  $\mathbb{CP}^N$- oscillator by the potential \eqref{ros}, in what follows  referred to as the
``$\mathbb{CP}^N$-Rosochatius system". It is defined by the Hamiltonian \eqref{Hamr0} and  Poisson brackets  \eqref{com} with $r_0=1$.
Equivalently, this system can be defined by the symplectic structure \eqref{hn} and the Hamiltonian

\be\label{Ham}
{\cal H}_{Ros}={g^{a\bar b}\pi_a\bar\pi_b} + (1+z\bar z)\left(\omega_0^2+ \sum_{a=1}^N\frac{\omega_a^2}{z^a{\bar z}^a}\right)-\sum_{i=0}^N\omega^2_i,
\ee
where $g^{ a \bar b}=(1+z\bar z)(\delta^{a \bar  b}+  z^a \bar z^b)$ is the inverse Fubini-Study metrics.

The model has $N$ manifest (kinematical) $U(1)$   symmetries  with the generators
\be
J_{a\bar a}=\imath\pi_a z^a-\imath\bar{\pi}_a\bar{z}^a-B\frac{ z^a{\bar z}^a}{1+z{\bar z}}\; :\quad\{J_{a\bar a}, \mathcal{H}\}=0,
\ee
and hidden symmetries with the  second-order generators $I_{ij}=(I_{0a},I_{ab})$ defined as
\be
I_{0a}=J_{0a}{\bar J}_{0\bar a} +\omega_0^2 z^a{\bar z}^a +\frac{\omega_a^2}{{\bar z}^a z^a} ,\qquad I_{ab}=J_{a\bar b}J_{b\bar a}+\omega_a^2\frac{z^b{\bar z}^b}{z^a{\bar z}^a} +
\omega_b^2\frac{z^a{\bar z}^a}{z^b{\bar z}^b}\; :  \quad\{I_{i\bar j}, \mathcal{H}\}=0\,.
\label{fradkinRos}\ee
In the homogeneous coordinates, the hidden symmetry  generators can be cast in a more succinct form
\be
I_{ij}=J_{i\bar j}J_{j\bar i}+ \omega_i^2\frac{u^j\bar u^j}{u^i\bar u^i}+\omega_j^2 \frac{u^i\bar u^i}{u^j\bar u^j} .
\ee
The relevant symmetry algebra is given by the brackets
\be
 \{J_{a\bar a}, I_{ij}\}=0, \qquad \{I_{ij}, I_{kl}\}=\delta_{jk}T_{ijl}+\delta_{ik}T_{jkl}-\delta_{jl}T_{ikl}-\delta_{il}T_{ijk}\,,
 \ee
 with
 \be
( T_{ijk})^2=2(I_{ij}-J_{i\bar i}J_{j \bar j})(I_{jk}-J_{j\bar j}J_{k \bar k})(I_{ik}-J_{i\bar i}J_{k\bar k})
 +2I_{ij}I_{ik}I_{jk}+J_{i\bar i}^2J_{j \bar j}^2J_{k\bar k}^2
\nonumber\ee
\be
-4(\omega_k^2I_{ij}(I_{ij} - J_{i\bar i}J_{j\bar j}) + \omega_i^2I_{jk}(I_{jk} - J_{j \bar j}J_{k \bar k}) +
    \omega_j^2I_{ik}(I_{ik} - J_{i\bar i}J_{k \bar k}))
\nonumber\ee
\be
 + 4\omega_j^2 \omega_k^2J_{i\bar i}^2 +
 4\omega_i^2 \omega_k^2J_{j \bar j}^2 + 4\omega_i^2 \omega_j^2J_{k\bar k}^2 + 16\omega_i^2 \omega_j^2 \omega_k^2
-(I_{jk}^2J_{i \bar i}^2+I_{ij}^2J_{k \bar k}^2+I_{ik}^2J_{j\bar j}^2)
 \ee
The  Hamiltonian is expressed via these generators as follows
\be
\mathcal{H}=\frac12\sum_{i=1}^{N+1}I_{ij}+\sum_{a=1}^N \omega_a^2+\frac{J_0^2-B^2}{2}=\sum_{a=1}^N I_{0a}+\sum_{a,b=1}^N\frac{I_{ab}}{2}+\sum_{a=1}^N \omega_a^2+\frac{J_0^2-B^2}{2}.
\ee

This consideration actually proves the superintegrability of the $\mathbb{CP}^N$-Rosochatius system. The number of the functionally independent
constants of motion will be counted in the end of this Section.

For sure, the   symmetry algebra written above can be found by a direct calculation of the Poisson brackets between the symmetry generators.
However, there is a more elegant and simple way to construct it. Namely, one has to consider the symmetry algebra of $\mathbb{C}^{N+1}$-Smorodinsky-Winternitz system (Part III)
with {\sl vanishing} magnetic field, and  to reduce it, by action of the generators $\imath(p_iu^i-\bar p_i\bar u^i)$, $u^i\bar u^i\,$(see the previous Section),
to the symmetry algebra of $\mathbb{CP}^N$-Rosochatius system.

\subsection{Reduction to (spherical) Rosochatius system}

In order to understand the relationship with the standard Rosochatius system (defined on the sphere) let us pass to the real canonical variables $y_a,\varphi^a$, $p_a, p_{\varphi_a}$
\be
z^a={ y_a{\rm e}^{\imath\varphi_a}},\quad \pi_a=\frac{1}{{2}}\left(p_a-{\imath}\Big(\frac{p_{\varphi_{a}}}{y_a}+\frac{B y_a}{1+y^2}\Big) \right){\rm e}^{-\imath\varphi_a} \; :\qquad\Omega=dp_a\wedge dy_a
+ dp_{\varphi_{a}}\wedge d\varphi_a\,.
\label{realcoor}\ee
In these variables the Hamiltonian \eqref{Ham} is rewritten as
\be
\mathcal{H}_{Ros}=\frac{1}{4}(1+y^2)\left[ \sum_{a,b=1}^N( \delta_{ab} +y_ay_b) p_ap_b+ 4{\widetilde\omega}^2_0+ 4\sum_{a=1}^N\frac{{\widetilde\omega}^2_a}{y^2_a} \right]-E_0\,,
\label{realHam}\ee
where
\be
{\widetilde\omega}^2_a=\omega^2_a+\frac14 {p^2_{\varphi_a}},\quad {\widetilde\omega}^2_0=\omega^2_0+\frac14\left(B+\sum_{a=1}^Np_{\varphi_a}\right)^2,\qquad
E_0=\frac{B^2}{4}+\sum_{i=0}^N \omega^2_i.
\label{tilde}\ee
Then, performing the reduction by cyclic variables $\varphi^a$ ({\it i.e.}, by  fixing the momenta $p_\varphi^a$), we arrive at the Rosochatius system
on the sphere with $y_a=x_a/x_0$, where
  $(x_0, x_a)$ are ambient Cartesian coordinates, $\sum_{i=0}^Nx^2_i=1$:
  \bea
  x_a=\frac{y_a}{\sqrt{1+y^2}},\quad x_0=\frac{1}{\sqrt{1+y^2}}\,.
  \eea
  As was already noticed, the $\mathbb{S}^N$-Rosochatius system   is maximally superintegrable, {\it i.e.} it has $2N-1$ functionally independent constants of motion.
  From the above reduction we conclude that the $\mathbb{CP}^N$-Rosochatius system has
  $2N-1+ N=3N-1$ functionally independent integrals. Hence, it lacks $N$ integrals needed for the maximal superintegrability.

\section{Classical Solutions}
To obtain the  classical  solutions of $\mathbb{CP}^N$-Rosochatius system we   introduce the spherical coordinates through the recursion
\be
y_N=r\cos\theta_{N-1},\quad y_{\alpha}=r\sin\theta_{N-1}u_\alpha, \qquad {\rm with} \quad r=\tan\theta_{N},\qquad \sum_{\alpha=1}^{N-1}u^2_\alpha=1,
\label{sphcoor}\ee
where $y_a$ were defined by \eqref{realcoor}.
In terms of these coordinates the Hamiltonian  \eqref{realHam} takes the form
\be\label{sphRos}
\mathcal{H}_{Ros}\equiv \mathcal{I}_{N}-E_0=\frac14(1+r^2)\left((1+r^2)p^2_r +\frac{4\mathcal{I}_{N-1}(\theta)}{r^2} +4\widetilde\omega^2_0\right)-E_0,
\nonumber\ee
\be
 \mathcal{I}_{a}=\frac{p^2_{\theta_{a}}}{4}+\frac{\mathcal{I}_{a-1}}{\sin^2\theta_{a}} +\frac{\widetilde\omega^2_{a+1}}{\cos^2\theta_{a}},
\ee
with $E_0$, $\omega_N\equiv {\tilde\omega}_0$ defined in \eqref{tilde} and $a=1, \ldots,N$.

 Thus we singled out the complete set  of Liouville integrals $(\mathcal{H}_{Ros}, \mathcal{I}_{\alpha}, p_{\varphi_{a}})$, and separated the variables.
 It is by no means the unique choice of Liouville integrals and of the coordinate frame in which  the Hamiltonian admits the separation of variables.
 However, for our purposes it is enough to deal with any particular choice.

 With the above expressions at hand,
 we  can derive classical solutions of the system by solving  the  Hamilton-Jacobi equation
 \be
 \mathcal{H}(p_a=\frac{\partial S}{\partial x^\mu} , x^\mu)=E,\qquad {\rm with}\quad  x^\mu=(\theta_a,\varphi_a ),
 \quad p_\mu=(p_a,p_{\varphi_{a}}).
 \ee
 To this end, we introduce the generating function  of the form
 \be
 S_{tot}=2\sum_{a=1}^{N}S_{a}(\theta_a)+\sum_{a=1}^N p_{\varphi_a}\varphi_a\,.
 \ee
 Substituting this ansatz in the Hamilton-Jacobi equation, we immediately separate the variables and arrive at  the set of ordinary differential equations:
\be
\left(\frac{dS_{{a}}}{d\theta_{a}}\right)^2+\frac{c_{a-1}}{\sin^2\theta_{a}} +\frac{\widetilde\omega^2_{a+1}}{\cos^2\theta_{a}}
=c_a,\qquad a = 1, \ldots,N, {\quad c_N := E+E_0, \quad \widetilde\omega^2_{N+1} := \widetilde\omega^2_{0}}\,.
\ee
Solving these equations, we obtain
\be
S_{{a}}=\int d\theta_{a}\sqrt{c_{a}-\frac{c_{a-1}}{\sin^2\theta_{a}} -\frac{\widetilde\omega^2_{a+1}}{\cos^2\theta_{a}}}\,.
\ee
Thus we  have found the general solution of the Hamilton-Jacobi equation ({\it i.e.}, the solution depending on $2N$ integration constants
$c_a, \, p_{\varphi_{a}}$).

In order to get the solutions of the classical equations of motion,  we should differentiate the generating functions with respect to these integration constants
and then equate the resulting functions to some constants $t_0,\kappa_\alpha,$ and $\varphi^a_0 $,
\be
\frac{\partial S_{tot}}{\partial E}=t-t_0,\qquad \frac{\partial S_{tot}}{\partial c_\alpha}=2\sum_{b=1}^N \frac{\partial S_b}{\partial c_\alpha}=\kappa_\alpha,\quad \alpha=1,\ldots, N-1, \qquad
\nonumber\ee
\be
\frac{\partial S_{tot}}{\partial p_{\varphi_{a}}}=\varphi^a+\sum_{b=1}^{N}2\frac{\partial S_b}{\partial p_{\varphi_a}}=\varphi^a_0\,. \label{EqDif}
\ee
Introducing
\be
\xi_a :=\sin^2\theta_a,\quad \mathcal{A}_a :=\frac{c_{a}+c_{a-1}-{\widetilde\omega}^2_{a+1} }{2c_a}\,,
\ee
we obtain from \eqref{EqDif}
\be
\xi_{N}-\mathcal{A}_N=\sqrt{\mathcal{A}_{N}^2-\frac{c_{N-1}}{c_{N}}}\sin 2\sqrt{c_N}(t-t_0),\ee
  \be
  \xi_{\alpha}=\sqrt{{\cal A}_{\alpha}^2-\frac{c_{\alpha-1}}{c_{\alpha}}}\Bigg(\frac{\sin \kappa_{\alpha}(\xi_{\alpha+1}{\cal A}_{\alpha+1}
  -\frac{c_{\alpha}}{c_{\alpha+1}})+\cos \kappa_{\alpha}\sqrt{-\xi_{\alpha+1}^2
  +2\xi_{\alpha+1}{\cal A}_{\alpha+1}-\frac{c_{\alpha}}{c_{\alpha+1}}}}{\xi_{\alpha+1}\sqrt{\frac{c_{\alpha+1}}{c_{\alpha}}{\cal A}_{\alpha+1}^2-1}}\Bigg)
  +{\cal A}_{\alpha},
\ee
\be
\varphi^a-\varphi^a_0=-\frac{ p_{\varphi_a}}{4\widetilde\omega_{a+1}}\arctan \frac{2 \widetilde\omega_{a+1} \sqrt{c_{a-1} \left(\xi_{a }-1\right)-\xi_{a } \left(c_a \left(\xi _{a
   }-1\right)+\widetilde\omega_{a+1}^2\right)}}{-c_{a-1} \left(\xi_{a
   }-1\right)+c_a \left(\xi_{a }-1\right)\widetilde\omega_{a+1}^2\left(\xi_{a }+1\right)}\,.
\ee
Thereby we have derived the  explicit classical solutions  of our $\mathbb{CP}^N$-Rosochatius system.

\section{Quantization}
In order to quantize the
$\mathbb{CP}^N$-Rosochatius system
we  replace the Poisson brackets \eqref{com} by the commutators (with $r_0=1$)
\be
 [\widehat{\pi}_a, z^b]=-\imath\hbar \delta^b_a,\qquad [\widehat{\pi}_a, \widehat{\bar\pi}_b]=\hbar B\left(\frac{\delta_{a\bar b}}{1+z\bar z}-\frac{\bar z^a z^b}{(1+z\bar z)^2}\right).
\ee
The appropriate quantum realization of the momenta operators reads
\be
\widehat{\pi}_a=-\imath\left(\hbar\frac{\partial }{\partial z^a} +\frac{B}{2}\frac{\bar z^a}{1+z\bar z}\right),\quad \widehat{\bar\pi}_a=-\imath\left(\hbar\frac{\partial }{\partial\bar z^a} -\frac{B}{2}\frac{\bar z^a}{1+z\bar z}\right).
\ee
Then we define the quantum Hamiltonian
\be\label{qH}
\widehat{\mathcal{H}}_{Ros}=\frac{1}{2} g^{a\bar b}\left( \widehat{\pi}_a\widehat{\bar\pi}_b+\widehat{\bar\pi}_b\widehat{\pi}_a\right) +\hbar^2(1+z\bar z)\left(\omega_0^2+ \sum_{a=1}^N\frac{\omega_a^2}{z^a{\bar z}^a}\right)-\hbar^2\sum_{i=0}^N\omega_i^2.
\ee
The kinetic term in this Hamiltonian  is written as the Laplacian on K\"ahler manifold (coupled to a magnetic field)
defined with respect to the volume element $dv_{\mathbb{CP}^N}=(1+z\bar z)^{-(1+N)}[dzd\bar z]$, while  in  the potential term we
have made the replacement $\omega_i\to\hbar \omega_i\,$.

In terms of the real coordinates $z^a=y_a{\rm e}^{\imath\varphi_a}$ this Hamiltonian reads (cf. \eqref{realHam})
\be\label{rqH}
\widehat{\mathcal{H}}_{Ros}=(1+y^2)\Bigg[-\frac{\hbar^2}{4}\Big(\sum_{a,b=1}^{N}(\delta_{ab}+y_a y_b)
\frac{\partial^2}{\partial y_a\partial y_b}+\sum_{a=1}^N\Big(y_a+\frac{1}{y^a}\Big)\partial_{y_a}\Big)+\widehat{\mathbb{\tilde{\omega}}}_{N+1}^2
+ \sum_{a=1}^N\frac{ \widehat{\mathbb{\tilde{\omega}}}_\alpha^2}{4y_a^2}\Bigg]-\tilde{E_0}\,.
\ee
Here  we introduced the operators
\be
\widehat{\mathbb{\tilde{\omega}}}_{N+1}^2=\Big(\frac{B}{\hbar}+\frac{1}{\hbar}\sum_{a=1}^N\widehat{p}_{\varphi_a}\Big)^2+4\omega_0^2,\qquad
\widehat{\mathbb{\tilde{\omega}}}_\alpha^2=4{\omega}_\alpha^2+\frac{\widehat{p}^2_{\varphi_\alpha}}{ \hbar^2 }
\ee
with
\be
\widehat{p}_{\varphi_a}=\widehat{J}_{a\bar a}=-\imath\hbar\frac{\partial}{\partial\varphi^a}
\qquad
 \tilde{E_0}=\frac{B^2}{
4}+\hbar^2\sum_{i=0}^N\omega_i^2.
\ee
Clearly, these operators are quantum analogs of the classical quantities \eqref{tilde}.
In  the spherical coordinates \eqref{sphcoor} the Hamiltonian \eqref{rqH} takes the form
\be
\widehat{\mathcal{H}}_{Ros}=\widehat{\mathcal{I}}_N - \tilde{E}_{0},
\nonumber\ee
\be
 \widehat{\mathcal{I}}_a=-\frac{\hbar^2}{4}\Bigg((\sin\theta_{a})^{1-a}\frac{\partial}{\partial \theta_{a}}\Big((\sin \theta_a)^{a-1}\frac{\partial}{\partial \theta_{a}}\Big)
 +(a \cot \theta_{a}-\tan \theta_{a})\frac{\partial}{\partial \theta_{a}}\Bigg)+\frac{ \widehat{\mathcal{I}}_{a-1}}{\sin^2 \theta_\alpha}
 +\frac{\hbar^2 \widehat{\mathbb{\tilde{\omega}}}_{a+1}^2}{4\cos^2\theta_{a}},
\label{spher}
\ee
where $a =1,...,N$.

This prompts  us to consider the spectral problem
\be
 \widehat{J}_{a\bar a}\Psi=\hbar m_a\Psi, \qquad \widehat{\mathcal{I}}_a\Psi=\frac{\hbar^2}{4}l_a(l_a+2a)\Psi,
\label{sp}\ee
and separate the variables  by the choice of  the wavefunction in such a way that it resolves  first $N$ equations in the above problem,
\be
\Psi=\frac{1}{({2\pi})^{N/2}}\prod_{a=1}^N\psi_{a}(\theta_{a})e^{\imath m_a\varphi_a},\quad m_a=0,\pm 1, \pm 2, \ldots
\ee
Then, passing to the variables  $\xi_{a}=\sin^2 \theta_{a}$,  we transform the reduced  spectral problem to the system of $N$ ordinary differential equations
\be
-\xi_a(1-\xi_a)\psi_{a}''+\big((a+1)\xi-a\big)\psi_{a}'+\frac{1}{4}\Bigg(\frac{l_{a-1}(l_{a-1}+2a-2)}{\xi_a}+\frac{\tilde\omega_{a+1}^2}{1-\xi_a}-l_{a}(l_{a}+2\alpha)\Bigg)\psi_a=0.
\ee
These equations can be cast in the form of a  hypergeometric equation through the following substitution
\be
\psi(\xi_a)=\xi_a^{\frac{l_{a-1}}{2}}\Big(1-\xi_a\Big)^{\frac{\omega_{a+1}}{2}}f(\xi_a):
\ee
\be
\xi_a(1-\xi_a)f''+\Big(l_{a-1}+a-\xi_a\Big(l_{a-1}+a+\tilde{\omega}_{a+1}+1\Big)\Big)f'-\frac{1}{4}\Big(l_{a-1}+\tilde{\omega}_{a+1}-l_a)(l_{a-1}+\tilde{\omega}_{a+1}+l_a+2a) \Big)f=0.
\ee
Introducing the following notions
\be
A=\frac{l_{a-1}+\tilde{\omega}_{a+1}-l_a}{2}, \qquad
B= \frac{l_{a-1}+\tilde{\omega}_{a+1}+l_a+2a}{2},
\qquad
C=l_{a-1}+a
\ee
 the equation reduces to the hypergeometric equation.
\be
\xi(1-\xi)f''+\Big(C-\xi\Big(A+B+1)\Big)f'-AB f=0.
\ee

The regular solution of this equation is the hypergeometric function \cite{flu}
\be
f(\xi)=C_0F(A;B;C; \xi)
\ee
Moreover there is requirement for the constants, which yields discrete energy spectrum
\be
 A=-n_a , \qquad n_a=0,1,2,...
\ee
So the solution will have the following form
\be
f_a(\xi)=C_0F(-n_a; l_{a-1},+\tilde{\omega}_{a+1}+a+n_a; l_{a-1}+a; \xi_a),
\ee
\be
 l_{a}=2n_a+l_{a-1}+\tilde{\omega}_{a+1} ,  \label{HyperG}
\ee
with
\be
  \tilde{\omega}_{a}=\sqrt{4\omega_a^2+m_a^2}.
\ee
Therefore,
$
l_N=\sum_{a=1}^N\left( 2 n_a + \tilde{\omega}_{a}\right)$,  so that
 the energy spectrum is given  by the expressions
\be
E_{n,\{m_a\}}=\frac{\hbar^2}{4}\Bigg(2n+N+\sqrt{({B}/{\hbar}+\sum_{a=1}^N m_a)^2+4\omega_0^2}+\sum_{a=1}^{N}\sqrt{4\omega_a^2+m_a^2}\Bigg)^2-
\nonumber
\ee
\be
- \frac{B^2+\hbar^2N^2}{
4}-\hbar^2\sum_{i=0}^N\omega_i^2,
\ee
where $n=\sum_{a=1}^N n_a=0,1,\ldots$
In fact, for the integer parameters $n_a$
the hypergeometric function \eqref{HyperG} is reduced to Jacobi polynomials.

Thus the spectrum of quantum $\mathbb{CP}^N$-Rosochatius system depends on $N+1$ quantum numbers. This is  in full agreement with the fact that
this system has $3N-1$  functionally independent constants of motion
(let us  remind that the spectrum of $D$-dimensional quantum mechanics with $D+K$ independent integrals of motion depends on $D-K$ quantum numbers.
E.g, the spectrum of maximally superintegrable system depends on the single (principal) quantum number).

Let us also write down the explicit expressions for the non-normalized wavefunctions and the  $\mathbb{CP}^N$ volume element
\be
\Psi_{\{n_a\},\{m_a\}}=\frac{C_0}{({2\pi})^{N/2}}\prod_{a=1}^Ne^{\imath m_a\varphi_a} F(-n_a; l_{a-1},+\tilde{\omega}_{a+1}+a+n_a; l_{a-1}+a; \xi_a),
\nonumber\ee
\be
dv_{\mathbb{CP}^N}=\frac{1}{(1+y^2)^{N+1}}\prod_{a=1}^N  y_a d y_a d\varphi_a\,,
\label{PvC}\ee
where
\be
\xi_a=\frac{y_a^2}{y_a^2+y_{a+1}^2}.
\ee
One can write these solutions in the initial complex coordinates using the following relations
\be
y_a=z^a\bar z^a ,\qquad \phi_a =\frac{i}{2} log \frac{\bar z^a}{z^a}
\ee

\subsection{Reduction to quantum (spherical)\newline Rosochatius system}
From the above consideration it is clear that, by fixing the eigenvalues of $\widehat{J}_{a\bar a}=\widehat{p}_{\varphi_{a}}$, we can reduce the Hamiltonians \eqref{qH} and \eqref{rqH}
to those of the quantum (spherical) Rosochatius system, the classical counterpart of which is defined by eq. \eqref{realHam}.

However, the quantization  of \eqref{realHam} through replacing the kinetic term by the Laplacian yields a slightly different expression for the Hamiltonian
\be
\widehat{H}_{Ros}=-\frac{\hbar^2}{4}(1+y^2)\Bigg[\sum_{a,b=1}^{N}(\delta_{ab}+y_a y_b)\frac{\partial^2}{\partial y_a\partial y_b}+\sum_{a=1}^N\left(2y_a\partial_{y_a}+\frac{g_a^2}{y_a^2}\right)+g_0^2\Bigg].
\label{qsRH}\ee
This is because the volume element on $N$-dimensional sphere is different from that reduced from  $\mathbb{CP}^N$:
\be
dv_{S^N}=\frac{1}{(1+y^2)^{(N+1)/2}}\prod_{a=1}^N d y_a ,
\ee
and it gives rise to a different Laplacian as compared to that directly obtained by reduction of the Laplacian on $\mathbb{CP}^N\,$.

As a result,  the relation between wavefunctions of the (spherical) Rosochatius system and those of  $\mathbb{CP}^N$-Rosochatius system is as follows,
\be
\Psi_{sph}=\sqrt{\frac{(1+y^2)^{(N+1)}}{{\prod_{a=1}^N y_a}}}\Psi\,.
\ee
So in order to transform the reduced $\mathbb{CP}^N$-Rosochatius Hamiltonian to the spherical one \eqref{qsRH}, we have to redefine the wavefunctions presented in \eqref{PvC}
 and perform the respective similarity transformation of the Hamiltonian.

\section{Concluding remarks}
In this chapter we proposed  the superintegrable $\mathbb{CP}^N$-analog of Rosochatius and Smorodinsky-Winternitz systems  which is specified by the presence of constant magnetic field
and is form-invariant under transition from one chart of $\mathbb{CP}^N$ to others accompanied by the appropriate permutation  of the characteristic parameters $\omega_i$.
We  showed that the system  possesses $3N-1$ functionally independent constants of motion
and explicitly constructed its classical and quantum solutions. In the generic case this model admits an extension with $SU(2|1)$ supersymmetry,
which is reduced,  under the special choice
of the characteristic parameters and in the absence of magnetic field, to the ``flat'' $\mathcal{N}=4, d=1$ Poincar\'e supersymmetry.

 When all constants $\omega_i$ are equal, the system is covariant under the above transitions between charts and so becomes globally defined
on the whole $\mathbb{CP}^N$ manifold. This covariance implies $N$ discrete symmetries,
\be
z^a\to \frac{1}{z^a},\quad z^\alpha\to\frac{z^\alpha}{z^a},\quad {\rm with }\quad\alpha\neq a.
\ee
Moreover, in this special case the model always admits (in the absence of magnetic field) $\mathcal{N}=4, d=1$ Poincar\'e supersymmetrization. This will be discussed in the next chapter.
The model with equal $\omega_i$ can be also interpreted as  a model of $N$ {\sl interacting } particles with an effective position-dependent mass
located in the quantum ring.  This agrees with the property that, in the flat limit, the model under consideration  can be interpreted  as an ensemble of  $N$  free
particles  in a single quantum ring interacting with a constant magnetic field orthogonal to the plane.
Thus the property of the exact solvability/superintegrability of the suggested model in the presence of constant magnetic field (equally as of the superextended
model implying the appropriate
inclusion of spin) makes it interesting  also from this point of view.

The obvious next tasks are the  Lax pair formulation  of the proposed model and the study of its $SU(2|1)$    supersymmetric extension, both on the classical
and the quantum levels.

Two important possible generalizations of the proposed system are the following ones:
\begin{itemize}
\item  An analog of $\mathbb{CP}^N$-Rosochatius system  on the quaternionic projective space $\mathbb{HP}^N$ in the presence of BPST instanton.

    Presumably, it can be defined by the Hamiltonian \eqref{Hamr0} and the symplectic structure \eqref{hn}, in which  $\pi_a,  z^a$ are replaced by quaternionic variables,
    and  the last term  in \eqref{hn} by terms responsible  for interaction
    with BPST instanton \cite{duval} (see also \cite{KoSmi}, \cite{IKoSmi} and \cite{hpn}). The phase space of this system is expected to be
    ${T}^*\mathbb{HP}^N\times \mathbb{CP}^1$, due to the isospin nature of instanton. We can hope that
    this system is also superintegrable and that  an interaction with BPST instanton  preserves the superintegrability. On this way we can also expect
    intriguing links with the recently explored  Quaternion-K\"ahler deformations of ${\cal N}=4$ mechanics \cite{ILuc}. These models also admit homogeneous
    $\mathbb{HP}^N$ backgrounds.

\item   $\mathbb{CP}^N$-analog of Coulomb problem.

     Such an extension could  be possible, keeping in mind the existence of superintegrable spherical analog  of Coulomb problem with
   additional $\sum_i g^2_i/x^2_i$  potential, as well as the observation that  the (spherical) Rosochatius system  is a real section of
 $\mathbb{CP}^N$-Rosochatius system.

\end{itemize}
One of the key  motivations of the present study  was to derive the superintegrable $\mathbb{CP}^N$- and $\mathbb{C}^N$- generalizations
of rational Calogero model. Unfortunately, until now we succeeded in constructing only trivial extensions of such kind.
We still hope to reach the general goal just mentioned in the future.

\newpage
\chapter{Supersymmetric extensions}

\setcounter{equation}{0}
\setcounter{section}{0}

\section{Introduction}

The following  chapter is based on the article mentioned in the previous chapter \cite{shmros} and another paper which is in progress (with Armen Nersessian, Evgeny Ivanov and  Stepan Sidorov) .

The (planar) Landau problem, that is the planar motion of electrons
in the presence of a constant perpendicular magnetic field,
has been an issue in physics textbooks for a long time~\cite{landau}.
It is extremely simple and relates to various mathematical constructions.
Also, it provides the first physical realization of supersymmetry
(see, e.g.~\cite{GK}). The compact(spherical) analog  of the planar Landau problem is defined as  a particle on the two-sphere in the constant magnetic field generated by a Dirac monopole located in the center and  enjoys an SO(3) invariance.
Similarly, the Landau problem on complex projective spaces is defined as a particle moving on
$\mathbb{CP}^n$ in the presence of constant magnetic field and enjoys the $SU(n+1)$ invariance due to the first Hopf map realized as $S^{2n+1}/S^1=CP^n$.
Quantum mechanically, the inclusion of constant magnetic field cuts the spectrum from below and provide the system by the degenerate ground state.
Thanks to this degeneracy the quantum-mechanical Landau became the  base  of the theory of quantum Hall effect \cite{hall1,hall2} and of its higher-dimensional
generalization  on complex projective spaces \cite{karabali}.

Thus, it is not surprising that there exists ``quaternionic Landau problem''  pertaining to the second Hopf map
$S^7/S^3=S^4$ , which is defined as an isospin particle on a four-sphere
in the field of a BPST instanton (the harmonic part of $SU(2)$ Yang monopole located
at the center of four-dimensional sphere).
Like in the conventional Landau problem, the gauge field configuration
is compatible with the spherical symmetry, in this case $SO(5)$.
It can be further generalized to the Landau problem on quaternionic projective spaces defined as a  particle moving on quaternionic projective space in the presence of constant $SU(2)$-instanton (BPST-instanton) field \cite{hpn}. Due to relation with the second Hopf map realized as a fibration $S^{4n+3}/S^3=HP^n$ this system is $Sp(n+1)$ invariant one.
Some two decades ago, Zhang and Hu proposed a model of the four-dimensional Hall effect
based on  quaternionic Landau problem~\cite{4hall}. Their theory
possesses some qualitatively new features and admits a stringy interpretation
\cite{Fabinger}. It inspired further generalizations of the Hall
effect, for instance on complex projective spaces~\cite{cpn} and on the
eight-sphere (using the third Hopf map $S^{15}/S^7=S^8$)~\cite{octonion}.
There were  numerous publications devoted  to supersymmetric extensions of the Landau problem, and more generally, to the systems on complex  projective spaces interacting with constant magnetic field\cite{hasebe,hasebe2,FIO,KO,BK,SM,klns}.
However, even ${\cal N}=4$ supersymmetric extensions of (two-dimensional) spherical Landau problem are not studied in details \cite{nlin}, while quantum-mechanical ${\cal N}=4$  supersymmetric Landau problem on complex projective spaces is not still considered, to our knowledge, except simplest case of $\mathbb{CP}^1$ \cite{kor}.

{\sl Moreover, all listed ${\cal N}=4$ supersymmetric Landau problems have an important luck: the supersymmetry transformations does not respect the initial $su(n+1)$ symmetry of the  Landau problem on complex/quaternionic. Thus, supersymmetries seemingly decreases the degeneracy of ground state which plays the key role in the construction of Hall effect theory. Thus, one may ask a question:

How one should supersymmetrize the Landau problem, or, more generally, the systems on K\"ahler manifolds
interacting with  constant magnetic fields, in order to preserve their initial symmetries?}

Some preliminary attempts in this direction were performed some fifteen years ago \cite{Kahlerosc}, when it was observed that the oscillator and Landau problem on  complex projective space admit the so-called "weak ${\cal N}=4$  supersymmetry"\cite{smilga} which preserves the initial symmetries of that system.
These results were recently recovered  within curved superfield approach to supersymmetric mechanics  \cite{sidorov1,sidorov2,sidorov3,sidorov4,sidorov5}, where  "weak ${\cal N}=4$  supersymmetry algebra" was identified there with $su(2|1)$ superalgebra. Having in mind the "practical importance" of supersymmetrization respecting initial symmetries, and
field-theoretical importance of "curved superspace approach" \cite{csa1,csa2}, we present here the Hamiltonian approach to the  supersymmetrization of systems in the constant magnetic field.

Namely, we suggest to construct  the ${\cal N}{=}4$ supersymmetric extensions of   Landau problem, including that  on complex  projective spaces  which is
based on the symplectic coupling of the external gauge field to the
supersymmetric system in question.
We find that in the  case of ${\cal N}{=}4$ it yields $SU(2|1)$ supersymmetric system.

We will show that  $\mathbb{CP}^N$-Rosochatius system belongs to the class of  ``K\"ahler oscillators" \cite{cpn,Kahlerosc}  which
admit $SU(2|1)$ supersymmetrization (or a `weak $\mathcal{N}=4$" supersymmetrization, in terminology of Smilga \cite{smilga}).
A few years ago it was found that these systems naturally arise within  the appropriate $SU(2|1), d=1$ superspace formalism
developed in a series of papers. This research was partly motivated by
the study of  the field theories with curved rigid analogs of Poincar\'e supersymmetry \cite{csa1,csa2}.
In the absence of the background magnetic field  and for the special choice of the parameters $\omega_i$, the  $\mathbb{CP}^N$-Rosochatius system
admits $\mathcal{N}=4, d=1$ Poincar\'e  supersymmetric extension.

This chapter is organised in the following way

{\sl Section 5.2} is devoted to the general discussion of  ${\cal N}{=}4$  supersymmetry in K\"ahler manifolds. Namely the structure supersymplectic structure,  Killing potentials for supersymmetric mechanics on  generic K\"ahler manifolds and corresponding Hamiltonian vector fields.

In {\sl Section 5.3} we discuss the free particle in presence of a constant magnetic field (Landau problem) and the related superalgebra.

In {\sl Section 5.4} we extend the discussion via adding  potential. This system is the K\"ahler superoscillator and this formalism is used for constructing the supersymmetric extensions of the systems discussed in previous parts.

In {\sl Section 5.5} we focus on specific examples of K\"ahler superoscillator, namely supersymmetric generalizations of   $\mathbb{C}^N$-Smorodinsky-Winternitz and   $\mathbb{CP}^N$-Rosochatius systems are discussed.

\section{Supersymmetry on K\"ahler manifolds}
To  describe the motion of charged particle on $M$  with the constant magnetic field of  strength $B$  we have to  equip  the cotangent bundle $T^*M$
with the following  symplectic structure and Hamiltonian
\begin{equation}
\omega_B=
d\pi_a\wedge dz^a +
d{\bar\pi}_{a}\wedge d{\bar z}^{a} +\imath B g_{a\bar b}dz^a\wedge
d{\bar z}^b,\qquad { H}_0= g^{a \bar b}\pi_a{\bar \pi}_b  .
\label{ssB}\end{equation}
The isometries of a K\"ahler
 structure define the
Noether's constants of motion of a free particle
 \begin{equation}
{J}_{\mu}=V_\mu^{a}\pi_a +
 {\bar V}_{\mu}^{\bar a} {\bar\pi}_{\bar a} +B\ch_\mu, \qquad  V^a_{\mu}=-\imath g^{a\bar{b}}\partial_{\bar{b}}h_{\mu} \; :\quad
\left\{
\begin{array}{c}
  \{{ H}_0, J_{\mu}\}=0, \\
  \{J_\mu, J_\nu\}=C_{\mu\nu}^\lambda J_\lambda .
\end{array}\right.\label{jmu}\end{equation}
Notice that the vector fields
 generated by ${ J}_\mu$ are independent on $B$
\begin{equation}
{\bf{\tilde V}_{\mu}}=\{J_{\mu},\quad\}_{B}=V_{(\mu)}^a(z)\frac{\partial}{\partial z^a}-V^a_{(\mu),b}\pi_a\frac{\partial}{\partial \pi_b}+\bar{V}_{(\mu)}^{\bar{a}}(\bar{z})\frac{\partial}{\partial \bar{z}^a}-\bar{V}^{\bar{a}}_{(\mu),\bar{b}}\bar{\pi}_a\frac{\partial}{\partial \bar{\pi}_b} .
\label{mommap}\end{equation}
Hence, the inclusion of a
 constant magnetic field preserves
the whole symmetry algebra of a free
particle moving on a K\"ahler manifold, i.e. the Landau problem can be properly defined on the generic K\"ahler manifold.\\
To construct supersymmetric counterpart of the above construction
let us consider  a $(2N.MN)_{C}$-dimensional  phase space
equipped with the symplectic structure
\begin{equation}
\begin{array}{c}
\Omega=d\pi_a\wedge dz^a+ d{\bar\pi}_a\wedge d{\bar z}^a
+\imath(B g_{a\bar b}+\imath R_{a{\bar b}c\bar d}\eta^{c\alpha}\bar\eta^d_\alpha)
dz^a\wedge d{\bar z}^b+
 g_{a\bar b}D\eta^{a\alpha}\wedge{D{\bar\eta}^b_\alpha}\quad,
\end{array}
\label{ss1}\end{equation}
where $D\eta^{a\alpha}
=d\eta^{a\alpha}+\Gamma^a_{bc}\eta^{b\alpha} dz^c, \alpha=1,\ldots M$, and
 $\Gamma^a_{bc},\; R_{a\bar b c\bar d}$ are,
respectively, the components of connection and curvature of
the K\"ahler structure

The Poisson brackets  defining by \eqref{ss1} are given by the expression
\be
\{f,g\}=\frac{\partial f}{\partial\pi_a}\wedge\nabla_a g +\frac{\partial f}{\partial\bar\pi_a} \wedge{\bar\nabla}_a g +\imath(Bg_{a\bar b}+
\imath R_{a\bar b c\bar d}\eta^{c\alpha}{\bar\eta}^d_\alpha)\frac{\partial f}{\partial\pi_a}\wedge\frac{\partial f}{\partial\bar\pi_b}+g^{a\bar b}\frac{\partial^r f}{\partial\eta^{a\alpha}}\wedge\frac{\partial^l g}{\partial\bar\eta^b_\alpha},
\ee
where
\be
\nabla_a\equiv \frac{\partial}{\partial z^a}-\Gamma^c_{ab}\eta^b_\alpha\frac{\partial^r}{\partial \eta^c_\alpha},\qquad f\wedge g= fg-(-1)^{p(f)p(g)}gf
\label{sPB}\ee


\be
\{\pi_a, z^b\}=\delta^b_a,\quad
\{\pi_a,\eta^b_\alpha\}=-\Gamma^b_{ac}\eta^{c\alpha},\quad
\{\pi_a,\bar\pi_b\}=i(Bg_{a\bar b}+
i R_{a\bar b c\bar d}\eta^{c\alpha}{\bar\eta}^d_\alpha),
\nonumber
\ee
\be
\{\eta^{a\alpha}, \bar\eta^b_\beta\}=
g^{a\bar b}\delta^{\alpha}_{\beta}.
\ee
The symplectic structure \eqref{ss1} and Poisson brackets \eqref{sPB}
are manifestly invariant with respect to transformations
\be
{\widetilde z}^a={\widetilde z}^a(z), \qquad {\widetilde\pi}_a=\frac{\partial z^b}{\partial{\widetilde z}^a}\pi_b, \qquad{\widetilde\eta}^{a\alpha}=\frac{\partial{\widetilde z}^a}{\partial z^b}\eta^b_\alpha.
\ee
Hence
we can  lift the isometries \eqref{mommap} to this supermanifold and define the following vector fields, which are Hamiltonian
with respect to Poisson brackets  \eqref{sPB}
\be
 \mathbf{V}_\mu=\{ \mathcal{J}_\mu, \;\}= V^a_\mu (z)\frac{\partial}{\partial z^a}-V^a_{(\mu),b}\pi_a
\frac{\partial}{\partial \pi_b}+V^{a}_{(\mu),b}\eta^{b\alpha}\frac{\partial}{\partial \eta^{a\alpha}}\;+\;{\rm c.c.}\;,
%
 \label{Vs}\ee
with
  \be
{\cal J}_{\mu}={J}_{\mu}-\imath
\frac{\partial^2 {\ch}_{\mu}}{\partial z^c\partial
 {\bar z}^d}\eta^{c\alpha}\bar\eta^{d}_\alpha 
\label{sNeth} \ee
 where $J_\mu$ is defined by \eqref{jmu}.

 With these expressions at hand we are ready to perform the supersymmetrization of Landau problems on K\"ahler manifolds.

\section{$SU(2|1)$ Landau Problem}
For the construction of $\mathcal{N}=4$ Landau problem we  choose standard "chiral" supercharges   $Q^\alpha, {\overline Q}_\alpha $ with $\alpha=1,2$ by the same Ansatz as in the absence of magnetic field  and the generators of additional  $SU(2)$ symmetry given by the $R$-charges
\be
Q^\alpha=\pi_a\eta^{a\alpha}\;,\quad {\overline Q}_\alpha=\bar\pi_a\bar\eta^a_\alpha, \quad\mathcal{R}^\alpha_\beta={\imath} g_{a\bar b}\eta^{a\alpha}\bar\eta^b_\beta-\frac{\imath}{2}\delta^\alpha_\beta g_{a\bar b}\eta^{a\gamma}\bar\eta^b_\gamma\;.
\label{TR}\ee
Closure of their Poisson brackets reads
\be
 \{Q^\alpha, Q^\beta\}=0, \quad   \{\mathcal{R}^\alpha_\beta,\mathcal{R}^\gamma_\delta\}=\imath\delta^\gamma_\beta \mathcal{R}^\alpha_\delta-\imath \delta^\alpha_\delta\mathcal{R}^\gamma_\beta,
\nonumber
\ee
\be
 \{Q^\alpha, \mathcal{R}^\beta_\gamma\}= - \imath\delta^{\alpha}_{\gamma}  Q^\beta+\frac{\imath}{2}\delta_{\gamma}^{\beta}  Q^{\alpha} , \qquad
\nonumber
 \{\overline{Q}_\alpha, \mathcal{R}^\beta_\gamma\}=\imath\delta_{\alpha}^{\beta} {\overline Q}_\gamma-\frac{\imath}{2}\delta_{\gamma}^{\beta} {\overline Q}_{\alpha}
\ee
\be
\{Q^\alpha,{\overline Q}_\beta\}=\delta^{\alpha}_{\beta}\mathcal{H}_0+ B \mathcal{R}^{\alpha}_{\beta},
\quad
\{Q^\alpha,\mathcal{H}_0\}=\frac{\imath B}{2} Q^\alpha,\quad\{\mathcal{R}^\alpha_\beta,\mathcal{H}_0\}=0
\label{R}
\ee
where
 \be
{\cal H}_{0}=g^{ a \bar b}\pi_a \bar\pi_b
-\frac12 R_{a\bar b c\bar d}\eta^{a\alpha}\bar\eta^b_\alpha\eta^{c\beta}\bar\eta^d_\beta +\frac B2
{\imath g_{a\bar b}\eta^{a\alpha}{\bar\eta}^b_\alpha }. \label{hosup1}
\ee
Hence, extending the set \eqref{TR} by the above generator \eqref{hosup1} we  get the $su(2|1)$ superalgebra, or weak $\mathcal{N}=4$ superalgebra.
These generators are obviously invariant under action of \eqref{Vs}
\be
\{Q^\alpha, \mathcal{J}_\mu\}=\{{\overline Q}_\alpha, \mathcal{J}_\mu\}=\{\mathcal{R}^\alpha_\beta,\mathcal{J}_\mu\}=\{\mathcal{H}_0,\mathcal{J}_\mu\}=0,
\ee
i.e. constructed supersymmetric system  inherits all  kinematical symmetries of the initial system. In particular, for the  $\mathbb{CP}^N$-Landau problem the system has a $SU(N+1)$ symmetry.
 Moreover, the last term in the Hamiltonian \eqref{hosup1} is obviously Zeeman term describing interaction of spin with external magnetic field, i.e. our choice of Hamiltonian is physically relevant.
Thus, {\it the generator  \eqref{hosup1} could be considered as a  well defined Hamiltonian of
 "weak  $\mathcal{N}=4$ supersymmetric"  Landau problem on K\"ahler manifold.}

Finally via modification of the initial Hamiltonian we can get the Hamiltonian which is commutative with the supercharges
\be
\widetilde{\mathcal{H}}_0=\mathcal{H}_0+\frac{B}{2}\imath g_{a\bar b}\eta^{a\alpha}\bar\eta^b_{\alpha}\;:\quad \{Q^\alpha, \widetilde{\mathcal{H}}_0\}=\{\mathcal{R}^\alpha_\beta, \widetilde{\mathcal{H}}_0\}=0.
\ee

\section{ $SU(2|1)$ K\"ahler superoscillator}
The K\"ahler oscillator  is defined by the Hamiltonian\cite{Kahlerosc}
\be\label{Kahosc}
{H}_{osc}=g^{a \bar b}\left(\pi_a{\bar \pi}_b +|\omega|^2\partial_a K \;\partial_{\bar{b}}K\right),
\ee
and by the symplectic structure \eqref{ssB}. It is  distinguished system by its respect  to supersymmetrization: inclusion of "oscillator potential" leads minor changes in the supersymmetrization described above.
Preserving the expressions of  $R$-charges \eqref{TR}, we choose the  "dynamical supercharges"
\be
\Theta^\alpha=\pi_a \eta^{a\alpha}+ \imath\bar\omega \bar\partial_a K \epsilon^{\alpha \beta}{\bar \eta}^{a}_{\beta},\quad \overline{\Theta}_{\alpha}=
\bar\pi_a \bar\eta^a_\alpha +\imath\omega\partial_a K  \epsilon_{\alpha \beta} {\eta}^{a \beta},
\ee
where
\be
\overline{\epsilon_{\alpha \beta}}=-\epsilon^{\alpha \beta},\qquad \epsilon^{\alpha\beta}=-\epsilon^{\beta \alpha}, \qquad \epsilon^{12}=1,\qquad  \epsilon^{\alpha\beta} \epsilon_{\beta\gamma}=\delta^{\alpha}_{\gamma}.
\ee
Another important identity should be noted.
\be
\epsilon^{\alpha\beta}\epsilon_{\gamma\delta}=\delta^{\alpha}_{\delta}\delta^{\beta}_{\gamma}-\delta^{\alpha}_{\gamma}\delta^{\beta}_{\delta}
\ee
Calculating  Poisson brackets of supercharges, we  get
\be
 \{\Theta^\alpha,\overline\Theta_\beta\}=\delta^{\alpha}_{\beta}\mathcal{H}_{SUSY}+ B \mathcal{R}^{\alpha}_{\beta}.
\ee

where the Hamiltonian has the following form
\be
{\cal H}_{osc}=g^{ a\bar  b}(\pi_a \bar\pi_b+|\omega|^2\partial_a K \partial_{\bar b} K)
-\frac 12  R_{a\bar b c\bar d}\eta^{a\alpha}\bar\eta^b_\alpha\eta^{c\beta}\bar\eta^d_\beta
\nonumber
\ee
\be
+\frac{\imath}{2}\omega K_{a;b}\eta^{a\alpha}\eta^b_\alpha+
\frac{\imath}{2}\bar \omega K_{\bar a;\bar b}\bar\eta^{a}_\alpha\bar\eta^{b\alpha}+ \frac B2
{\imath g_{a\bar b}\eta^{a\alpha}{\bar\eta}^b_\alpha }, \label{hosup}
\ee

We can compute other commutators
\be
 \{\Theta^\alpha,\Theta^\beta\}=\bar\omega(\epsilon^{\beta \gamma}\mathcal{R}^{\alpha}_{\gamma}+\epsilon^{\alpha \gamma}\mathcal{R}^{\gamma}_{\beta}),
\qquad
 \{\overline \Theta_\alpha,\overline\Theta_\beta\}=-\omega(\epsilon_{\beta \gamma}\mathcal{R}^{\gamma}_{\alpha}+\epsilon_{\alpha \gamma}\mathcal{R}^{\gamma}_{\beta})
\ee
\be
 \{\Theta^\alpha, \mathcal{R}^\beta_\gamma\}= - \imath\delta^{\alpha}_{\gamma}  \Theta^\beta+\frac{\imath}{2}\delta_{\gamma}^{\beta}  \Theta^{\alpha} , \qquad
\nonumber
 \{\overline{\Theta}_\alpha, \mathcal{R}^\beta_\gamma\}=\imath\delta_{\alpha}^{\beta} {\overline \Theta}_\gamma-\frac{\imath}{2}\delta_{\gamma}^{\beta} {\overline \Theta}_{\alpha}
\ee
Here again $\mathcal{R}^{\alpha}_{\beta}$ are $SU(2)$ generators of $R$-symmetry

\be
\mathcal{R}^\alpha_\beta={\imath} g_{a\bar b}\eta^{a\alpha}\bar\eta^b_\beta-\frac{\imath}{2}\delta^\alpha_\beta g_{a\bar b}\eta^{a\gamma}\bar\eta^b_\gamma, \qquad \{\mathcal{R}^\alpha_\beta,\mathcal{R}^\gamma_\delta\}=\imath\delta^\gamma_\beta \mathcal{R}^\alpha_\delta-\imath \delta^\alpha_\delta\mathcal{R}^\gamma_\beta.
\ee

To present this superalgebra in more conventional (and convenient) form let rotate the supercharges as follows
\be
Q^{\alpha}=e^{i\nu/2}\cos\lambda \Theta^{\alpha}+e^{-i\nu/2}\sin\lambda \epsilon^{\alpha \gamma}\overline \Theta_{\gamma} ,\qquad \overline Q_{\alpha}=e^{-i\nu/2}\cos\lambda \overline\Theta_{\alpha}-e^{i\nu/2}\sin\lambda \epsilon_{\alpha \gamma}\Theta^{\gamma}
\ee

where
\be
 \cos 2\lambda=\frac{B}{\sqrt{4|\omega|^2+B^2}},\quad
\sin 2\lambda=-\frac{2|\omega|}{\sqrt{4|\omega|^2+B^2}},
\qquad
\omega=|\omega|e^{i\nu}
\ee
In these terms the symmetry algebra reads

\be
\{Q^{\alpha},Q^{\beta}\}=0, \qquad \{\overline Q_{\alpha},\overline Q_{\beta}\}=0,
\ee
\be
\{Q^{\alpha},\overline Q_{\beta}\}=\delta^{\alpha}_{\beta}{\cal H}_{osc}+\sqrt{4|\omega|^2+B^2} \  \mathcal{R}^{\alpha}_{\beta}
\qquad
\{Q^{\alpha},\mathcal{R}^{\beta}_{\gamma}\}= - \imath\delta^{\alpha}_{\gamma}  Q^\beta+\frac{\imath}{2}\delta_{\gamma}^{\beta}  Q^{\alpha}
\ee
\be
\{Q^\alpha,\mathcal{H}_{osc}\}=\imath \sqrt{|\omega|^2+\frac{B^2}{4}}\ Q^\alpha
\qquad
\{\mathcal{R}^\alpha_\beta,\mathcal{H}_{osc}\}=0
\ee
This is the $SU(2|1)$ supersymmetry algebra.

Let us remind  that K\"ahler potential is defined up to (anti-)holomorphic function, so that the above supersymmetrization involves, not a single Hamiltonian, but a   family of Hamiltonians parameterized by arbitrary holomorphic function.
Namely, replacing the initial K\"ahler potential by the equivalent one,
\be
K(z,\bar z)\to K(z,\bar z)+\frac{1}{\omega}U(z)+\frac{1}{\bar \omega}{\bar U}(\bar z),
\label{gauge}\ee
we will get the family of Hamiltonians formulated on given background,
\begin{gather}
\mathcal{H}_{SUSY}\to{\cal H}_{SUSY}
+
g^{a\bar b}\partial_a U  \partial_{\bar b} U
+\frac{\imath}{2} \;U_{a;b}\eta^{a\alpha}\eta^b_{\alpha}+\frac
{\imath}{2} {\bar U}_{\bar a;\bar b}\bar\eta^{a}_{\alpha}\bar\eta^{b\alpha} 
+ g^{a\bar  b}\left(\bar\omega \partial_{a} K \partial_{\bar b} {U} +\omega\partial_a U \partial_{\bar b} K\right)
. \label{hosup}
\end{gather}
In the  limit $\omega=0$ we  arrive to the well-known Hamiltonian which admits, in the absence of magnetic field,
the $\mathcal{N}=4$ supersymmetry (see, e.g. \cite{PRDrapid}). It is given by the first line in the above expression.

\section{Examples of $SU(2|1)$ K\"ahler\newline  superoscillator }

\subsection{supersymmetric $\mathbb{C}^N$-Harmonic oscillator}
At the first let us consider the system defined by
the K\"ahler potential
\be
K(z,\bar z)=\sum_{a=1}^N\left(z^a\bar z^a+\frac{g_az^az^a}{2\omega}+ \frac{\bar g_a\bar z^a\bar z^a }{2\bar \omega} \right).
\ee
It yields the  K\"ahler oscillator  defined by the Hamiltonian.
\be
\mathcal{H}_{osc}=\sum_{a=1}^N\Big(\pi_a\bar\pi_a+ (\omega\bar\omega+ g_a\bar g_a)z^a\bar z^a+\bar\omega g_az^a z^a+\omega \bar g_a\bar z^a\bar z^a
\ee
\be
+\frac{i}{2}g_a \eta^{a\alpha}\eta^{a}_{\alpha}+\frac{i}{2}\bar{g}_a\bar \eta^{a}_{\alpha}\bar \eta^{a\alpha} +\frac B2 \imath\eta^{a \alpha}\bar\eta^{a}_\alpha\Big)
\ee
Supercharges and $R$-charges have the following form.
\be
\Theta^\alpha=\sum_{a}\Big(\pi_a \eta^{a\alpha}+ \imath(\bar g_a \bar z^a+\bar \omega z^a) \epsilon^{\alpha \beta}{\bar \eta}^{a}_{\beta}\Big) \qquad
\mathcal{R}^\alpha_\beta={\imath} \eta^{a\alpha}\bar\eta^a_\beta-\frac{\imath}{2}\delta^\alpha_\beta \eta^{a\gamma}\bar\eta^a_\gamma
\ee
The canonical Poisson brackets are as follows
\be
\{\pi_a, z^b\}=\delta^b_a,\quad \{\bar \pi_a, \bar z^b\}=\delta^b_a,\quad\{\pi_a, \bar\pi_b\}=\imath B \delta_{a\bar b}, \quad\{\eta^{a\alpha},\bar\eta^b_\beta\}=\delta^{a\bar b}\delta^{\alpha}_{\beta}.
\ee
Diagonalizing this quadratic form we get the potential of $2N$-dimensional oscillator.

For  $\omega=0$ it yields the sum of two-dimensional isotropic oscillators with frequencies $|g_a|$.
Hence,  in the absence of magnetic field is possible to construct the exact $\mathcal{N}=4$ supersymmetric extension  only for the sum of $N$ two-dimensional  oscillators with frequencies $|g_a|$.

Supersymmetric extension of isotropic oscillator is just a sum of bosonic and fermionic parts, so that all constants of motion of the bosonic Hamiltonian become those of fermionic one.
When the ration of frequencies is rational, the hidden symmetries appears in this system, which conserved  in supersymmetric extension as well. Moreover, additional symmetry generators could appear in supersymmetric system depending on fermionic variables only.
Let us illustrate these issues for  the case of isotropic superoscillator.  defined by the potential $K=z\bar z$ and for $\omega=\bar \omega$.
 Its Hamiltonian, dynamical supercharges and $R$-charges decouples to those of two-dimensional isotropic oscillator
\be
\mathcal{H}=\sum_{a=1}^N \mathcal{H}_a,\quad  \Theta^\alpha=\sum_{a=1}^N\Theta^{a\alpha},\qquad R^{\alpha}_{\beta}=\sum_{a=1}^N R^{a\alpha}_{\beta}
\ee
with
\be
\mathcal{H}_a=\pi_a\bar\pi_a+\omega^2z^a\bar z^a+\frac B2 \imath \eta^{a\alpha}\bar \eta^a_\alpha
, \qquad \Theta^{a\alpha}=\pi_a\eta^{a\alpha}+\imath \omega z^a\epsilon^{\alpha\beta}\bar\eta^a_\beta.
\ee
This system has kinematical  $SU(N)$ symmetries acting in the bosonic sector,  $su(N)$ symmetries acting in fermionic sector (which includes, as a subset, the $su(2)$ R-symmetries)
\be
 R_{a\bar b}=\sum_{\alpha}\imath\eta^{b\alpha}\bar \eta^q_\alpha:
 \qquad \{R_{a\bar b}, R_{c\bar d}\}=\imath\delta_{a\bar d}R_{c\bar b} -\imath \delta_{c\bar b }R_{a\bar d},
\ee
and the hidden symmetries
given by the so-called  ``Fradkin tensor":
\be
I_{a\bar b}=\pi_a\bar\pi_b+\omega^2\bar z^a z^b\; :
\ee
\be
\{I_{a\bar b},I_{c\bar d}\}=\imath \delta_{a\bar d}J_{c\bar b}-
\imath \delta_{c\bar b}J_{a\bar d},\qquad \{I_{a\bar b},J_{c\bar d}\}=\imath\omega \delta_{a\bar d}I_{c\bar b}-
\imath\omega \delta_{c\bar b}I_{a\bar d}.
\ee

Now, we are ready to consider less trivial example of $SU(2|1)$ supersymmetric K\"ahler oscillator with hidden symmetry.
\subsection{Supersymmetric  $\mathbb{C}^N$-Smorodinsky-Winternitz }
Let us consider  K\"ahler superoscillator   underlined by the $\mathbb{C}^N$-Smorodinsky-Winternitz system. We define it by the K\"ahler potential
\be
K=z\bar z+\frac{ g_a}{\omega}\log{z^a}+\frac{{\bar g}_a}{\bar\omega}\log {{\bar z}^a}.
\ee
In that case the Hamiltonian decouples to the sum of $N$  weak supersymmetric  $\mathbb{C}^{1}$-Smorodinsky-Winternitz systems,
\be
\mathcal{H}_{SW}=\sum_{a=1}^N\mathcal{I}_a,
\ee
where
\be
 \mathcal{I}_a= \pi_a\bar{\pi}_a+|\omega|^2 z^a\bar{z}^a+\frac{|g_a|^2}{z^a\bar z^a}+ \omega \bar g_a+\bar\omega g_a-\frac{ig_a}{2}\frac{\eta^{a\alpha}\eta^{a}
_{\alpha}}{z^az^a}-
 \frac{i\bar g_a}{2}\frac{\bar\eta^a_{\alpha}\bar\eta^{a\alpha}}{{\bar z}^a\bar z^a}+\frac{B}{2}\imath\eta^{a\alpha}\bar\eta^a_\alpha
\ee
We can also present the expressions for supercharges and  $su(2)$ supercharges.
\be
 \Theta^{\alpha}=\sum_a\Big(\pi_a\eta^{a\alpha}+\imath\Big(\bar \omega z^a+\frac{\bar g_a}{\bar z^a}\Big)\epsilon^{\alpha\beta}\bar\eta^a_\beta\Big),
\qquad
\mathcal{R}^\alpha_\beta={\imath} \eta^{a\alpha}\bar\eta^a_\beta-\frac{\imath}{2}\delta^\alpha_\beta \eta^{a\gamma}\bar\eta^a_\gamma
\ee
In this case supersymplectic structure  has the same form as for the previous system.
\be
\{\pi_a, z^b\}=\delta^b_a,\quad \{\bar \pi_a, \bar z^b\}=\delta^b_a,\quad\{\pi_a, \bar\pi_b\}=\imath B \delta_{a\bar b}, \quad\{\eta^{a\alpha},\bar\eta^b_\beta\}=\delta^{a\bar b}\delta^{\alpha}_{\beta}.
\ee
Clearly, that $\mathcal{I}_a$ commutes with each other, and defines the constants of motion of the supersymmetric $\mathbb{C}^{N}$-Smorodinsky-Winternitz system.
The system possesses $N$ manifest $U(1)$ symmetries  $z^a\to {\rm e}^{\imath\kappa}, \eta^a_\alpha\to {\rm e}^{\imath\kappa}\eta^a_\alpha$ given by the generators
\be
\mathcal{J}_{a\bar a}=J_{a\bar a}+\imath \bar\eta^a_\alpha\eta^{a\alpha}\; :\qquad \{\mathcal{J}_{a\bar a}, \mathcal{J}_{b\bar b}\}=\{\mathcal{J}_{a\bar a}, \mathcal{I}_{b}\}=0
\ee
where
\be
J_{a\bar a}=\imath\pi_a z^a-\imath\bar{\pi}_a\bar{z}^a-B\frac{ z^a{\bar z}^a}{1+z{\bar z}}
\ee

\newpage

\subsection{Supersymmetric $\mathbb{CP}^N$-Rosochatius}

Let us briefly discuss the possibility of supersymmetrization of $\mathbb{CP}^N$-Rosochatius system.
The $\mathbb{CP}^N$-Rosochatius system belongs to the class of the so-called  ``K\"ahler oscillators" \cite{cpn,Kahlerosc} (up to a constant shift of the Hamiltonian),
and therefore, admits $SU(2|1)$ (or, equivalently, ``weak $\mathcal{N}=4$") supersymmetric extension.
Namely, its Hamiltonian \eqref{Ham} can be cast in the form
\be
\mathcal{H}_{Ros}=g^{a\bar b}\left(\pi_a\bar\pi_b + |{\omega}|^2\partial_{a}K\partial_{\bar a}K\right)-E_{0},
\label{Kosc}\ee
with
\be
 K=\log(1+z\bar z)-\frac{1}{|\omega|}\sum_{a=1}^N ({\omega_a}\log z^a +{{\bar \omega}_a}\log {\bar z}^a),\quad \omega={\omega_0+\sum_{a=1}^N\omega_a},
\label{suprel}\ee
\be
 E_{0}=|\sum_{i=0}^N\omega_i|^2-\sum_{i=0}^N|\omega_i|^2
\ee
Here, as opposed to the previous Sections, we assume that $\omega_i$ are complex numbers, {\it i.e.} we replaced
\be
\omega_i\to\omega_i{\rm e}^{\imath\nu_i},
\ee
 with $\nu_i$ being arbitrary real constants.

The $SU(2|1)$ superextension is reduced to that with $\mathcal{N}=4, d=1 $  Poincar\'e supersymmetry under the conditions.
\be
B=0\,,\quad \omega=\sum_{i=0}^N\omega_i=0.
\label{Bo0}\ee
From the viewpoint of $SU(2|1)$ mechanics, $B$ is just the parameter of contraction to $\mathcal{N}=4$ $ d=1$ supersymmetry.
One could expect that the second constraint corresponds to the vanishing   potential.
 However, it is not the case: looking at the explicit expression for  the Hamiltonian, one can see that the  parameter $\omega$ does not appear
 in denominators anymore.

 Indeed, the second constraint above leads the relation $|\omega_0|^2=|\sum_{a=1}^N\omega_a|^2$, which allows
  to represent  the Hamiltonian \eqref{Ham} in the following form
 \be
 \mathcal{H}_{Ros}=\sum_{a,b=1}^N g^{\bar a b}\left(\bar\pi_a\pi_b +\partial_{\bar a}{\bar U}\partial_{b}U\right)-\sum_{i=0}^N|\omega_i|^2
 \label{hHam}\ee
and  $U(z)$ be  the holomorphic function (``superpotential")
\be
{U}(z)=\sum_{a=1}^N\omega_a\log z^a.
\ee
It is well-known that the systems with such Hamiltonian admit  the $\mathcal{N}=4$ supersymmetric extension  in the absence  of magnetic field (see, e.g., \cite{PRDrapid}).
Explicitly it looks as follows.

Let us consider  a $(2N.4N)_{C}$-dimensional  phase space
equipped with the symplectic structure
\begin{equation}
\begin{array}{c}
\Omega=d\pi_a\wedge dz^a+ d{\bar\pi}_a\wedge d{\bar z}^a
-\frac{1}{2}R_{a{\bar b}c\bar d}\eta^c_\alpha\bar\eta^{d\alpha}
dz^a\wedge d{\bar z}^b+
\frac12 g_{a\bar b}D\eta^a_\alpha\wedge{D{\bar\eta}^b_\alpha}\quad,
\end{array}
\label{ss}\end{equation}

The Poisson brackets  defined by \eqref{ss} are given
by the following non-zero
relations and their complex-conjugates:
\be
\{\pi_a, z^b\}=\delta^b_a,\quad
\{\pi_a,\eta^b_\alpha\}=-\Gamma^b_{ac}\eta^c_\alpha,\quad
\{\pi_a,\bar\pi_b\}=- R_{a\bar b c\bar d}\eta^c_\alpha{\bar\eta}^{d\alpha},
\quad
\{\eta^a_\alpha, \bar\eta^{b\beta}\}=
g^{a\bar b}\delta_{\alpha}^{\beta}.
\ee
We can define the Hamiltonian and the supercharges
\be
Q^\alpha=  \pi_a \eta^{a\alpha}+ \imath {\bar U}_{,\bar a}{\bar \eta}^{a\alpha},\quad {\overline Q}_{\alpha}=
\bar\pi_a \bar\eta^a_\alpha +\imath U_{,a} {\eta}^a_\alpha,
\nonumber
\ee
\be
\mathcal{H}_{SUSY}=\mathcal{H}_{Ros}
-\frac12 R_{a\bar b c\bar d}\eta^{a\alpha}\bar\eta^b_\alpha\eta^{c\beta}\bar\eta^d_\beta
+\frac{\imath}{2} U_{,a;b}\eta^{a\alpha}\eta^b_\alpha+
\frac{\imath}{2} {\bar U}_{,\bar a;\bar b}\bar\eta^{a\alpha}\bar\eta^b_\alpha\quad
\ee
Straightforward calculations show that the following supercharges and Hamiltonian obey the $(\mathcal{N}=4, d=1)$ Poincare superalgebra
\be
\{Q^{\alpha},{\overline Q}_\beta\}=\delta^\alpha_{\beta}\left(\mathcal{H}_{SUSY}+ \sum_{i=0}^N|\omega_i|^2\right),
\nonumber
\ee
\be
\{Q^{\alpha},Q^\beta\}= \{{\overline Q}_{\alpha},{\overline Q}_\beta\}=\{Q^{\alpha},\mathcal{H}_{SUSY}\}=\{ {\overline Q}_{\alpha},\mathcal{H}_{SUSY}\}=0,
\ee
Hence, with the  constraint \eqref{Bo0} imposed,   we can construct the  $\mathcal{N}=4$ supersymmetric extension of $\mathbb{CP}^N$-Rosochatius system.

An interesting question is the symmetries of constructed supersymmetric system. Writing down the explicit expressions for the Hamiltonian and supercharges one can see
that they are explicitly invariant under $U(1)$-transformations $z^a\to{\rm e}^{\imath\kappa}z^a, \pi_a\to{\rm e}^{-\imath\kappa}\pi_a, \eta^{a\alpha}
\to{\rm e}^{\imath\kappa}\eta^{a\alpha}$ which are obviously, canonical transformations.Hence, one can easily construct the ``supersymmetric counterpart" of $U(1)$  generators. However, to the moment we are unable to answer the question weather hidden symmetries of the  system can be lifted to the supersymmetric extension of the model.

Let us emphasize that the restriction $\omega=0$ can be graphically represented  as a planar  polygon  with the edges $|\omega_i|$ (see Fig.1),
which leads to the inequality
\be
|\omega_i|\leq \sum_{j\neq i}|\omega_j|. \label{InEq}
\ee
\begin{tikzpicture}
 \draw[thick,->] (0,0) -- (0,3);

 \draw[thick,<-] (2,0) -- (2,3);

\draw (-0.5,1.7) node {$\omega_0$};
\draw (1.5,1.7) node {$\omega_1$};

  \draw[thick,->] (5,0.3) -- (6.5,3) ;

    \draw[thick,->]  (6.5,3)--(7,0.3) ;

  \draw[thick,->] (7,0.3) -- (5,0.3) ;

\draw (5.2,1.7) node {$\omega_0$};
\draw (7.2,1.7) node {$\omega_1$};
\draw (6,0) node {$\omega_2$};

 \draw[thick,->] (10.5,0.3) -- (11.5,2);

    \draw[thick,->]  (11.5,2)--(13.5,3);

  \draw[thick,->] (13.5,3) -- (14.5,0.3) ;

 \draw[thick,->] (14.5,0.3) -- (10.5,0.3);

\draw (10.8,1.4) node {$\omega_0$};
\draw (12.5,2.8) node {$\omega_1$};
\draw (14.3,1.7) node {$\omega_2$};
\draw (12.5,0) node {$\omega_4$};
\end{tikzpicture}
\vspace{0.2cm}
\begin{center}
Fig.1
\end{center}
\noindent This implies that:
\begin{itemize}
\item For $N=1$  the constraint $\omega=0$ uniquely fixes the values of parameters in the case of $\mathbb{CP}^1$: $ \nu_0=-\nu_1$  and $|\omega_0|=|\omega_1|$.
 The latter property leads to the appearance of discrete symmetry
\be
z\to \frac{1}{z}.
\label{discrete}
\ee

\item For  $N=2$ the above constraints amount to a triangle, which  fixes the parameters $\nu_a$ as follows
\be
\cos{(\nu_2-\nu_0)}=\frac{|\omega_1|^2-|\omega_0|^2 -|\omega_2|^2}{2|\omega_0||\omega_2|},\quad\cos{(\nu_1-\nu_0)}
=\frac{|\omega_2|^2-|\omega_0|^2 - |\omega_2|^2}{2|\omega_0||\omega_1|}\,.\quad
\ee
\item  For $N>2$ the parameters $\nu_a$ are not uniquely fixed, so that we obtain a family of $\mathcal{N}=4$ supersymmetric Hamiltonians depending on up to $N-1$ parameters.
\end{itemize}
We observe that for any value of $N$  at least one parameter  $\nu_i$ remains unfixed. But this does not affect our consideration  since
such parameter can be absorbed into a redefinition of  fermionic variables.

Finally, note that
the constraint $\sum_{i=1}^N\omega_i=0$  also appeared  in constructing the $\mathcal{N}=4$ supersymmetric extension of  $\mathbf{S}^N$-Rosochatius system
\cite{wdvv}, but  with $\omega_i$ being real numbers. The above trick   with complexification of the parameters $\omega_i$ is  certainly
applicable to the $\mathbf{S}^N$-Rosochatius system, giving rise to a less restrictive form of the $\mathcal{N}=4$ superextension of the latter.

\section{Concluding remarks}

In this chapter we have discussed Supersymmetric generalizations of $\mathbb{C}^N$-Smorodinsky-Winternitz and   $\mathbb{CP}^N$-Rosochatius models. For this purpose we have introduced   $SU(2|1)$ supersymmetrization which allows to construct weak ${\cal N}=4$ superextensions of systems on K\"ahler manifolds interacting with an external magnetic field.
First of all we have discussed $SU(2|1)$-Landau problem (system without an external potential). After this we have introduced $SU(2|1)$-K\"ahler oscillator. Using this formalism  One can find many supersymmetric models on K\"ahler manifolds using the fact that all these systems can be viewed as $SU(2|1)$-K\"ahler oscillator with different K\"ahler potentials. Then we have shown  K\"ahler potentials which give rise to $SU(2|1)$-Supersymmetric $\mathbb{C}^N$-Smorodinsky-Winternitz and  $SU(2|1)$-Supersymmetric $\mathbb{CP}^N$-Rosochatius systems.

\chapter*{Conclusion }

To sum up we will briefly discuss the main results of this thesis.

{\sl First Chapter} is an introduction and some general concepts are discussed. First of all we give a brief discussion of Hamiltonian mechanics. We discuss  well known examples of maximally superintegrable models, namely the oscillator and Coulomb systems. We discuss mechanical models interacting with an external magnetic field, and introduce action angle variables.
Moreover we give a short review on K\"ahler manifolds and discuss maximally symmetric examples of it, namely complex  Euclidean and complex projective spaces. Finally  we give  a short description of supersymmetric mechanics.

{\sl Second Chapter} is devoted to holomorphic factorization formalism. This formalism allows to describe generalizations of Coulomb and oscillator models via introduction of complex variables. First of all we discuss this scheme on well known examples of TTW and PW systems. Then we  do this for higher dimensional cases. We  do the so called oscillator-Coulomb reduction procedure using the holomorphic factorization formalism. Moreover we discuss also curved spaces namely the spherical and pseudospherical generalizations. Finally we describe some examples of superintegrable models using this formalism.

In the {\sl Third Chapter} we concentrate on the complex analogue of the Smorodinsky-Winternitz system interacting with an external magnetic field. Firstly we discuss the usual real $N$-dimensional Smorodinsky-Winternitz system. The main result
we have obtained for the real case is the  convenient form of the symmetry algebra.  Then we introduce the complex analogue of this system, and write down the its hidden symmetries. We also obtain important result for this model, namely the symmetry algebra and quantum solutions.  Eventually  we compute the symmetry algebra for the generalized MICZ-Kepler system using the results we have obtained before for the  $\mathbb{C}^2$-Smorodinsky-Winternitz system.

In the {\sl Fourth Chapter}  we introduce the complex projective analogue of the Rosochatius system in an external magnetic field.  Here again we see that it is superintegrable, since it has hidden constants of motion. We write have found also its symmetry algebra, classical and quantum solutions. Namely we find solutions for the classical equations of motion, wavefunctions and the  energy spectrum.

Finally in the {\sl Fifth Chapter} we formulate the $SU(2|1)$-Supersymmetric mechanics. We describe the  $SU(2|1)$-Landau problem (supersymmetric particle moving on a K\"ahler manifold with en external magnetic field). Then  $SU(2|1)$-Superoscillator is discussed.   Via this we construct  ${\cal N}=4$ supersymmetric extensions of the   $\mathbb{C}^N$-Smorodinsky-Winternitz  and $\mathbb{CP}^N$-Rosochatius models.

\end{document}